%% file: paper.tex
\documentclass[aps,prd,amsmath,amssymb,nofootinbib,showpacs,reprint]{revtex4-1}
\input{header}
\begin{document}
\title{Critical flavour number of the Thirring model in three dimensions}

\newcommand{\FSU}{Theoretisch-Physikalisches Institut, Friedrich-Schiller-Universit{\"a}t Jena, 
07743 Jena, Germany}
\newcommand{\JLU}{Institut f\"ur Theoretische Physik, Justus-Liebig-Universit\"at Giessen, 35392 Giessen, Germany}

\author{Bj\"orn H. Wellegehausen}
\email{bjoern.wellegehausen@uni-jena.de}
\affiliation{\JLU}
\affiliation{\FSU}

\author{Daniel Schmidt}
\email{d.schmidt@uni-jena.de}
\affiliation{\FSU}

\author{Andreas Wipf}
\email{wipf@tpi.uni-jena.de}
\affiliation{\FSU}

\begin{abstract}
\noindent 
The Thirring model is a four-fermion theory with a current-current interaction and $U(2\Nr)$ chiral symmetry. 
It is closely related to three-dimensional QED and other models used to describe properties of graphene. 
In addition it serves as a toy model to study chiral symmetry breaking. 
In the limit of flavour number $\Nr \to 1/2$ it is equivalent to the Gross-Neveu model, which shows a parity-breaking discrete phase transition.
The model was already studied with different methods, including Dyson-Schwinger
equations, functional renormalisation group methods and lattice
simulations.
Most studies agree that there is a phase transition from a symmetric phase to a spontaneously broken phase for a small number of fermion flavours, but no symmetry breaking for large $\Nr$.
But there is no consensus on the critical flavour number $\Nrc$
above which there is no phase transition anymore and on
further details of the critical behaviour.
Values of $\Nr$ found in the literature vary between $2$ and $7$.

All earlier lattice studies were performed with staggered fermions. Thus it is questionable if in the continuum limit the lattice model recovers the internal symmetries of the continuum model.
We present new results from lattice Monte Carlo simulations of the Thirring model 
with SLAC fermions which exactly implement \emph{all} internal
symmetries of the continuum model 
even at finite lattice spacing.
If we reformulate the model in an irreducible representation of the Clifford algebra, we find, in contradiction to earlier results, that the behaviour for even and odd flavour numbers is very different:
For even flavour numbers, chiral and parity symmetry are always unbroken. 
For odd flavour numbers parity symmetry is spontaneously broken below the critical flavour number $\Nic=9$ while chiral symmetry is still unbroken.

\end{abstract}


\maketitle

\input{intro.tex}
\input{symm.tex}

\input{eff_pot.tex}
\input{symm_pot.tex}
\input{strong_coupling.tex}
\input{results.tex}
\input{conclusions.tex}

\begin{acknowledgments}
We are gratefull to Shailesh Chandrasekharan, Holger Gies, Lukas Janssen, Simon Hands,
Rajamani Narayanan and Urs Wenger for helpful discussions and comments.
This work was supported by the Helmholtz International Center for FAIR within the LOEWE initiative of the State of Hesse.
 D.\ S.\ was supported by the graduate school GRK 1523/2. Simulations were performed on the LOEWE-CSC at the University of Frankfurt and on the HPC cluster at the University of Jena.
 
\end{acknowledgments}

\appendix 
\input{appendix}

\renewcommand{\eprint}[1]{ \href{http://arxiv.org/abs/#1}{[arXiv:#1]}}


\bibliography{paper}


\end{document}

%% file: header.tex
\usepackage[utf8]{inputenc}

\usepackage{graphicx}
\usepackage{slashed}
\usepackage{tabularx}
\usepackage{stackrel}
\usepackage[british]{babel}
\usepackage[hyperref]{xcolor}

\usepackage{amsfonts}
\usepackage{amsmath}
\usepackage{amssymb}
\usepackage{amsthm}
\usepackage[fixamsmath, disallowspaces]{mathtools}
\usepackage{fixmath}		
\usepackage{bbm}		
\usepackage{mleftright}		

\usepackage{microtype}
\usepackage[pdftex]{hyperref}
\usepackage{nicefrac}

\definecolor{darkgreen}{rgb}{0,0.73,0}

\newcommand{\dx}{\mathrm{d}}

\newcommand{\ii}{\mathrm{i}}
\newcommand{\Nr}{N}
\newcommand{\Ni}{N_\text{ir}}
\newcommand{\Nrc}{{N^\text{cr}}}
\newcommand{\Nic}{{N_\text{ir}^\text{cr}}}
\newcommand{\lc}{\lambda_\text{c}}
\newcommand{\ex}[1]{\mathrm{e}^{#1}}
\newcommand{\srm}[1]{_\text{#1}}

\newcommand{\erw}[1]{\left \langle #1 \right \rangle}

\newcommand{\abs}[1]{\left \vert #1 \right \vert}

\newcommand{\algebra}[1]{\mathfrak{#1}}

\newcommand{\id}{\mathbbm{1}}

\newcommand{\cP}{\mathcal{P}}
\newcommand{\cQ}{{\hskip-.25mm\mathcal{Q}}}
\newcommand{\cS}{\mathcal{S}}

\newcommand{\ft}[2]{{\textstyle\frac{#1}{#2}}}
\DeclareMathOperator{\tr}{tr}

\DeclareMathOperator{\diag}{diag}

\newcommand{\vecf}[1]{\mathbf{#1}}

\DeclareMathOperator{\Vol}{Vol}

\graphicspath{{./}{./plots/}{./publPlots/}}

%% file: intro.tex
\section{Introduction}
\noindent
The Thirring model \cite{Thirring1958}  is a fermionic quantum field theory
with a current-current interaction.
While it was originally studied in two spacetime dimensions, a lot of recent works concern its three-dimensional version with a varying number of $\Nr$ flavours.
This model is renormalisable in a $\nicefrac{1}{\Nr}$-expansion for $2<d<4$ \cite{Parisi1975,Gawedzki:1985ed,Rosenstein:1990nm,Hands1995}.
Its Lagrangian in Euclidean spacetime is given by
\begin{equation}
  \mathcal{L}=\bar\Psi_a \ii\Gamma^\mu\partial_\mu\Psi_a-\frac{g^2}{2\Nr}\left(\bar\Psi_a\Gamma_\mu\Psi_a\right)^2 \quad a = 1,\dots,\Nr
  \label{e:th_orig_lagrangian}
\end{equation}
with summation over fermion flavours.
In three dimensions, an irreducible representation of the Clifford algebra is two-dimensional, 
but we will start with a reducible representation here and take the well-known 
$\Gamma_\mu$-matrices of the four-dimensional theory with four-component spinor fields
$\Psi_a$.
This is motivated by a strong similarity to three-dimensional QED~\cite{Hands1995,Itoh1995},  that is often used to model electronic properties of materials like graphene~\cite{Semenoff1984,Hands2008} or high-temperature superconductors~\cite{Herbut2002,Franz2002}.

The Thirring model is also interesting on its own, because it has a large continuous chiral symmetry.
It is believed, that spontaneous breaking of this symmetry can happen with the pattern
\begin{equation}
  U(2 \Nr)\rightarrow U(\Nr)\otimes U(\Nr),
  \label{e:th_breaking_pattern}
\end{equation}
but only for a small number of flavours.
A critical flavour number $\Nrc$ should exist, such that chiral symmetry breaking (csb) only occurs for $\Nr < \Nrc$.
The main focus of our work is to find the value of $\Nrc$ for the reducible Thirring model.

While there is a broad agreement on this general behaviour in the literature, the predictions for $\Nrc$ vary to some extent.
Numerous works employing a large-$\Nr$ expansion are available: 
There are different studies using Dyson-Schwinger equations (DSE), the first \cite{Gomes1991} reporting $\Nrc\approx3.24$.
Later work \cite{Itoh1995, Sugiura1997} found $\Nrc\approx 4.32$ in the limit $g^2\rightarrow\infty$.
By constructing an effective potential by an inversion method, \textcite{Kondo1995} 
reports $\Nrc=2$ for infinite Thirring coupling.
Additionally, these works report relations between chiral condensate, 
$\Nr$ and $g^2$, that are qualitatively not in agreement with each other.

A recent extensive study \cite{Gies2010, Janssen2012a} of four-fermion theories with
functional renormalisation group methods spotted a structure with three 
interacting fixed points in the plane spanned by Thirring and Gross-Neveu coupling.
The fixed point governing the critical behaviour of the Thirring model is only on the axis of pure Thirring interaction for $\Nr\rightarrow\infty$, while it is off the axis for any finite $\Nr$.
Contrary, for small $\Nr$ this fixed point is dominated by another four-fermion interaction, showing dynamical generation of a fermion mass.
This is not the case for large $\Nr$.
Balancing this competition, the authors find $\Nrc\approx 5.1(7)$.

Regarding lattice field theory, many simulations with staggered fermions and a small mass were performed so far.
With a setup using the standard Hybrid Monte Carlo (HMC) algorithm, simulations are only possible with an integer number of flavours. Since for staggered fermions each lattice flavour corresponds 
to two continuum flavours \cite{Burden1987}, only simulations with even $\Nr$ are possible.
The first results \cite{Kim1996} report a change of the chiral behaviour for $2 < \Nrc < 6$.
Another series of publications \cite{DelDebbio1996,DelDebbio1997,DelDebbio1999} used the same algorithm with a slightly different action and found chiral symmetry breaking with a second order phase transition for $\Nr=2$ and $\Nr=4$, while the transition for $\Nr=6$ is different.
The authors claim that in the latter case there is a first order transition with coexisting symmetric and broken phases.
They conclude that $\Nrc$ is between 4 and 6.
Simulations with a hybrid molecular dynamics (HMD) algorithm were performed in
\cite{Hands1999}, allowing also for odd and non-integer values of $\Nr$.
The authors present a phase diagram in the $(\Nr,g^2)$ plane with a critical line.
Along this line, the critical exponents and the order of phase transition changes from a second order transition at $\Nr=4$ to a first order transition at $\Nr=6$. 
For $\Nr=5$ the simulations did not lead to a conclusive result.

In a more recent lattice study \cite{Christofi2007} with this setup, simulations in the limit
$g^2\to\infty$ were performed, in order to compare the results with those from DSE approaches.
To ensure transversality of the vacuum polarisation tensor for large $\Nr$
on the lattice, a renormalisation of the coupling was necessary.
The renormalised coupling at leading order in $\nicefrac{1}{\Nr}$ is
\begin{equation}
  g_\mathrm{R}^2 = \frac{g^2}{1-g^2 J(m)} \quad \text{with}\quad J(m)\rightarrow \frac{2}{3},
  \label{e:renorm_coupling}
\end{equation}
where the integral $J(m)$ is given in \cite{DelDebbio1997}.
Obviously, $g_\mathrm{R}^2$ gets negative, if the bare coupling $g^2$ is larger than $\nicefrac{3}{2}$. Thus the strong coupling limit is reached for finite bare coupling.
For stronger bare couplings, an unphysical phase is present.
All these works found a non-monotonic behaviour of the chiral condensate and it is argued, that its maximum corresponds to the point where the renormalised coupling becomes negative, although the coupling at the maximum does not match the value of $\nicefrac{3}{2}$.
Looking at the maximal value of the chiral condensate for different $\Nr$ at small masses, 
the authors concluded that $\Nrc=6.6(1)$.

The fermion bag approach was applied by \textcite{Chandrasekharan2010} 
to study the Thirring model  with a single staggered flavour, corresponding to $\Nr=2$, 
and to obtain critical exponents for the csb phase transition~\cite{Chandrasekharan2011,Chandrasekharan2012}.
This was the first lattice work in the chiral limit $m=0$.
But since staggered fermions do not preserve all internal symmetries it is not clear, if the correct symmetry breaking pattern \eqref{e:th_breaking_pattern} is recovered in the continuum limit.
In a subsequent work the authors \cite{Chandrasekharan2013a} observe, that their lattice version 
of the Thirring model has the same symmetry and critical exponents as the Gross-Neveu model.
This seems to contradict our knowledge about the continuum models.
More recently, a first study with domain wall fermions was done \cite{Hands2016}.
Contrary to the older works, no remnant of csb was found in the extrapolation $m\rightarrow 0$ for $\Nr=2$. Very recent preliminary results \cite{Hands2017} for $\Nr=1$ showed no csb either.

In the present work we follow an alternative route and simulate the Thirring model 
with chiral  SLAC fermions \cite{Drell1976,Drell1976a}. It is well-known that these
fermions should not be used in lattice gauge theories \cite{Karsten1979}.
But they have been successfully applied to simulate supersymmetric Yukawa models \cite{Kirchberg2005, Bergner2008, Kastner2008, Bergner2010,Bergner2009a}, where also the renormalisability of lattice perturbation theory with SLAC fermions up to one-loop was established.
With these fermions the step scaling function of the non-linear O(3) sigma model
has been calculated to high accuracy on moderately large lattices, see \cite{Flore:2012xj}.
Thus there are good reasons to believe that they work well for four-fermion theories.
SLAC fermions admit the exact internal symmetry at any finite lattice spacing.
Nevertheless, it was difficult to establish csb at any $\Nr$ in first investigations
\cite{Schmidt2015,Schmidt2016}.
Including an explicit breaking, a physical phase transition was observed, but it seems to merge with the artefact transition explained around \eqref{e:renorm_coupling}, leaving no reliable trace of csb when performing the limit to the massless Thirring model.
An attempt to use Fierz identities to reformulate the Thirring model showed a very strong sign problem preventing HMC simulations \cite{Schmidt2016}.
The present work will present an approach to circumvent these problems and
gives a definite answer about the existence and values of the critical flavour number.
We shall see that there is no symmetry breaking in all models with integer $\Nr$.
This is in complete agreement with the absence of bilinear condensation in three-dimensional
QED with an even number of massless irreducible flavours \cite{Karthik:2015sgq}.
In contrast, we present numerical evidence that all models with half-integer $\Nr\leq \Nrc=9/2$ 
show a symmetry breaking.

To ease the computations, we also consider the Thirring model in the irreducible representation, where the chiral symmetry is merely a flavour symmetry.
Then, dynamical generation of a fermion mass is associated to a spontaneous breaking of parity.
This was studied by DSE in the large-$\Nr$ limit, where a dynamically generated mass was found \cite{Hong1994}, implying $\Nrc\rightarrow \infty$.
Employing Fierz identities when computing the effective potential 
a different parity breaking pattern emerged in \cite{Ahn1994,Ahn1998}:
a dynamical mass generation for two and three irreducible flavours
is seen, whereas the potential becomes unbounded from below
for $\Nr\rightarrow\infty$.
Similarly, in the functional Schrödinger picture no symmetry breaking in the 
large $\Nr$ limit was found, while it appears when higher-order corrections in $\nicefrac{1}{\Nr}$ 
are included \cite{Hyun1994}. 

With the help of an auxiliary vector field $v_\mu$ the Lagrangian (\ref{e:th_orig_lagrangian}) can be written in the equivalent form
 \begin{equation}
  \mathcal{L}=\bar\Psi_a\,\ii \Gamma^\mu D_\mu\Psi_a+\lambda v_\mu v^\mu,\quad \lambda=\frac{\Nr}{2g^2}\label{eqLagrangian}
 \end{equation}
with 'covariant derivative' $D_\mu=\partial_\mu-\ii v_\mu$.
In an adapted base the reducible matrices
$\Gamma_\mu$ and reducible spinors take the form
\begin{equation}
 \Gamma_\mu=\sigma_3\otimes\gamma_\mu,\quad
 \Psi_a=\begin{pmatrix}\psi_{1,a}\\ \psi_{2,a}\end{pmatrix},\;\;
 \bar\Psi_a=\big(\bar\psi_{1,a},\bar\psi_{2,a}\big)\,,\label{red:matrices}
\end{equation}
where  the two-dimensional $\gamma_\mu$ form an irreducible 
representation of the Clifford algebra in three dimensions. For massless fermions the overall sign 
of $\slashed{D}=\gamma^\mu(\partial_\mu-\ii v_\mu)$ in 
\begin{equation}
  \mathcal{L}=\bar\psi_{1,a}\,\ii \slashed{D}\psi_{1,a}-\bar\psi_{2,a}\,\ii\slashed{D}\,\psi_{2,a}
 +\lambda v_\mu v^\mu
  \label{redLagrangian}
\end{equation}
is irrelevant and the model with $\Nr$ reducible flavours is equivalent
to the model with $\Ni=2\Nr$ irreducible flavours, for which the second term 
in (\ref{redLagrangian}) has a positive sign. For massive fermions this is not true anymore:
when integrating out irreducible fermions a parity violating (imaginary)
Chern-Simons-like term is generated\footnote{More accurately: an imaginary Chern-Simons term
is generated in the limit $m\to\infty$. For small $m$ the imaginary part of the effective
action is proportional to the $\eta$-invariant.}. In contrast, the fermionic determinant is
real for reducible massive fermions and we do not expect parity breaking
in models with reducible fermions.

This paper is organised as follows: In \autoref{s:symm}, we summarize the symmetries 
of the Thirring model in the reducible and an irreducible representation.
The relation between both formulations is further discussed.
Then in \autoref{s:fierz} we use a Fierz identity  to rewrite the Lagrangian
for irreducible fermions.
The main derivation of the effective potential for local condensates can be found in \autoref{s:eff_pot}, while its symmetries are studied in \autoref{s:symm_pot}.
We also give the explicit forms of the potential for $\Ni=1$ and $2$.
Next we study the effective potential in the strong coupling limit 
in \autoref{s:strong_coupling}, before we present our main results from 
numerical simulations in \autoref{s:results}.
Finally, a discussion of our findings and a comparison with previous results can be found in \autoref{s:conclusions}.

%% file: symm.tex
\section{Continuum symmetries}
\label{s:symm}
\noindent
First we discuss the internal symmetries of the Thirring model in three dimensions. 
The results look different for $\Nr$ fermions in the $4$-dimensional reducible 
representation of the Lorentz group and $\Ni$ fermions in a $2$-dimensional irreducible representation.
The flavour numbers are related as $\Ni=2 \Nr$.

In the reducible representation $\Psi_a$ is a four-component Dirac spinor where
$a=1, \dots, \Nr$ labels the flavours. The theory possesses a $U(2 \Nr)$ chiral symmetry 
generated by
\begin{equation}
 T= T_\text{f} \otimes\{\id,\Gamma_4,\Gamma_5,\ii\, \Gamma_4\,\Gamma_5\}
\end{equation}
where $T_\text{f}\in \algebra{u}(\Nr)$ generates rotations in flavour space.
A Dirac spinor transforms under the chiral and flavour rotations as
\begin{equation}
\begin{aligned}
 \Psi \to & e^{\ii\, T^a_\text{f} \otimes \left(t^a_1\, \id+t^a_2\, \Gamma_4+t^a_3\, \Gamma_5+\ii \, t^a_4\,  \Gamma_4\,\Gamma_5\right)}\,\Psi\\
 \bar{\Psi} \to & \bar{\Psi}\, e^{-\ii\,  T^a_\text{f} \otimes \left(t^a_1\, \id-t^a_2\, \Gamma_4-t^a_3\, \Gamma_5+\ii \, t^a_4\,  \Gamma_4\,\Gamma_5\right)}.
 \end{aligned}
\end{equation}
Furthermore, the theory is invariant under a discrete $\mathbbm{Z}_2$ parity transformation
\begin{equation}
 \Psi(x)\to \ii \,\Gamma_1 \,\Gamma_4\, \Psi(x') \quad \text{with} \quad x'=(x_0,-x_1,x_2),
 \label{e:red_parity}
\end{equation}
where alternative formulations using $\Gamma_5$ instead of $\Gamma_4$ are possible.
A detailed discussion of both continuous and discrete symmetries can be found 
in \cite{Gomes1991,Gies2010}.
In total, the global symmetry group is $U(2\Nr) \otimes \mathbbm{Z}_2$.
Now, we can define a parity-even chiral condensate $\Sigma=\bar{\Psi} \Psi$ and a parity-odd condensate $\Sigma_{45}=\bar{\Psi} \, \ii\, \Gamma_4\,\Gamma_5 \,\Psi$.
The chiral condensate $\Sigma$ is an order parameter for spontaneous breaking of the 
continuous chiral symmetry according to
\begin{equation}
 U(2\Nr) \otimes \mathbbm{Z}_2 \stackrel{\Sigma}{\to} U(\Nr)\otimes U(\Nr) \otimes \mathbbm{Z}_2
\end{equation}
while the parity condensate $\Sigma_{45}$ serves as an order parameter for discrete parity breaking
\begin{equation}
  U(2\Nr) \otimes \mathbbm{Z}_2 \stackrel{\Sigma_{45}}{\to}  U(2\Nr).
\end{equation}

For calculations it is often more convenient to reformulate the Thirring model in an irreducible representation. 
A useful reduction is given by
\begin{equation}
\begin{aligned}
 \Gamma_\mu=&\sigma_3 \otimes \gamma_\mu\,,&\Gamma_4&=\sigma_2 \otimes \id\,,\\ \quad \Gamma_5=&\sigma_1 \otimes \id \quad \text{and} &\ii\, \Gamma_4\,\Gamma_5&=\sigma_3 \otimes \id\,.
\end{aligned}
 \end{equation}
In order to obtain a \emph{standard} kinetic term in the irreducible representation, we decompose the
$\Nr$ four-component $\Psi_a$ as
\begin{equation}
 \Psi_a=\begin{pmatrix}\psi_{1,a}\\\psi_{2,a}\end{pmatrix} \quad \text{and} \quad
 \bar{\Psi}_a=
 (\bar{\psi}_{1,a},-\bar{\psi}_{2,a})\,,
 \end{equation}
where $\psi_{i,a}$ are two-component spinors in a fixed irreducible representation.
With this decomposition, the Lagrangian is given by
\begin{equation}
 \mathcal{L}=\bar{\psi}_\alpha \, \ii \,\slashed{\partial} \,\psi_\alpha-\frac{g^2}{2 N_f} \left(\bar{\psi}_\alpha \gamma_\mu \psi_\alpha\right)^2\,,
 \label{e:th_irred_lagrangian}
\end{equation}
where the irreducible flavour index $\alpha=(i,a)$ assumes 
$2\Nr=\Ni$ different values.
Note the difference to the decomposition (\ref{red:matrices}).
The condensates are
\begin{align}
 \Sigma&=\sum \limits_{a=1}^{\Nr} \left(\bar{\psi}_{1,a}\psi_{1,a}-\bar{\psi}_{2,a}\psi_{2,a}\right)\quad \text{and}\\
 \pi&=\Sigma_{45}=\sum \limits_{\alpha=1}^{\Ni}\bar{\psi}_\alpha\psi_\alpha.
\end{align}
The parity condensate in the reducible representation $\pi$ turns into a \emph{chiral condensate} in the irreducible representation, while for even $\Ni$ the former chiral condensate now has a flavour-staggered structure. 
This already indicates that in the irreducible representation the behaviour is different for even and odd flavour numbers $\Ni$.
We will further investigate this difference in the next sections.

The continuous chiral symmetry of the reducible Thirring model becomes a pure flavour symmetry in the irreducible representation, i.e. the theory is invariant under $U(\Ni)$ flavour transformations given by
\begin{equation}
 \psi \to U\, \psi \quad \text{and} \quad \bar{\psi} \to \bar{\psi} \,U^\dagger
\end{equation}
with a unitary matrix $U=\exp \mleft(\ii \,t_i\, T_i\mright)$ acting in flavour space only.
Here, the $T_i$ are generators of the algebra $\algebra{u}(\Ni)$.
Due to its relation to the chiral transformations of the 4-component spinors, this flavour symmetry will still be called \emph{chiral symmetry} in the following, although the concept of chirality does not exist
in odd dimensions.

The parity transformation of the 4-component spinors turns into a combination of flavour rotations and a parity transformation in the irreducible representation, that is given by
\begin{equation}
 \psi(x) \to \gamma_1 \psi(x') \quad \text{and} \quad \bar{\psi}(x) \to -\bar{\psi}(x')\,\gamma_1.
\end{equation}
The \emph{irreducible chiral condensate} $\pi$ is invariant under chiral transformations (i.e. 2-component flavour rotations) while it breaks the discrete $\mathbbm{Z_2}$ parity symmetry.


\section{Fierz identities}
\label{s:fierz}
\noindent
To integrate over the fermionic fields in the functional integral, it is useful to 'linearise' the 
four-fermion term by a Hubbard Stratonovich transformation, which 
transforms the Lagrangian (\ref{e:th_irred_lagrangian}) into the equivalent form
\begin{equation}
 \mathcal{L}=\bar{\psi}\left(\ii \slashed{\partial}+\gamma_\mu v_\mu\right) \psi+\lambda \, v_\mu^2\label{lagrange:vector}
\end{equation}
with $\lambda=\nicefrac{\Nr}{2 g^2}$ and three real scalar fields $v_\mu$.
After integration over the $\psi_\alpha$ we obtain the effective action
\begin{equation}
 S_\text{eff}=-\Ni \ln \det\left(\ii \slashed{\partial}+\gamma_\mu v_\mu\right)+\lambda \int d^3 x \, v_\mu(x)^2. 
\end{equation}
Note that this action is not gauge invariant and $v_\mu$ should not be viewed 
as gauge potential%
\footnote{It could be promoted to a gauge potential after introducing
a Stückelberg field \cite{Itoh1995}.}.

A technical problem arises if one discretises this formulation on a (hypercubic) lattice 
to perform Monte Carlo simulations. Since $v_\mu$ is invariant under
chiral transformations there is no natural order parameter for chiral 
symmetry breaking in the massless theory. 
But the advantage of the vector formulation is that except for $\Ni=1$ it 
is free of a fermion sign problem. 
This is obvious for even $\Ni$ since $\ii\slashed{\partial}+\gamma_\mu v_\mu$ is 
hermitian.
For odd $\Ni$ we never observed a negative sign in our Monte Carlo 
simulations except for $\Ni=1$.

We can circumvent the technical problem by applying a Fierz transformation to the four-fermion interaction 
\begin{equation}
 \left(\bar{\psi}_\alpha \gamma_\mu \psi_\alpha\right)^2=-2 \left(\bar{\psi}_\alpha \psi_\beta\right)\left(\bar{\psi}_\beta \psi_\alpha\right)-\left(\bar{\psi}_\alpha \psi_\alpha\right)^2.\label{fierz_id}
\end{equation}
This identity was for example also applied in \cite{Ahn1994,Ahn1998}.
The transformed Lagrangian reads
\begin{equation}
 \mathcal{L}=\bar{\psi} \,\ii \,\slashed{\partial}\,\psi+\frac{g^2}{N_f} \left(\bar{\psi}_\alpha \psi_\beta\right)\left(\bar{\psi}_\beta \psi_\alpha\right)+\frac{g^2}{2 N_f} \left(\bar{\psi} \psi\right)^2.
\end{equation}
Now applying the Hubbard Stratonovich transformation, we can reproduce the four-fermion
terms with the help of a matrix-valued field,
\begin{equation}
 \mathcal{L}=\bar{\psi} \Big( \ii \,\slashed{\partial} +\ii \, T +\,\frac{\ii}{2}\tr T\Big)\psi+\frac{\lambda}{2} \tr T^2+\frac{\lambda}{4} (\tr T)^2\,,
 \label{e:L-Fierz}
\end{equation}
where $T=T^\dagger$ is a generic $\algebra{u}(\Ni)$-algebra valued field  
i.e.
\begin{equation}
 T=t_i \, T_i \quad \text{with} \quad i=1 \dots {\Ni}^2.
 \label{e:t_def}
\end{equation}
Under a chiral transformation the components of $T$ transform according to
\begin{equation}
 T \to U \, T \, U^\dagger,\quad U\in U(\Ni)\,,
\end{equation}
such that their expectation values serve as order parameters for chiral symmetry breaking. 
They are related to fermionic condensates by DSEs.
Unfortunately, this formulation of the model suffers from a severe sign problem on the lattice \cite{Schmidt2016}.

%% file: eff_pot.tex
\section{Effective Potential}
\label{s:eff_pot}
\noindent
The vector formulation (\ref{lagrange:vector}) and matrix formulation
(\ref{e:L-Fierz}) each have their own advantages and disadvantages.
The former can, except for $\Ni=1$, be simulated without sign problem, 
but information about chiral symmetry is not directly accessible.
On the other hand, in the matrix formulation we have
direct access to order parameters for csb, but there is a 
strong sign problem that prevents reliable simulations.
We proceed with an analytical treatment in the matrix formulation
and calculate the resulting expectation values in the vector formulation.
We begin with splitting the matrix field in \eqref{e:L-Fierz} as
\begin{equation}
  T(x)=T^\text{c}(x)+T^\perp(x),
\end{equation}
where the first term is in the Cartan subalgebra of $\algebra{u}(\Ni)$ and
the second in its orthogonal complement.
We shall introduce a dual variables formulation in \autoref{ss:dual_variables} and 
afterwards present a calculation of the (constraint) effective potential
 \begin{equation*}
  \begin{aligned}
   V_\text{eff}(T^\text{c})=&-\ln \int \mathcal{D} T\, \mathcal{D}\phi \, e^{-S_\text{eff}(T,\phi)}\, \delta(T^\text{c}-T^\text{c}(x_0))\\
   =& -\ln \sum \limits_{n=0}^{2 \Ni} \sum \limits_{i=1}^{\Ni} a_{n,i} \left(t_i\right)^n\,,\quad T^\text{c}=t_i \, H^i\,,
  \end{aligned}
 \end{equation*}
in which the constraint fixes the field at an arbitrary point $x_0$ to 
the prescribed value $T^\text{c}$ in \autoref{ss:eff_theory}.
The matrix field $T(x_0)$ is hermitian and can be 
diagonalized by a global chiral transformation. Hence it is sufficient to calculate
the effective potential for a field in the Cartan subalgebra with generators $H^i$.
In \autoref{ss:rel_obs} we will relate the coefficients $a_{n,i}$ to 
expectation values of observables $O_{n,i}$ in the vector field formulation,
\begin{equation}
  a_{n,i}=\erw{O_{n,i}}_{v_\mu}.
\end{equation}
This allows us to employ Monte Carlo simulations in the vector field formulation
to calculate the effective potential defined in the matrix formulation.

\subsection{The partition function in the dual variables formulation}
\label{ss:dual_variables}
\noindent
First we reformulate the partition function of the Thirring model with
massive fermions in terms of discrete \emph{spin} (or dual) variables.
Then we take derivatives with respect to the mass to relate
observables in the dual formulation to powers of fermionic bilinears.
Actually we introduce an $x$-dependent and diagonal fermion mass matrix
\begin{equation}
 M(x)=\diag(m_1(x),\dots,m_{\Ni}(x)).
 \label{e:mass_def}
\end{equation}
The partition function for the massive model is now given by (integral over spacetime is assumed in the exponent)
\begin{equation}
 Z(\lambda,M)=\left(\frac{\pi}{\lambda}\right)^\frac{V}{2}\int \mathcal{D}T \, Z_F[T,M] \, e^{-\frac{\lambda}{2} \tr T^2-\frac{\lambda}{4} (\tr T)^2}\,,\label{partfunc2}
\end{equation}
where the fermionic partition function $Z_F[T,M]$ is given by the fermionic integral
\begin{equation}
 Z_F[T,M]=\int \mathcal{D}\bar{\psi}\mathcal{D}\psi \,e^{-\bar{\psi} \, \ii \, \slashed{\partial} \,\psi} e^{-\bar{\psi} \, \ii \, H \, \psi}.
\end{equation}
Here, the shifted field $H=T+\frac{1}{2} \tr T+M$ was introduced.
With spinor index $i$ and flavour indices $\alpha$ and $\beta$
the product expansion of the second exponent leads to
\begin{equation}
  Z_F[T,M]=\int \mathcal{D}\bar{\psi}\mathcal{D}\psi \,e^{-\bar{\psi} \, \ii \,\slashed{\partial} \, \psi} \prod_{xi\alpha\beta} \left(1-\bar{\psi}_{xi}^\alpha \, \ii \, H_x^{\alpha\beta} \, \psi_{xi}^\beta\right),
\end{equation}
where we used that $\psi_{xi}^\alpha$ and $\bar{\psi}_{xi}^\alpha$
are Grassmann variables.
A similar expansion is met in the fermion bag approach,
where one expands the integrand directly in powers of the four-fermion term
\cite{Chandrasekharan2012}.
After integrating over the auxiliary field the two expansions
yield the same results, although the intermediate expressions are different.
A related expansion is also encountered in attempts to dualise gauge theories \cite{Gattringer:2016lml}.

At this point we introduce a spin field $k_{xi}^{\alpha\beta} \in \{0,1\}=\mathbbm{Z}_2$, rearrange the weight function
as a sum over configurations of the $k$-field and perform the integration over the fermions to get
\begin{equation}
  Z_F[T,M]=\sum_{\{k_{xi}^{\alpha\beta}\}} (-\ii)^k \det \left(\ii \,\slashed{\partial}[k]\right) \prod_{x\alpha\beta} \left(H_x^{\alpha\beta}\right)^{k_{x}^{\alpha\beta}}\,,
  \label{wtm2}
  \end{equation}
 where we introduced the abbreviations
 \begin{equation} k_x^{\alpha\beta}=\sum \limits_i k_{xi}^{\alpha\beta}=k_{x0}^{\alpha\beta}+k_{x1}^{\alpha\beta}\,,\quad k
 =\sum \limits_{xi\alpha\beta} k_{xi}^{\alpha\beta}\label{kxi}
 \end{equation}
and $\slashed{\partial}[k]$ is the matrix in which the $\{xi\alpha\}$-th row 
and $\{xi\beta\}$-th column of $\slashed{\partial}$ are removed whenever $k_{xi}^{\alpha\beta}=1$.
A similar expansion in terms of minors of a fermionic
matrix was recently presented
within a transfer matrix approach in \cite{Wenger:2016dpm}.
It is important to note that in the minor expansion
there are constraints on the spin field $k_{xi}^{\alpha\beta}$ 
in order to get non-vanishing contributions to the weight function: Due to 
  $\psi_{xi}^\alpha \psi_{xi}^\alpha=\bar{\psi}_{xi}^\alpha \bar{\psi}_{xi}^\alpha=0$ for fixed $x,i$ the sum over rows and columns of the matrix $k^{\alpha\beta}$ has to be zero or one,
  \begin{equation}
   \sum_\alpha k_{xi}^{\alpha\beta} \in \{0,1\} \quad \text{and} \quad \sum_\beta k_{xi}^{\alpha\beta} \in \{0,1\}.
  \end{equation}
Summing over the spinor index $i$ leads to the following local constraints
on the elements of the matrix $k_x=(k_x^{\alpha\beta})$ in (\ref{kxi}):
 \begin{equation}
   \sum_\alpha k_{x}^{\alpha\beta} \in \{0,1,2\} \quad \text{and} \quad \sum_\beta k_{x}^{\alpha\beta} \in \{0,1,2\}.\label{constr2}
  \end{equation}
  We will summarise these constraints in a local constraints function 
  \begin{equation}
   \delta_\text{constr}(k_x)=\begin{cases} 1 \quad \text{all local constraints are fulfilled,} \\ 0 \quad \text{else.}\end{cases}
  \end{equation}
  Inserting (\ref{wtm2}) into the partition function (\ref{partfunc2})
 we observe that all contributions, with the exception of the minors,
 are given by a product over the lattice sites.
  After rescaling the variables $\tilde{T}=\sqrt{\lambda}\, T$ 
  and $\tilde{M}=\sqrt{\lambda}\, M$  the partition function reads
  \begin{equation}
  \begin{aligned}
   Z(\lambda,\tilde{M})&=C\sum_{\{k_{x i}^{\alpha\beta}\}}\lambda^{-\frac{k}{2}} \det \left(\,\slashed{\partial}[k]\right) \\ 
   &\hskip3mm \cdot \prod_x \delta_\text{constr}(k_x) \,W_\text{loc}
   \big(k_x,\tilde{M}(x)\big)
   \label{e:Z_constr_Wloc}
  \end{aligned}
  \end{equation}
  with overall factor $C=(\pi/\lambda)^{\gamma/2}$, where
  $\gamma=V({\Ni}^2+1)$. The local weight function is defined for any $k_x$ by
  \begin{equation}
  \begin{aligned}
   W_\text{loc}(k,\tilde{M})&=\int \prod \limits_{i=1}^{{\Ni}^2}\left(\frac{\dx\tilde{t}_{i}}{\sqrt{\pi}}\right) \,e^{-\frac{1}{2} \tr \tilde{T}^2-\frac{1}{4} (\tr \tilde{T})^2}
   \\
   &\cdot\prod \limits_{\alpha\beta}\Big(\tilde{T}^{\alpha\beta}+\frac{1}{2} \tr \tilde{T} \,\delta^{\alpha\beta}+\tilde{M}^{\alpha\beta}\Big)^{k^{\alpha\beta}}\,.
   \label{wloc1}
  \end{aligned}
  \end{equation}
  In the following we drop the tilde above variables again to simplify our notation.
Because of (\ref{constr2}) the exponent $k^{\alpha\beta}$ only takes the 
values $0,1$ and $2$. The integration variables are the expansion
coefficients in $T = t_i T_i$. 
   
The integration over the non-Cartan fields is performed in appendix~\ref{s:fermion_bag} and leads to the final form of the local weight function
\begin{equation}
 W_\text{loc}(k,M)=w_\text{o}(k_\perp) \, W_{\vecf{p}(k)}(M)\label{eq:fermion_bag_form}
\end{equation}
 where $\vecf{p}(k)=(k^{11}, k^{22}, \dots)$ is the $\Ni$-component vector 
 that collects the diagonal entries of $k^{\alpha\beta}$.
 Recall that these entries can assume the values 0, 1 or 2. In addition, the 
 first factor $w_\text{o}(k_\perp)$ is a non-negative integer depending on
 the non-diagonal entries of the matrix $(k^{\alpha\beta})$.
 The explicit form of $w_\text{o}$ is given in appendix \ref{s:fermion_bag}.
 The second factor is an integral over the Cartan subalgebra,
 \begin{equation}
 \begin{aligned}
  W_{\vecf{p}}(M)=&\int \prod \limits_{i=1}^{{\Ni}}\left(\frac{\dx t_{i}}{\sqrt{\pi}}\right) \,e^{-t\,A\,t} \\
  &\cdot \prod \limits_{\alpha}\Big(T^{\alpha\alpha}+\frac{1}{2} \tr T +m_\alpha\Big)^{p_\alpha}.
  \label{e:wnkm}
 \end{aligned}
 \end{equation}
The symmetric matrix $A$ in the exponent can be written as
 \begin{equation}
  A_{ij}=\frac{1}{2} \tr \left(H_i H_j\right)+\frac{1}{4} \tr H_i \tr H_j\,\label{ee:wnkm}
 \end{equation}
 and the integration is only over the Cartan subalgebra with generators $H_i$.
Although the final result does not depend on a specific choice of the generators,
we use the generators given in appendix~\ref{s:fermion_bag} for the explicit calculations.
$W_\vecf{p}$ is a polynomial in $m_\alpha$ of degree
$p_\alpha$.
The explicit form of $W_\text{loc}(k,M)$ is also
given in appendix~\ref{s:fermion_bag}.

In this form it looks like the partition function can be simulated with a 
fermion bag like algorithm.
Unfortunately the minors induce a severe sign problem. So far we could solve this problem
for $\Ni=1$, where we applied the fermion bag algorithm to the model
with chiral SLAC fermions \cite{Schmidta}. 
 In the following we refer to the integer number $k$ in (\ref{kxi}) 
 as lattice filling factor, because it counts how many fermions take part in the local interaction. Due to
 Pauli blocking it can only take the values $k=0, \dots, 2 V\Ni$.
 In the strong coupling limit ($\lambda \to 0$) we find $k=2V\Ni$ on 
 every configuration such that every lattice site is occupied by the maximal
 number of fermions.
 Here we expect strong lattice artefacts due to saturation effects.
 In the weak coupling limit ($\lambda \to \infty$) we have $k=0$ and the theory reduces to a theory of free fermions on the lattice.

\subsection{Effective potential}
\label{ss:eff_theory}
\noindent
In this section we derive an effective theory for the local chiral condensates $\langle\bar{\psi}_\alpha \psi_\alpha\rangle$ (no sum), or equivalently the 
expectation values of scalar fields in the Cartan subalgebra, that are related to the condensates by DSEs.
Therefore, we constrain the local mass matrix defined in \eqref{e:mass_def} to
\begin{equation}
 M=\text{diag}(m_1,\dots,m_{\Ni})\, \delta_{x,x_0}
\end{equation}
with a fixed lattice point $x_0$ and
determine the dependence of the partition function on the parameters $m_\alpha$. 
This allows us to relate contributions to the partition function $Z(\lambda,M)$
in the dual formulation to expectation values of chirally invariant local 
observables in the original 
vector formulation by differentiating with respect to this local mass parameters. Formally, the $M$-dependent partition function is given by
\begin{equation}
 Z(\lambda,M)=\sum \limits_{\cP} \, \sum \limits_{\vecf{p} \in \cS_\cP} a_{\vecf{p}} \,W_\vecf{p}(M)\,,\label{double_sum}
\end{equation}
where the first sum extends over the $(\Ni+1)(\Ni+2)/2$
triplets 
\begin{equation}
\cP=(P_0,P_1,P_2).
\end{equation}
The elements of $\cP$ obey
\begin{equation}
 P_k\in\{0,1,\dots,\Ni\}\quad\text{and}\quad \sum_{k=0}^2 P_k=\Ni\,.
\end{equation}
The second sum in \eqref{double_sum} extends over the
\begin{equation}
 \Vol(\cS_\cP)=\frac{\Ni!}{P_0!\,P_1! \,P_2!}
\end{equation}
permutations $\cS_\cP$ of the $\Ni$-tuple $\vecf{p}$ with $P_0$ elements equal to zero, 
$P_1$ elements equal to one and $P_2$ elements equal to two. 
Altogether the double sum in (\ref{double_sum}) consists of $3^{\Ni}$ terms.
The coefficients $a_\vecf{p}$ depend on $\lambda$, on the volume and on further details 
of the lattice formulation as for example the choice of lattice fermion derivative.

When all flavours have the same mass, $(m_\alpha=m)$, 
the coefficients $a_\vecf{p}$ do not depend on a specific permutation (permutation of flavours) but only on the three numbers $P_k$ and we can write the partition function as
\begin{equation}
 Z(\lambda,m)=\sum \limits_{\cP} a_{\cP} 
 \cdot \Vol(\cS_\cP) \, W_\cP(m)\,,
 \end{equation}
where 
\begin{equation}
\begin{aligned}
 W_\cP(m)=&\left.\,W_{\vecf{p}_\text{sort}}(M)\right \vert_{M(m_\alpha\to m)} \\
 \text{with} \quad & \vecf{p}_\text{sort}=(\underbrace{0, \dots,0}_{P_0\;\text{times}},\underbrace{1, \dots,1}_{P_1\;\text{times}},\underbrace{2, \dots,2}_{P_2\;\text{times}})
\end{aligned}
 \end{equation}
is the weight of a particular representative in an orbit
of the permutation group.

The constraint effective potential in the limit of a constant mass in 
flavour space is then given by the negative logarithm of the distribution function, i.e.
 \begin{equation}
 \begin{aligned}
  V_\text{eff}(\lambda,T,m)=&\frac{1}{2} \tr ({T}^\text{c})^2+\frac{1}{4} (\tr T^\text{c})^2\\
  -\ln \Bigg(\sum\limits_{\cP} \, a_{\cP}\,&\sum \limits_{\vecf{p} \in 
  \cS_{\cP}} \,\prod \limits_{\alpha=1}^{\Ni}\Big(T^{\alpha\alpha}+\frac{1}{2} \tr T+m\Big)^{p_\alpha}\Bigg)\,.
 \end{aligned}
 \end{equation}
Here, we have to sum over all permutations again, because no integration over the 
diagonal scalar fields is performed in the effective potential.

\subsection{Relation to observables}
\label{ss:rel_obs}
\noindent
In order to relate the coefficients $a_{\cP}$ to expectation values, 
we take derivatives of the partition function (\ref{double_sum}) with respect to 
the local masses $m_\alpha$ and afterwards set $m_\alpha=m$.
With the definition of moments of the Gaussian weights
\begin{equation}
 W_{\vecf{p},\vecf{q}}(m)=
 \prod_{\alpha=1}^{\Ni} \Bigg(\frac{\partial^{\,q_\alpha}}{\partial m_\alpha^{q_\alpha}}\Bigg)W_\vecf{p}( M)\Big\vert_{ m_\alpha\to m}
 \label{multi_index}
\end{equation}
we can write the partition function and its derivatives with respect to the local masses as
\begin{equation}
  Z_\vecf{q}(\lambda,m)=\sum \limits_{\cP} a_{\cP} \sum \limits_{\vecf{p} \in 
  \cS_\cP} \,W_{\vecf{p},\vecf{q}}(m).
  \label{e:Zd_def}
\end{equation}
Since $W_\vecf{p}(M)$ is a polynomial of 
degree $p_\alpha$ in $m_\alpha$ we
only get a non-zero result in (\ref{multi_index}) when 
$q_\alpha\leq p_\alpha$.
In the limit of a constant mass in flavour space, the expectation values $Z_\vecf{q}$ do not depend on the ordering of the flavours, but only on the number 
$Q_0$ of zeroth, $Q_1$ of first and $Q_2$ of second derivatives. Denoting the 
triple $\{Q_0,Q_1,Q_2\}$ by $\cQ$ we can write
\begin{equation}
 \sigma_{\cQ}
 =\frac{1}{\Vol(\cS_\cQ)}\sum \limits_{\vecf{q} \in \cS_\cQ}Z_{\vecf{q}}(\lambda,m).
\label{e:sigma_dd_def}
 \end{equation}
The index sets $\cQ$ and $\cP$ are identical, with the 
constraints that their three entries sum to $\Ni$.
The sum in (\ref{e:sigma_dd_def}) is over 
the $\Vol(\cS_\cQ)=\Ni!/(Q_0!\,Q_1!\,Q_2!)$ permutations 
$\cS_\cQ$ of $\vecf{q}$.
The derivatives of the partition sum are directly related to expectation values via
\begin{equation}
\sigma_\cQ=\lambda^{-\frac{|\vecf{q}|}{2}} Z \cdot\Big\langle \prod \limits_{\alpha}\, \left(\bar{\psi}_\alpha \psi_\alpha\right)^{q_\alpha}\Big\rangle\,.
\label{e:sigma_condensate}
\end{equation}
Here, the right-hand side can be computed, up to the arbitrary normalization factor $Z=\sigma_{\Ni,0,0}$, with conventional simulations of the Thirring model in the vector field formulation.
Inserting \eqref{e:Zd_def} in \eqref{e:sigma_dd_def}, we obtain
\begin{equation}
 \sigma_\cQ=\sum \limits_{\cP} a_{\cP} 
 \sum \limits_{\vecf{p} \in \cS_\cP} \frac{1}{\Vol(\cS_\cP)}\sum \limits_{\vecf{q} 
 \in\cS_\cQ} \,W_{\vecf{p},\vecf{q}}(m).
 \end{equation}
We can also write this equation as a matrix equation
\begin{equation}
 \begin{aligned}
  \sigma_\cQ=& K_{\cQ \cP}\, a_\cP \quad \text{with}\\
  K_{\cQ\cP}=&\sum \limits_{\vecf{p} \in \cS_\cP} 
  \frac{1}{\Vol(\cS_\cQ)}\sum \limits_{\vecf{q} \in \cS_\cQ} \,W_{\vecf{p},\vecf{q}}(m)
 \end{aligned}
\end{equation}
where $K$ is a square matrix that can always be represented in an upper triangular form (due to $q_\alpha\leq p_\alpha$) with non-zero diagonal elements.
Therefore a unique solution for the coefficients $a_\cP$ is given by
\begin{equation}
 \vec{a}=K^{-1} \, \vec{\sigma}.
\end{equation}
For symmetry reasons, the double sum for the matrix coefficients simplifies to
\begin{equation}
 K_{\cQ\cP}=\frac{\Vol(\cS_\cP)}{\Vol(\cS_\cQ)}
 \sum \limits_{\vecf{q} \in \cS_\cQ} \,W_{\vecf{p}(\cP),\vecf{q}}(m),
\end{equation}
where $\vecf{p}(\cP)$ is one representative in the equivalence class
defined by $\cP$.
In this way, we can uniquely relate the coefficients $\vec{a}$ to expectation values $\vec{\sigma}$ of the Thirring model.
Alternatively, the coefficients $\vec{a}$ can be calculated directly with a fermion bag simulation of the partition function \eqref{e:Z_constr_Wloc}.

%% file: symm_pot.tex
\section{Symmetries of the effective potential}
\label{s:symm_pot}
\noindent
A suitable order parameter for chiral symmetry breaking is the position of the 
global minimum of the effective potential.
We can simplify the discussion of the potential, if we locally apply a chiral transformation such that the local condensates are in the Cartan subalgebra of $\algebra{u}(\Ni)$.
As part of the remaining symmetry of the Cartan subalgebra, it is possible to exchange flavours and the sign of the condensate, without changing physics.
Therefore, minima of the effective potential can only occur in the directions
\begin{equation}
 T^\text{c}_\text{min}=\frac{2\,x}{\Ni}\begin{pmatrix} \pm 1 &  &  \\  & \ddots &  \\  &  & \pm 1 \end{pmatrix} \in \mathbbm{Z}_2^{\Ni}\,,
\end{equation}
where $x \sim (\bar{\psi}\psi)_\alpha$ is a free parameter.
These directions where also spotted by the simulation results for the full effective 
potential. Once a direction is fixed we are left with the problem of finding the 
minimizing $x$.
Physically equivalent solutions are related by a reflection $x \to -x$ or
a permutation of flavours. The latter is given by the action of the Weyl group 
of $U(\Ni)$. We conclude that physically 
distinct solutions are characterized by the trace of $T^\text{c}_\text{min}$
\begin{equation}
\begin{aligned}
 \tr T^\text{c}_\text{min}=&\frac{2\,x}{\Ni}\,n \quad \text{with} \\
 n=&\begin{cases} 0,2,4, \dots,\Ni & \Ni \quad \text{even} \\ 1,3,5, \dots,\Ni & \Ni \quad \text{odd}. \end{cases}
 \end{aligned}
\end{equation}
This leads to $n_\text{sol}=\left[\ft12\Ni\right]+1$ different solutions
for the potential. Every solution gives rise to a different breaking pattern
of chiral symmetry, leaving different subgroups intact. A non-vanishing expectation value in direction $n$ breaks the symmetry down to
\begin{equation}
 U(\Ni) \to U(n_+) \otimes U(n_-),\quad n_\pm=\frac{\Ni \pm n}{2}\,.
\end{equation}
A symmetric breaking with $n_+=n_-$ is only possible for even $\Ni$ with $n=0$. 
This is the proposed breaking of the reducible Thirring model. For the solution with $n=\Ni$
and $x\neq 0$, only parity symmetry is broken. 
Therefore, the solution with $n=0$ is called a \emph{Thirring}-like breaking
while the solution with $n=\Ni$ is called a \emph{Gross-Neveu}-like
breaking.

Along the different directions, the effective potential in the massless case is
\begin{align}
 &V_\text{eff}(x,n)=\frac{2\Ni+n^2}{{\Ni}^2} x^2 \label{effpotml}\\
  &-\ln \Bigg(\sum \limits_{\cP} a_{\cP} \sum \limits_{\vecf{p} \in \cS_\cP} 
  (n+2)^{\vert\vecf{p}\vert_<} (n-2)^{\vert\vecf{p}\vert_>}
  \left(\frac{x}{\Ni}\right)^{\vert\vecf{p}\vert}\Bigg)\,,\nonumber
 \end{align}
where the exponents are given by the (partial) sums,
\begin{equation}
\vert\vecf{p}\vert_<=\!\sum_{\alpha \leq n_+}p_\alpha,\hskip3mm
\vert\vecf{p}\vert_>=\!\sum_{\alpha > n_+}p_\alpha,\hskip3mm
\vert\vecf{p}\vert=\!\sum_{\alpha=1}^{\Ni} p_\alpha\,.
\end{equation}
In the massive case $n\pm 2$ in \eqref{effpotml} is replaced
by $n\pm2+m\Ni$.
Evaluating the above potential in the Thirring-like direction, we get
 \begin{equation}
 \begin{aligned}
  V^\text{Th}_\text{eff}(x)&=\frac{2\, x^2}{\Ni}\\
  &-\ln \Bigg(\sum \limits_{\cP} \, a_{\cP}\, \sum \limits_{\vecf{p} \in \cS_\cP} \,(-1)^{\vert\vecf{p}\vert_>}\left(\frac{2\,x}{\Ni}\right)^{\vert\vecf{p}\vert}\Bigg)
 \end{aligned}
 \end{equation}
 and in the Gross-Neveu-like direction
 \begin{equation}
 \begin{aligned}
  V^\text{GN}_\text{eff}(x)&=\frac{\Ni+2}{\Ni}x^2\\
  &-\ln \Bigg(\sum \limits_{\cP} \, a_{\cP}\, \sum \limits_{\vecf{p} \in \cS_\cP} \,\left(\frac{\Ni+2}{\Ni}x\right)^{\vert\vecf{p}\vert}\Bigg).
 \end{aligned}
 \end{equation}
An important quantity to determine chiral symmetry breaking is the curvature $\kappa$ of the effective potential at the origin, 
\begin{equation}
 \kappa(n)=\left.\frac{d^2 V_\text{eff}(x,n)}{d x^2}\right \vert_{x=0}.
\end{equation}
Possible phase transitions are expected to be second order.
Thus, we conclude that chiral symmetry is unbroken if all curvatures $\kappa(n)$ are positive, while it is spontaneously broken if at least one of the $\kappa(n)$ is negative.

\subsection{Effective theory for \texorpdfstring{$\Ni=1$}{Nfir=1}}
\noindent
For one irreducible flavour the Thirring model
is equivalent to the Gross-Neveu model.
This follows from the Fierz identity~\eqref{fierz_id} for one 
flavour. For massless fermions the integral~\eqref{e:wnkm} vanishes
for $\vecf{p}=(1)$ such that only $\vecf{p}=(0)$ and $\vecf{p}=(2)$ remain.  
Likewise only multi-derivatives with $\vecf{q}=(0)$ and $\vecf{q}=(2)$
appear. The $K$-matrix is then given by
\begin{equation}
 K=\begin{pmatrix}1 & \frac{3}{2} \\ 0 & 2\end{pmatrix}\,.
\end{equation}
The explicit form of the matrix depends on the normalization of the generators $H_i$ given in appendix~\ref{s:fermion_bag}.
With the arbitrary normalization $\sigma_{1,0,0}=1$, we find for the coefficients $\vec{a}$
\begin{equation}
 a_{1,0,0}=1-\frac{3}{4}\sigma_{0,0,1}\quad \text{and} \quad a_{0,0,1}=\frac{1}{2}\sigma_{0,0,1}\,,
 \end{equation}
 and obtain for the effective potential
 \begin{equation}
   V_\text{eff}(x)=3\, x^2-\ln\Big(1-\frac{3}{4}\sigma_{0,0,1}+\frac{9}{2} \sigma_{0,0,1} x^2\Big)\,.
 \end{equation}
The curvature of the potential at the origin is then 
\begin{equation}
\kappa=18\,\frac{\sigma_{0,0,1}-4/9}{\sigma_{0,0,1}-4/3},\quad \sigma_{0,0,1}=\frac{1}{\lambda}\erw{\big(\bar{\psi}\psi\big)^2}_{m=0}.
\end{equation} 
At the critical inverse coupling $\lambda_c$ the curvature vanishes and we get $\sigma_{0,0,1}(\lambda_c)=4/9$.
The condensate $\big\langle(\bar{\psi}\psi)^2\big\rangle$ is later calculated within the strong coupling expansion since Monte Carlo simulations
in the vector formulation face a severe sign problem. 

The fermion bag approach, which is free of a sign problem and directly 
yields the coefficients $\vec a$, will be discussed in a follow-up paper \cite{Schmidta}.
 
\subsection{Effective theory for \texorpdfstring{$\Ni=2$}{Nfir=2}}
\noindent
For two massless irreducible flavours the Thirring model is equivalent 
to the massless reducible model with $\Nr=1$.
For a vanishing $m$ the non-vanishing $\vecf{p}$-configurations
are $\vecf{p}=\{(0,0),(1,1),(2,0),(0,2),(2,2)\}$. 
The configurations $\vecf{p}=(0,2)$ and $\vecf{p}=(2,0)$ are equivalent under flavour exchange and the corresponding
coefficients are the same.
The weights are listed in \autoref{confN2}.
\begin{table}[tb]
 \begin{tabular}{|c|c|c|c|c|c|c|c|}
 \hline
  $I$ & $a$ & $\vecf{p}$ & $W_{(0,0)}$ & $W_{(1,1)}$ & $W_{(2,0)}$&$W_{(0,2)}$ & $W_{(2,2)}$\\
  \hline\hline
  $1$ & $a_{2,0,0}$ & $(0,0)$ & $1$ & $0$ & $0$ & $0$ & $0$\\
  $2$ & $a_{0,2,0}$ & $(1,1)$ & $\frac{1}{2}$ & $1$ & $0$ & $0$& $0$\\
  $3$ & $a_{1,0,1}$ & $(2,0)$ & $\frac{3}{2}$ & $0$ & $2$ & $0$& $0$\\
  $3$ & $a_{1,0,1}$ & $(0,2)$ & $\frac{3}{2}$ & $0$ & $0$ & $2$& $0$\\
  $4$ & $a_{0,0,2}$ & $(2,2)$ & $\frac{11}{4}$ & $2$ & $3$ & $3$& $4$\\
  \hline
 \end{tabular}
 \caption{Configurations and weights for $\Ni=2$ in the massless limit.}
 \label{confN2}
\end{table}
The $K$-Matrix and its inverse are then given by
\begin{equation}
 K=\begin{pmatrix}1 & \frac{1}{2} & 3 & \frac{11}{4} \\ 0 & 1 & 0 & 2\\ 0 & 0 & 2 & 3 \\ 0 & 0 & 0 & 4\end{pmatrix},\;\;
 K^{-1}=\!\begin{pmatrix}1 & -\frac{1}{2} & -\frac{3}{2} & \frac{11}{16} \\ 0 & 1 & 0 & -\frac{1}{2}\\ 0 & 0 & \frac{1}{2} & -\frac{3}{8} \\ 0 & 0 & 0 & \frac{1}{4}\end{pmatrix}.
\end{equation}
We obtain with the normalization $\sigma_{2,0,0}=1$
\begin{equation}
 \begin{aligned}
  a_{2,0,0}=&1-\frac{1}{2}\sigma_{0,2,0}-\frac{3}{2}\sigma_{1,0,1}+\frac{11}{16}\sigma_{0,0,2}\,,\\
  a_{0,2,0}=&\sigma_{0,2,0}-\frac{1}{2} \sigma_{0,0,2}\,,\\
  a_{1,0,1}=&\frac{1}{2}\sigma_{1,0,1}-\frac{3}{8}\sigma_{0,0,2}\,,\\
  a_{0,0,2}=&\frac{1}{4}\sigma_{0,0,2}.
 \end{aligned}
\end{equation}
The Gross-Neveu-type potential is given by
\begin{align}
  V^\text{GN}_\text{eff}(x)&=2\,x^2\\ 
  -\ln &\left(a_{2,0,0}+4\,a_{0,2,0}\,x^2+8\,a_{1,0,1}\,x^2+16\,a_{0,0,2}\,x^4\right)
  \nonumber
 \end{align}
 and the Thirring-type potential by
 \begin{align}
  V^\text{Th}_\text{eff}(x)&=x^2\\
  -\ln &\left(a_{2,0,0}-a_{0,2,0}\,x^2+2\,a_{1,0,1}\,x^2+a_{0,0,2}\,x^4\right).\nonumber
 \end{align}
 In the following sections, the observables $\sigma$ are calculated first in the strong coupling expansion and after that with Monte Carlo simulations in the vector 
 formulation of the Thirring model which for $\Ni\geq 2$ has no sign problem.

%% file: strong_coupling.tex
  \section{Strong Coupling expansion}\label{s:strong_coupling}
\noindent
In this section we compute the effective potential in the strong coupling expansion.
The lattice partition function in the presence of fermion sources is given by
\begin{equation}
\begin{aligned}
 Z[\eta,\bar{\eta}]=\int \mathcal{D}v \mathcal{D}\psi \mathcal{D}\bar{\psi}\, 
 e^{-\sum \limits_x  \left(\lambda\, v^2-\bar{\psi}\left(\ii \slashed{\partial}+\slashed{v}\right)\psi-\bar{\psi}\eta-\bar{\eta}\psi\right)}\\
 =K\left[\frac{\delta}{\delta \eta},\frac{\delta}{\delta \bar{\eta}}\right]\int \!
 \mathcal{D}v \mathcal{D}\psi \mathcal{D}\bar{\psi}\, e^{-\sum \limits_x \left(\lambda\, v^2-
 \bar{\psi}\slashed{v}\psi-\bar{\psi}\eta-\bar{\eta}\psi\right)}\\
\label{e:strongCouplingPartition}
 \end{aligned}
\end{equation}
where the sum is over all lattice points $x$ and where we already use dimensionless fields and inverse coupling $\lambda$. The kinetic operator is given by
 \begin{equation}
   K\left[\frac{\delta}{\delta \eta},\frac{\delta}{\delta \bar{\eta}}\right]=e^{-\sum \limits_{x,y} \frac{\delta}{\delta \eta_x} \ii \slashed{\partial}_{xy} \frac{\delta}{\delta \bar{\eta}_y}}
 \end{equation}
 and $\slashed{\partial}_{xy}$ is a lattice regularised derivative operator.
 With a rescaling of the sources according to $\chi=\lambda^{1/4}\,\eta$ we shift the explicit $\lambda$ dependence to the kinetic operator. Furthermore, derivatives with respect to the sources $\chi$ exactly reproduce the expectation values \eqref{e:sigma_dd_def} needed in the effective potential.
 After performing the integration over the vector field and the fermions (details are given in appendix \ref{s:strong_details}), we obtain our master equation for the strong coupling expansion
 \begin{equation}
 \begin{aligned}
  Z[\chi,\bar{\chi}]&=K\left[\frac{\delta}{\delta \chi},\frac{\delta}{\delta \bar{\chi}},\lambda\right] \\
  &\prod \limits_x \sum \limits_{k=0}^{\Ni}\frac{\Gamma(\frac{3}{2}+\Ni-k)}{\Gamma(2k+2)}\left((\bar{\chi}_x\,\gamma_\mu\,\chi_x)^2\right)^{k}.
  \label{e:strongCouplingPartition1}
  \end{aligned}
 \end{equation}
In the following we are only interested in local fermionic observables. Therefore, 
\emph{after} applying the kinetic operator $K$  we set the source $\chi$ at all points
$x$ with the exception of $x_0$ to zero.
With the definitions
\begin{equation}
\begin{aligned}
 K^{(n)}=&(-1)^n\frac{\lambda^{n/2}}{n!}
 \left(\sum \limits_{x,y} \frac{\delta}{\delta \chi_x} \ii \slashed{\partial}_{xy} \frac{\delta}{\delta \bar{\chi}_y}\right)^n\,,\\
 F^{(k)}(x)=&\frac{\Gamma(\frac{3}{2}+\Ni-k)}{\Gamma(2k+2)}\left((\bar{\chi}_x\,\gamma_\mu\,\chi_x)^2\right)^{k}
 \end{aligned}
\end{equation}
we can write the partition function as
\begin{equation}
Z[\chi_{x_0},\bar{\chi}_{x_0}]=\sum \limits_n K^{(n)} \prod \limits_x \sum \limits_k F^{(k)}(x) 
\Big\vert_{\chi_{x \neq x_0}=0}\,.
\label{e:strongCouplingPartition2}
\end{equation}
In appendix \ref{s:strong_details} we show, that in the infinite volume limit 
and up to any finite order $n$ in the expansion of the kinetic operator, the 
partition function has the form
\begin{align}
 Z[\chi_{x_0},\bar{\chi}_{x_0}]&=C(\lambda) \sum \limits_k F^{(k)}(x_0)
 \label{e:strongCouplingPartition3}\\
 &\hskip-3mm=C(\lambda)\sum \limits_{k=0}^{\Ni}\frac{\Gamma(\frac{3}{2}+\Ni-k)}{\Gamma(2k+2)}\left((\bar{\chi}_{x_0}\,\gamma_\mu\,\chi_{x_0})^2\right)^{k},
\nonumber
 \end{align}
 where $C(\lambda)$ is an unknown function that cancels in expectation values.
Then the solution for the expectation values is
\begin{equation}
\begin{aligned}
\sigma_{\cQ}&=\lambda^{-Q_2-Q_1/2}\prod \limits_{\alpha=1}^{Q_1}\partial_{m_\alpha}\prod \limits_{\beta=Q_1+1}^{Q_1+Q_2} \partial^2_{m_\beta}\ln Z(M)\Big\vert_{M=0}\\
&= \frac{\Gamma(\frac{3}{2}+\Ni-Q_2)}{\Gamma(\frac{3}{2}+\Ni)}
\,\delta_{Q_1,0}
\,.
\end{aligned}
\end{equation}
For the vector interaction we obtain
\begin{equation}
 \frac{1}{4\,\lambda} \erw{\left(\bar{\psi} \gamma_\mu \psi\right)^2}=\Ni
\end{equation}
and it follows that the normalized lattice filling factor takes
its maximal value $\erw{k_\text{norm}}=1$.
In conclusion, within the strong coupling expansion we are not able to leave 
the strong coupling phase where strong lattice artefacts dominate due to 
complete Pauli blocking on every lattice site.

\subsection{Results for the effective potential}
\noindent
\begin{figure*}[htb]
 \begin{center}
  \scalebox{0.85}{\input{Potential_Strong_1.tex}}\hskip2mm
  \scalebox{0.85}{\input{Potential_Strong_2.tex}}\hskip2mm
   \scalebox{0.85}{\input{Potential_Strong_3.tex}}
    \scalebox{0.85}{\input{Potential_Strong_4.tex}}\hskip2mm
  \scalebox{0.85}{\input{Potential_Strong_5.tex}}\hskip2mm
   \scalebox{0.85}{\input{Potential_Strong_7.tex}}
  \end{center}
  \caption{Effective potentials in the strong coupling expansion for $\Ni=1,2,3,4,5,7$ 
with different values of $n$.}
  \label{fig:PotentialStrong}
 \end{figure*}
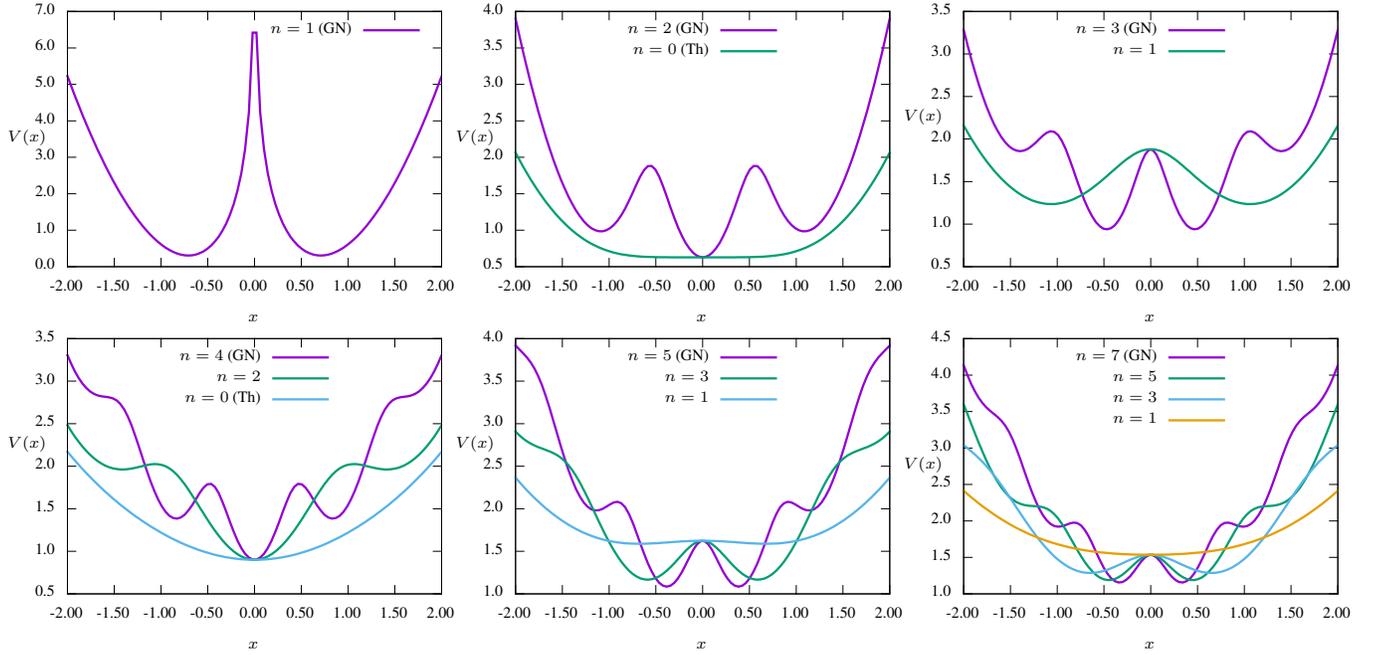%
With the observables from the strong coupling expansion we can calculate the effective potential in the lattice artefact phase, that was discussed around \eqref{e:renorm_coupling}, in the infinite volume limit.
The results for various flavour numbers $\Ni$ are shown in \autoref{fig:PotentialStrong}. For $\Ni=1$, the Gross-Neveu-type potential reads
\begin{equation}
 V_\text{eff}^\text{GN}(x)=2 x^2+\ln \left(\frac{1}{4 x^2}\right)=-2 \ln (2 x)+2 x^2.
\end{equation}
It has a global minimum at $x=\pm 1/\sqrt{2}$ and parity symmetry is spontaneously broken. Its curvature at the origin diverges.

For odd flavour numbers parity symmetry is always broken at strong coupling with 
a minimum of the potential in the Gross-Neveu-like direction. With increasing
$\Ni$ the curvature
at the origin decreases. Extrapolating the curvature of the Gross-Neveu-type potential to $\Ni=\infty$ predicts a broken symmetry in this limit. At strong lattice coupling there 
is no critical flavour number and parity symmetry is always broken for odd $\Ni$. For $\Ni=7$ we observe that the potential with $n=1$ has a positive curvature while this is not the case for smaller $\Ni$, indicating that the breaking of parity symmetry becomes weaker for larger flavour numbers.

For $\Ni=2$, the potentials at strong coupling read
\begin{equation}
\begin{aligned}
 V_\text{eff}^\text{GN}(x)=&2 x^2-\ln \left(8 x^4-4 x^2+1\right)+\ln \left(\frac{15}{8}\right)\\
 =&\ln \left(\frac{15}{8}\right)+6 x^2-\frac{32 x^6}{3}+O\left(x^8\right)\\
 V_\text{eff}^\text{Th}(x)=&x^2-\ln \left(x^4+2 x^2+2\right)+\ln \left(\frac{15}{4}\right)\\
 =&\ln \left(\frac{15}{8}\right)+\frac{x^6}{6}+O\left(x^8\right).
 \end{aligned}
\end{equation}
The curvature in the Gross-Neveu direction is positive.
In the Thirring-like direction the leading power is $x^6$ with a positive 
coefficient such  that the minimum of the potential is at $x=0$. Hence chiral symmetry 
is unbroken. For larger even $\Ni$ the curvature in all directions is positive. 
In conclusion, chiral and parity symmetry is unbroken for any even $\Ni$ and parity symmetry is broken for any odd $\Ni$ in the strong coupling limit.
In the next section we will investigate with Monte Carlo simulations whether these results hold outside the lattice artefact phase and in the continuum limit.

%% file: Potential_Strong_1.tex
\begingroup
\scriptsize
  \makeatletter
  \providecommand\color[2][]{%
    \GenericError{(gnuplot) \space\space\space\@spaces}{%
      Package color not loaded in conjunction with
      terminal option `colourtext'%
    }{See the gnuplot documentation for explanation.%
    }{Either use 'blacktext' in gnuplot or load the package
      color.sty in LaTeX.}%
    \renewcommand\color[2][]{}%
  }%
  \providecommand\includegraphics[2][]{%
    \GenericError{(gnuplot) \space\space\space\@spaces}{%
      Package graphicx or graphics not loaded%
    }{See the gnuplot documentation for explanation.%
    }{The gnuplot epslatex terminal needs graphicx.sty or graphics.sty.}%
    \renewcommand\includegraphics[2][]{}%
  }%
  \providecommand\rotatebox[2]{#2}%
  \@ifundefined{ifGPcolor}{%
    \newif\ifGPcolor
    \GPcolortrue
  }{}%
  \@ifundefined{ifGPblacktext}{%
    \newif\ifGPblacktext
    \GPblacktexttrue
  }{}%
  \let\gplgaddtomacro\g@addto@macro
  \gdef\gplbacktext{}%
  \gdef\gplfronttext{}%
  \makeatother
  \ifGPblacktext
    \def\colorrgb#1{}%
    \def\colorgray#1{}%
  \else
    \ifGPcolor
      \def\colorrgb#1{\color[rgb]{#1}}%
      \def\colorgray#1{\color[gray]{#1}}%
      \expandafter\def\csname LTw\endcsname{\color{white}}%
      \expandafter\def\csname LTb\endcsname{\color{black}}%
      \expandafter\def\csname LTa\endcsname{\color{black}}%
      \expandafter\def\csname LT0\endcsname{\color[rgb]{1,0,0}}%
      \expandafter\def\csname LT1\endcsname{\color[rgb]{0,1,0}}%
      \expandafter\def\csname LT2\endcsname{\color[rgb]{0,0,1}}%
      \expandafter\def\csname LT3\endcsname{\color[rgb]{1,0,1}}%
      \expandafter\def\csname LT4\endcsname{\color[rgb]{0,1,1}}%
      \expandafter\def\csname LT5\endcsname{\color[rgb]{1,1,0}}%
      \expandafter\def\csname LT6\endcsname{\color[rgb]{0,0,0}}%
      \expandafter\def\csname LT7\endcsname{\color[rgb]{1,0.3,0}}%
      \expandafter\def\csname LT8\endcsname{\color[rgb]{0.5,0.5,0.5}}%
    \else
      \def\colorrgb#1{\color{black}}%
      \def\colorgray#1{\color[gray]{#1}}%
      \expandafter\def\csname LTw\endcsname{\color{white}}%
      \expandafter\def\csname LTb\endcsname{\color{black}}%
      \expandafter\def\csname LTa\endcsname{\color{black}}%
      \expandafter\def\csname LT0\endcsname{\color{black}}%
      \expandafter\def\csname LT1\endcsname{\color{black}}%
      \expandafter\def\csname LT2\endcsname{\color{black}}%
      \expandafter\def\csname LT3\endcsname{\color{black}}%
      \expandafter\def\csname LT4\endcsname{\color{black}}%
      \expandafter\def\csname LT5\endcsname{\color{black}}%
      \expandafter\def\csname LT6\endcsname{\color{black}}%
      \expandafter\def\csname LT7\endcsname{\color{black}}%
      \expandafter\def\csname LT8\endcsname{\color{black}}%
    \fi
  \fi
  \setlength{\unitlength}{0.0500bp}%
  \begin{picture}(3840.00,2880.00)%
    \gplgaddtomacro\gplbacktext{%
      \csname LTb\endcsname%
      \put(408,595){\makebox(0,0)[r]{\strut{}0.0}}%
      \csname LTb\endcsname%
      \put(408,919){\makebox(0,0)[r]{\strut{}1.0}}%
      \csname LTb\endcsname%
      \put(408,1242){\makebox(0,0)[r]{\strut{}2.0}}%
      \csname LTb\endcsname%
      \put(408,1566){\makebox(0,0)[r]{\strut{}3.0}}%
      \csname LTb\endcsname%
      \put(408,1889){\makebox(0,0)[r]{\strut{}4.0}}%
      \csname LTb\endcsname%
      \put(408,2213){\makebox(0,0)[r]{\strut{}5.0}}%
      \csname LTb\endcsname%
      \put(408,2536){\makebox(0,0)[r]{\strut{}6.0}}%
      \csname LTb\endcsname%
      \put(408,2860){\makebox(0,0)[r]{\strut{}7.0}}%
      \csname LTb\endcsname%
      \put(510,409){\makebox(0,0){\strut{}-2.00}}%
      \csname LTb\endcsname%
      \put(925,409){\makebox(0,0){\strut{}-1.50}}%
      \csname LTb\endcsname%
      \put(1340,409){\makebox(0,0){\strut{}-1.00}}%
      \csname LTb\endcsname%
      \put(1754,409){\makebox(0,0){\strut{}-0.50}}%
      \csname LTb\endcsname%
      \put(2169,409){\makebox(0,0){\strut{}0.00}}%
      \csname LTb\endcsname%
      \put(2584,409){\makebox(0,0){\strut{}0.50}}%
      \csname LTb\endcsname%
      \put(2999,409){\makebox(0,0){\strut{}1.00}}%
      \csname LTb\endcsname%
      \put(3413,409){\makebox(0,0){\strut{}1.50}}%
      \csname LTb\endcsname%
      \put(3828,409){\makebox(0,0){\strut{}2.00}}%
      \csname LTb\endcsname%
      \put(162,1727){\makebox(0,0){\strut{}$V(x)$}}%
      \csname LTb\endcsname%
      \put(2169,130){\makebox(0,0){\strut{}$x$}}%
    }%
    \gplgaddtomacro\gplfronttext{%
      \csname LTb\endcsname%
      \put(3040,2693){\makebox(0,0)[r]{\strut{}$n=1$ (GN)}}%
    }%
    \gplbacktext
    \put(0,0){\includegraphics{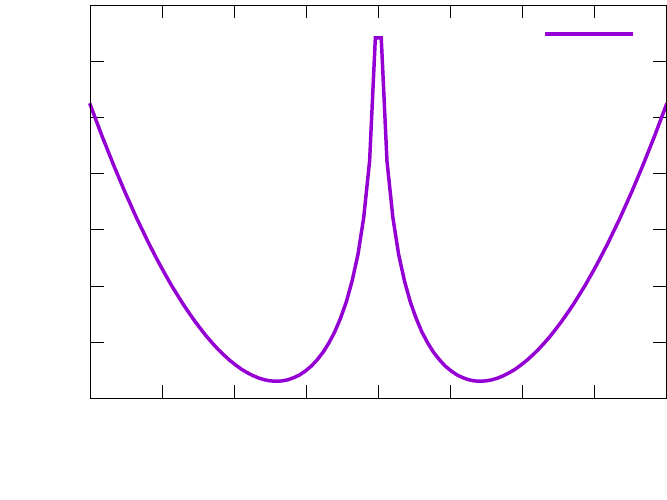}}%
    \gplfronttext
  \end{picture}%
\endgroup

%% file: Potential_Strong_2.tex
\begingroup
\scriptsize
  \makeatletter
  \providecommand\color[2][]{%
    \GenericError{(gnuplot) \space\space\space\@spaces}{%
      Package color not loaded in conjunction with
      terminal option `colourtext'%
    }{See the gnuplot documentation for explanation.%
    }{Either use 'blacktext' in gnuplot or load the package
      color.sty in LaTeX.}%
    \renewcommand\color[2][]{}%
  }%
  \providecommand\includegraphics[2][]{%
    \GenericError{(gnuplot) \space\space\space\@spaces}{%
      Package graphicx or graphics not loaded%
    }{See the gnuplot documentation for explanation.%
    }{The gnuplot epslatex terminal needs graphicx.sty or graphics.sty.}%
    \renewcommand\includegraphics[2][]{}%
  }%
  \providecommand\rotatebox[2]{#2}%
  \@ifundefined{ifGPcolor}{%
    \newif\ifGPcolor
    \GPcolortrue
  }{}%
  \@ifundefined{ifGPblacktext}{%
    \newif\ifGPblacktext
    \GPblacktexttrue
  }{}%
  \let\gplgaddtomacro\g@addto@macro
  \gdef\gplbacktext{}%
  \gdef\gplfronttext{}%
  \makeatother
  \ifGPblacktext
    \def\colorrgb#1{}%
    \def\colorgray#1{}%
  \else
    \ifGPcolor
      \def\colorrgb#1{\color[rgb]{#1}}%
      \def\colorgray#1{\color[gray]{#1}}%
      \expandafter\def\csname LTw\endcsname{\color{white}}%
      \expandafter\def\csname LTb\endcsname{\color{black}}%
      \expandafter\def\csname LTa\endcsname{\color{black}}%
      \expandafter\def\csname LT0\endcsname{\color[rgb]{1,0,0}}%
      \expandafter\def\csname LT1\endcsname{\color[rgb]{0,1,0}}%
      \expandafter\def\csname LT2\endcsname{\color[rgb]{0,0,1}}%
      \expandafter\def\csname LT3\endcsname{\color[rgb]{1,0,1}}%
      \expandafter\def\csname LT4\endcsname{\color[rgb]{0,1,1}}%
      \expandafter\def\csname LT5\endcsname{\color[rgb]{1,1,0}}%
      \expandafter\def\csname LT6\endcsname{\color[rgb]{0,0,0}}%
      \expandafter\def\csname LT7\endcsname{\color[rgb]{1,0.3,0}}%
      \expandafter\def\csname LT8\endcsname{\color[rgb]{0.5,0.5,0.5}}%
    \else
      \def\colorrgb#1{\color{black}}%
      \def\colorgray#1{\color[gray]{#1}}%
      \expandafter\def\csname LTw\endcsname{\color{white}}%
      \expandafter\def\csname LTb\endcsname{\color{black}}%
      \expandafter\def\csname LTa\endcsname{\color{black}}%
      \expandafter\def\csname LT0\endcsname{\color{black}}%
      \expandafter\def\csname LT1\endcsname{\color{black}}%
      \expandafter\def\csname LT2\endcsname{\color{black}}%
      \expandafter\def\csname LT3\endcsname{\color{black}}%
      \expandafter\def\csname LT4\endcsname{\color{black}}%
      \expandafter\def\csname LT5\endcsname{\color{black}}%
      \expandafter\def\csname LT6\endcsname{\color{black}}%
      \expandafter\def\csname LT7\endcsname{\color{black}}%
      \expandafter\def\csname LT8\endcsname{\color{black}}%
    \fi
  \fi
  \setlength{\unitlength}{0.0500bp}%
  \begin{picture}(3840.00,2880.00)%
    \gplgaddtomacro\gplbacktext{%
      \csname LTb\endcsname%
      \put(408,595){\makebox(0,0)[r]{\strut{}0.5}}%
      \csname LTb\endcsname%
      \put(408,919){\makebox(0,0)[r]{\strut{}1.0}}%
      \csname LTb\endcsname%
      \put(408,1242){\makebox(0,0)[r]{\strut{}1.5}}%
      \csname LTb\endcsname%
      \put(408,1566){\makebox(0,0)[r]{\strut{}2.0}}%
      \csname LTb\endcsname%
      \put(408,1889){\makebox(0,0)[r]{\strut{}2.5}}%
      \csname LTb\endcsname%
      \put(408,2213){\makebox(0,0)[r]{\strut{}3.0}}%
      \csname LTb\endcsname%
      \put(408,2536){\makebox(0,0)[r]{\strut{}3.5}}%
      \csname LTb\endcsname%
      \put(408,2860){\makebox(0,0)[r]{\strut{}4.0}}%
      \csname LTb\endcsname%
      \put(510,409){\makebox(0,0){\strut{}-2.00}}%
      \csname LTb\endcsname%
      \put(925,409){\makebox(0,0){\strut{}-1.50}}%
      \csname LTb\endcsname%
      \put(1340,409){\makebox(0,0){\strut{}-1.00}}%
      \csname LTb\endcsname%
      \put(1754,409){\makebox(0,0){\strut{}-0.50}}%
      \csname LTb\endcsname%
      \put(2169,409){\makebox(0,0){\strut{}0.00}}%
      \csname LTb\endcsname%
      \put(2584,409){\makebox(0,0){\strut{}0.50}}%
      \csname LTb\endcsname%
      \put(2999,409){\makebox(0,0){\strut{}1.00}}%
      \csname LTb\endcsname%
      \put(3413,409){\makebox(0,0){\strut{}1.50}}%
      \csname LTb\endcsname%
      \put(3828,409){\makebox(0,0){\strut{}2.00}}%
      \csname LTb\endcsname%
      \put(162,1727){\makebox(0,0){\strut{}$V(x)$}}%
      \csname LTb\endcsname%
      \put(2169,130){\makebox(0,0){\strut{}$x$}}%
    }%
    \gplgaddtomacro\gplfronttext{%
      \csname LTb\endcsname%
      \put(2234,2693){\makebox(0,0)[r]{\strut{}$n=2$ (GN)}}%
      \csname LTb\endcsname%
      \put(2234,2507){\makebox(0,0)[r]{\strut{}$n=0$ (Th)}}%
    }%
    \gplbacktext
    \put(0,0){\includegraphics{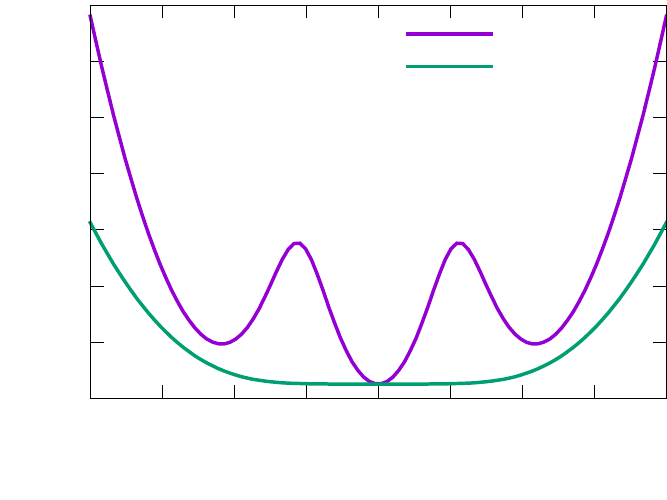}}%
    \gplfronttext
  \end{picture}%
\endgroup

%% file: Potential_Strong_3.tex
\begingroup
\scriptsize
  \makeatletter
  \providecommand\color[2][]{%
    \GenericError{(gnuplot) \space\space\space\@spaces}{%
      Package color not loaded in conjunction with
      terminal option `colourtext'%
    }{See the gnuplot documentation for explanation.%
    }{Either use 'blacktext' in gnuplot or load the package
      color.sty in LaTeX.}%
    \renewcommand\color[2][]{}%
  }%
  \providecommand\includegraphics[2][]{%
    \GenericError{(gnuplot) \space\space\space\@spaces}{%
      Package graphicx or graphics not loaded%
    }{See the gnuplot documentation for explanation.%
    }{The gnuplot epslatex terminal needs graphicx.sty or graphics.sty.}%
    \renewcommand\includegraphics[2][]{}%
  }%
  \providecommand\rotatebox[2]{#2}%
  \@ifundefined{ifGPcolor}{%
    \newif\ifGPcolor
    \GPcolortrue
  }{}%
  \@ifundefined{ifGPblacktext}{%
    \newif\ifGPblacktext
    \GPblacktexttrue
  }{}%
  \let\gplgaddtomacro\g@addto@macro
  \gdef\gplbacktext{}%
  \gdef\gplfronttext{}%
  \makeatother
  \ifGPblacktext
    \def\colorrgb#1{}%
    \def\colorgray#1{}%
  \else
    \ifGPcolor
      \def\colorrgb#1{\color[rgb]{#1}}%
      \def\colorgray#1{\color[gray]{#1}}%
      \expandafter\def\csname LTw\endcsname{\color{white}}%
      \expandafter\def\csname LTb\endcsname{\color{black}}%
      \expandafter\def\csname LTa\endcsname{\color{black}}%
      \expandafter\def\csname LT0\endcsname{\color[rgb]{1,0,0}}%
      \expandafter\def\csname LT1\endcsname{\color[rgb]{0,1,0}}%
      \expandafter\def\csname LT2\endcsname{\color[rgb]{0,0,1}}%
      \expandafter\def\csname LT3\endcsname{\color[rgb]{1,0,1}}%
      \expandafter\def\csname LT4\endcsname{\color[rgb]{0,1,1}}%
      \expandafter\def\csname LT5\endcsname{\color[rgb]{1,1,0}}%
      \expandafter\def\csname LT6\endcsname{\color[rgb]{0,0,0}}%
      \expandafter\def\csname LT7\endcsname{\color[rgb]{1,0.3,0}}%
      \expandafter\def\csname LT8\endcsname{\color[rgb]{0.5,0.5,0.5}}%
    \else
      \def\colorrgb#1{\color{black}}%
      \def\colorgray#1{\color[gray]{#1}}%
      \expandafter\def\csname LTw\endcsname{\color{white}}%
      \expandafter\def\csname LTb\endcsname{\color{black}}%
      \expandafter\def\csname LTa\endcsname{\color{black}}%
      \expandafter\def\csname LT0\endcsname{\color{black}}%
      \expandafter\def\csname LT1\endcsname{\color{black}}%
      \expandafter\def\csname LT2\endcsname{\color{black}}%
      \expandafter\def\csname LT3\endcsname{\color{black}}%
      \expandafter\def\csname LT4\endcsname{\color{black}}%
      \expandafter\def\csname LT5\endcsname{\color{black}}%
      \expandafter\def\csname LT6\endcsname{\color{black}}%
      \expandafter\def\csname LT7\endcsname{\color{black}}%
      \expandafter\def\csname LT8\endcsname{\color{black}}%
    \fi
  \fi
  \setlength{\unitlength}{0.0500bp}%
  \begin{picture}(3840.00,2880.00)%
    \gplgaddtomacro\gplbacktext{%
      \csname LTb\endcsname%
      \put(408,595){\makebox(0,0)[r]{\strut{}0.5}}%
      \csname LTb\endcsname%
      \put(408,973){\makebox(0,0)[r]{\strut{}1.0}}%
      \csname LTb\endcsname%
      \put(408,1350){\makebox(0,0)[r]{\strut{}1.5}}%
      \csname LTb\endcsname%
      \put(408,1728){\makebox(0,0)[r]{\strut{}2.0}}%
      \csname LTb\endcsname%
      \put(408,2105){\makebox(0,0)[r]{\strut{}2.5}}%
      \csname LTb\endcsname%
      \put(408,2483){\makebox(0,0)[r]{\strut{}3.0}}%
      \csname LTb\endcsname%
      \put(408,2860){\makebox(0,0)[r]{\strut{}3.5}}%
      \csname LTb\endcsname%
      \put(510,409){\makebox(0,0){\strut{}-2.00}}%
      \csname LTb\endcsname%
      \put(925,409){\makebox(0,0){\strut{}-1.50}}%
      \csname LTb\endcsname%
      \put(1340,409){\makebox(0,0){\strut{}-1.00}}%
      \csname LTb\endcsname%
      \put(1754,409){\makebox(0,0){\strut{}-0.50}}%
      \csname LTb\endcsname%
      \put(2169,409){\makebox(0,0){\strut{}0.00}}%
      \csname LTb\endcsname%
      \put(2584,409){\makebox(0,0){\strut{}0.50}}%
      \csname LTb\endcsname%
      \put(2999,409){\makebox(0,0){\strut{}1.00}}%
      \csname LTb\endcsname%
      \put(3413,409){\makebox(0,0){\strut{}1.50}}%
      \csname LTb\endcsname%
      \put(3828,409){\makebox(0,0){\strut{}2.00}}%
      \csname LTb\endcsname%
      \put(162,1913){\makebox(0,0){\strut{}$V(x)$}}%
      \csname LTb\endcsname%
      \put(2169,130){\makebox(0,0){\strut{}$x$}}%
    }%
    \gplgaddtomacro\gplfronttext{%
      \csname LTb\endcsname%
      \put(2234,2693){\makebox(0,0)[r]{\strut{}$n=3$ (GN)}}%
      \csname LTb\endcsname%
      \put(2234,2507){\makebox(0,0)[r]{\strut{}$n=1$}}%
    }%
    \gplbacktext
    \put(0,0){\includegraphics{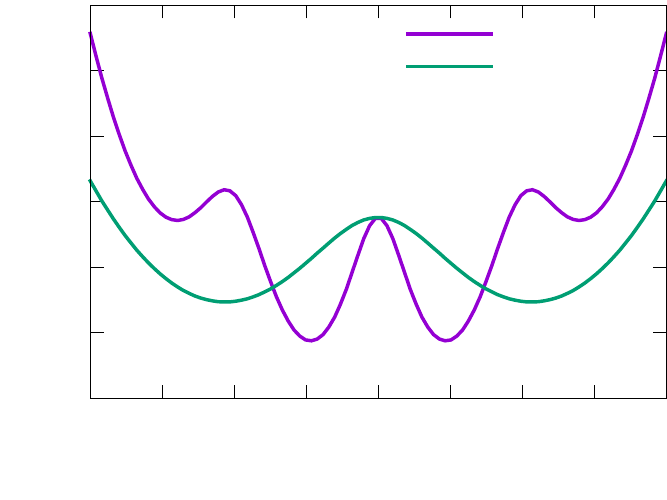}}%
    \gplfronttext
  \end{picture}%
\endgroup

%% file: Potential_Strong_4.tex
\begingroup
\scriptsize
  \makeatletter
  \providecommand\color[2][]{%
    \GenericError{(gnuplot) \space\space\space\@spaces}{%
      Package color not loaded in conjunction with
      terminal option `colourtext'%
    }{See the gnuplot documentation for explanation.%
    }{Either use 'blacktext' in gnuplot or load the package
      color.sty in LaTeX.}%
    \renewcommand\color[2][]{}%
  }%
  \providecommand\includegraphics[2][]{%
    \GenericError{(gnuplot) \space\space\space\@spaces}{%
      Package graphicx or graphics not loaded%
    }{See the gnuplot documentation for explanation.%
    }{The gnuplot epslatex terminal needs graphicx.sty or graphics.sty.}%
    \renewcommand\includegraphics[2][]{}%
  }%
  \providecommand\rotatebox[2]{#2}%
  \@ifundefined{ifGPcolor}{%
    \newif\ifGPcolor
    \GPcolortrue
  }{}%
  \@ifundefined{ifGPblacktext}{%
    \newif\ifGPblacktext
    \GPblacktexttrue
  }{}%
  \let\gplgaddtomacro\g@addto@macro
  \gdef\gplbacktext{}%
  \gdef\gplfronttext{}%
  \makeatother
  \ifGPblacktext
    \def\colorrgb#1{}%
    \def\colorgray#1{}%
  \else
    \ifGPcolor
      \def\colorrgb#1{\color[rgb]{#1}}%
      \def\colorgray#1{\color[gray]{#1}}%
      \expandafter\def\csname LTw\endcsname{\color{white}}%
      \expandafter\def\csname LTb\endcsname{\color{black}}%
      \expandafter\def\csname LTa\endcsname{\color{black}}%
      \expandafter\def\csname LT0\endcsname{\color[rgb]{1,0,0}}%
      \expandafter\def\csname LT1\endcsname{\color[rgb]{0,1,0}}%
      \expandafter\def\csname LT2\endcsname{\color[rgb]{0,0,1}}%
      \expandafter\def\csname LT3\endcsname{\color[rgb]{1,0,1}}%
      \expandafter\def\csname LT4\endcsname{\color[rgb]{0,1,1}}%
      \expandafter\def\csname LT5\endcsname{\color[rgb]{1,1,0}}%
      \expandafter\def\csname LT6\endcsname{\color[rgb]{0,0,0}}%
      \expandafter\def\csname LT7\endcsname{\color[rgb]{1,0.3,0}}%
      \expandafter\def\csname LT8\endcsname{\color[rgb]{0.5,0.5,0.5}}%
    \else
      \def\colorrgb#1{\color{black}}%
      \def\colorgray#1{\color[gray]{#1}}%
      \expandafter\def\csname LTw\endcsname{\color{white}}%
      \expandafter\def\csname LTb\endcsname{\color{black}}%
      \expandafter\def\csname LTa\endcsname{\color{black}}%
      \expandafter\def\csname LT0\endcsname{\color{black}}%
      \expandafter\def\csname LT1\endcsname{\color{black}}%
      \expandafter\def\csname LT2\endcsname{\color{black}}%
      \expandafter\def\csname LT3\endcsname{\color{black}}%
      \expandafter\def\csname LT4\endcsname{\color{black}}%
      \expandafter\def\csname LT5\endcsname{\color{black}}%
      \expandafter\def\csname LT6\endcsname{\color{black}}%
      \expandafter\def\csname LT7\endcsname{\color{black}}%
      \expandafter\def\csname LT8\endcsname{\color{black}}%
    \fi
  \fi
  \setlength{\unitlength}{0.0500bp}%
  \begin{picture}(3840.00,2880.00)%
    \gplgaddtomacro\gplbacktext{%
      \csname LTb\endcsname%
      \put(408,595){\makebox(0,0)[r]{\strut{}0.5}}%
      \csname LTb\endcsname%
      \put(408,973){\makebox(0,0)[r]{\strut{}1.0}}%
      \csname LTb\endcsname%
      \put(408,1350){\makebox(0,0)[r]{\strut{}1.5}}%
      \csname LTb\endcsname%
      \put(408,1728){\makebox(0,0)[r]{\strut{}2.0}}%
      \csname LTb\endcsname%
      \put(408,2105){\makebox(0,0)[r]{\strut{}2.5}}%
      \csname LTb\endcsname%
      \put(408,2483){\makebox(0,0)[r]{\strut{}3.0}}%
      \csname LTb\endcsname%
      \put(408,2860){\makebox(0,0)[r]{\strut{}3.5}}%
      \csname LTb\endcsname%
      \put(510,409){\makebox(0,0){\strut{}-2.00}}%
      \csname LTb\endcsname%
      \put(925,409){\makebox(0,0){\strut{}-1.50}}%
      \csname LTb\endcsname%
      \put(1340,409){\makebox(0,0){\strut{}-1.00}}%
      \csname LTb\endcsname%
      \put(1754,409){\makebox(0,0){\strut{}-0.50}}%
      \csname LTb\endcsname%
      \put(2169,409){\makebox(0,0){\strut{}0.00}}%
      \csname LTb\endcsname%
      \put(2584,409){\makebox(0,0){\strut{}0.50}}%
      \csname LTb\endcsname%
      \put(2999,409){\makebox(0,0){\strut{}1.00}}%
      \csname LTb\endcsname%
      \put(3413,409){\makebox(0,0){\strut{}1.50}}%
      \csname LTb\endcsname%
      \put(3828,409){\makebox(0,0){\strut{}2.00}}%
      \csname LTb\endcsname%
      \put(162,1913){\makebox(0,0){\strut{}$V(x)$}}%
      \csname LTb\endcsname%
      \put(2169,130){\makebox(0,0){\strut{}$x$}}%
    }%
    \gplgaddtomacro\gplfronttext{%
      \csname LTb\endcsname%
      \put(2234,2693){\makebox(0,0)[r]{\strut{}$n=4$ (GN)}}%
      \csname LTb\endcsname%
      \put(2234,2507){\makebox(0,0)[r]{\strut{}$n=2$}}%
      \csname LTb\endcsname%
      \put(2234,2321){\makebox(0,0)[r]{\strut{}$n=0$ (Th)}}%
    }%
    \gplbacktext
    \put(0,0){\includegraphics{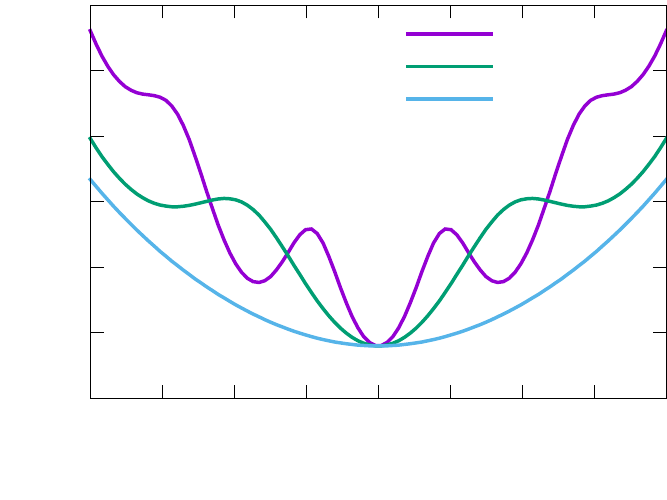}}%
    \gplfronttext
  \end{picture}%
\endgroup

%% file: Potential_Strong_5.tex
\begingroup
\scriptsize
  \makeatletter
  \providecommand\color[2][]{%
    \GenericError{(gnuplot) \space\space\space\@spaces}{%
      Package color not loaded in conjunction with
      terminal option `colourtext'%
    }{See the gnuplot documentation for explanation.%
    }{Either use 'blacktext' in gnuplot or load the package
      color.sty in LaTeX.}%
    \renewcommand\color[2][]{}%
  }%
  \providecommand\includegraphics[2][]{%
    \GenericError{(gnuplot) \space\space\space\@spaces}{%
      Package graphicx or graphics not loaded%
    }{See the gnuplot documentation for explanation.%
    }{The gnuplot epslatex terminal needs graphicx.sty or graphics.sty.}%
    \renewcommand\includegraphics[2][]{}%
  }%
  \providecommand\rotatebox[2]{#2}%
  \@ifundefined{ifGPcolor}{%
    \newif\ifGPcolor
    \GPcolortrue
  }{}%
  \@ifundefined{ifGPblacktext}{%
    \newif\ifGPblacktext
    \GPblacktexttrue
  }{}%
  \let\gplgaddtomacro\g@addto@macro
  \gdef\gplbacktext{}%
  \gdef\gplfronttext{}%
  \makeatother
  \ifGPblacktext
    \def\colorrgb#1{}%
    \def\colorgray#1{}%
  \else
    \ifGPcolor
      \def\colorrgb#1{\color[rgb]{#1}}%
      \def\colorgray#1{\color[gray]{#1}}%
      \expandafter\def\csname LTw\endcsname{\color{white}}%
      \expandafter\def\csname LTb\endcsname{\color{black}}%
      \expandafter\def\csname LTa\endcsname{\color{black}}%
      \expandafter\def\csname LT0\endcsname{\color[rgb]{1,0,0}}%
      \expandafter\def\csname LT1\endcsname{\color[rgb]{0,1,0}}%
      \expandafter\def\csname LT2\endcsname{\color[rgb]{0,0,1}}%
      \expandafter\def\csname LT3\endcsname{\color[rgb]{1,0,1}}%
      \expandafter\def\csname LT4\endcsname{\color[rgb]{0,1,1}}%
      \expandafter\def\csname LT5\endcsname{\color[rgb]{1,1,0}}%
      \expandafter\def\csname LT6\endcsname{\color[rgb]{0,0,0}}%
      \expandafter\def\csname LT7\endcsname{\color[rgb]{1,0.3,0}}%
      \expandafter\def\csname LT8\endcsname{\color[rgb]{0.5,0.5,0.5}}%
    \else
      \def\colorrgb#1{\color{black}}%
      \def\colorgray#1{\color[gray]{#1}}%
      \expandafter\def\csname LTw\endcsname{\color{white}}%
      \expandafter\def\csname LTb\endcsname{\color{black}}%
      \expandafter\def\csname LTa\endcsname{\color{black}}%
      \expandafter\def\csname LT0\endcsname{\color{black}}%
      \expandafter\def\csname LT1\endcsname{\color{black}}%
      \expandafter\def\csname LT2\endcsname{\color{black}}%
      \expandafter\def\csname LT3\endcsname{\color{black}}%
      \expandafter\def\csname LT4\endcsname{\color{black}}%
      \expandafter\def\csname LT5\endcsname{\color{black}}%
      \expandafter\def\csname LT6\endcsname{\color{black}}%
      \expandafter\def\csname LT7\endcsname{\color{black}}%
      \expandafter\def\csname LT8\endcsname{\color{black}}%
    \fi
  \fi
  \setlength{\unitlength}{0.0500bp}%
  \begin{picture}(3840.00,2880.00)%
    \gplgaddtomacro\gplbacktext{%
      \csname LTb\endcsname%
      \put(408,595){\makebox(0,0)[r]{\strut{}1.0}}%
      \csname LTb\endcsname%
      \put(408,973){\makebox(0,0)[r]{\strut{}1.5}}%
      \csname LTb\endcsname%
      \put(408,1350){\makebox(0,0)[r]{\strut{}2.0}}%
      \csname LTb\endcsname%
      \put(408,1728){\makebox(0,0)[r]{\strut{}2.5}}%
      \csname LTb\endcsname%
      \put(408,2105){\makebox(0,0)[r]{\strut{}3.0}}%
      \csname LTb\endcsname%
      \put(408,2483){\makebox(0,0)[r]{\strut{}3.5}}%
      \csname LTb\endcsname%
      \put(408,2860){\makebox(0,0)[r]{\strut{}4.0}}%
      \csname LTb\endcsname%
      \put(510,409){\makebox(0,0){\strut{}-2.00}}%
      \csname LTb\endcsname%
      \put(925,409){\makebox(0,0){\strut{}-1.50}}%
      \csname LTb\endcsname%
      \put(1340,409){\makebox(0,0){\strut{}-1.00}}%
      \csname LTb\endcsname%
      \put(1754,409){\makebox(0,0){\strut{}-0.50}}%
      \csname LTb\endcsname%
      \put(2169,409){\makebox(0,0){\strut{}0.00}}%
      \csname LTb\endcsname%
      \put(2584,409){\makebox(0,0){\strut{}0.50}}%
      \csname LTb\endcsname%
      \put(2999,409){\makebox(0,0){\strut{}1.00}}%
      \csname LTb\endcsname%
      \put(3413,409){\makebox(0,0){\strut{}1.50}}%
      \csname LTb\endcsname%
      \put(3828,409){\makebox(0,0){\strut{}2.00}}%
      \csname LTb\endcsname%
      \put(162,1913){\makebox(0,0){\strut{}$V(x)$}}%
      \csname LTb\endcsname%
      \put(2169,130){\makebox(0,0){\strut{}$x$}}%
    }%
    \gplgaddtomacro\gplfronttext{%
      \csname LTb\endcsname%
      \put(2234,2693){\makebox(0,0)[r]{\strut{}$n=5$ (GN)}}%
      \csname LTb\endcsname%
      \put(2234,2507){\makebox(0,0)[r]{\strut{}$n=3$}}%
      \csname LTb\endcsname%
      \put(2234,2321){\makebox(0,0)[r]{\strut{}$n=1$}}%
    }%
    \gplbacktext
    \put(0,0){\includegraphics{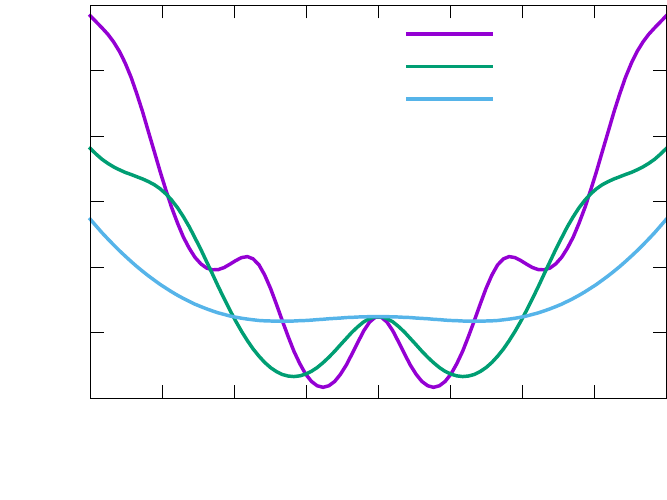}}%
    \gplfronttext
  \end{picture}%
\endgroup

%% file: Potential_Strong_7.tex
\begingroup
\scriptsize
  \makeatletter
  \providecommand\color[2][]{%
    \GenericError{(gnuplot) \space\space\space\@spaces}{%
      Package color not loaded in conjunction with
      terminal option `colourtext'%
    }{See the gnuplot documentation for explanation.%
    }{Either use 'blacktext' in gnuplot or load the package
      color.sty in LaTeX.}%
    \renewcommand\color[2][]{}%
  }%
  \providecommand\includegraphics[2][]{%
    \GenericError{(gnuplot) \space\space\space\@spaces}{%
      Package graphicx or graphics not loaded%
    }{See the gnuplot documentation for explanation.%
    }{The gnuplot epslatex terminal needs graphicx.sty or graphics.sty.}%
    \renewcommand\includegraphics[2][]{}%
  }%
  \providecommand\rotatebox[2]{#2}%
  \@ifundefined{ifGPcolor}{%
    \newif\ifGPcolor
    \GPcolortrue
  }{}%
  \@ifundefined{ifGPblacktext}{%
    \newif\ifGPblacktext
    \GPblacktexttrue
  }{}%
  \let\gplgaddtomacro\g@addto@macro
  \gdef\gplbacktext{}%
  \gdef\gplfronttext{}%
  \makeatother
  \ifGPblacktext
    \def\colorrgb#1{}%
    \def\colorgray#1{}%
  \else
    \ifGPcolor
      \def\colorrgb#1{\color[rgb]{#1}}%
      \def\colorgray#1{\color[gray]{#1}}%
      \expandafter\def\csname LTw\endcsname{\color{white}}%
      \expandafter\def\csname LTb\endcsname{\color{black}}%
      \expandafter\def\csname LTa\endcsname{\color{black}}%
      \expandafter\def\csname LT0\endcsname{\color[rgb]{1,0,0}}%
      \expandafter\def\csname LT1\endcsname{\color[rgb]{0,1,0}}%
      \expandafter\def\csname LT2\endcsname{\color[rgb]{0,0,1}}%
      \expandafter\def\csname LT3\endcsname{\color[rgb]{1,0,1}}%
      \expandafter\def\csname LT4\endcsname{\color[rgb]{0,1,1}}%
      \expandafter\def\csname LT5\endcsname{\color[rgb]{1,1,0}}%
      \expandafter\def\csname LT6\endcsname{\color[rgb]{0,0,0}}%
      \expandafter\def\csname LT7\endcsname{\color[rgb]{1,0.3,0}}%
      \expandafter\def\csname LT8\endcsname{\color[rgb]{0.5,0.5,0.5}}%
    \else
      \def\colorrgb#1{\color{black}}%
      \def\colorgray#1{\color[gray]{#1}}%
      \expandafter\def\csname LTw\endcsname{\color{white}}%
      \expandafter\def\csname LTb\endcsname{\color{black}}%
      \expandafter\def\csname LTa\endcsname{\color{black}}%
      \expandafter\def\csname LT0\endcsname{\color{black}}%
      \expandafter\def\csname LT1\endcsname{\color{black}}%
      \expandafter\def\csname LT2\endcsname{\color{black}}%
      \expandafter\def\csname LT3\endcsname{\color{black}}%
      \expandafter\def\csname LT4\endcsname{\color{black}}%
      \expandafter\def\csname LT5\endcsname{\color{black}}%
      \expandafter\def\csname LT6\endcsname{\color{black}}%
      \expandafter\def\csname LT7\endcsname{\color{black}}%
      \expandafter\def\csname LT8\endcsname{\color{black}}%
    \fi
  \fi
  \setlength{\unitlength}{0.0500bp}%
  \begin{picture}(3840.00,2880.00)%
    \gplgaddtomacro\gplbacktext{%
      \csname LTb\endcsname%
      \put(408,595){\makebox(0,0)[r]{\strut{}1.0}}%
      \csname LTb\endcsname%
      \put(408,919){\makebox(0,0)[r]{\strut{}1.5}}%
      \csname LTb\endcsname%
      \put(408,1242){\makebox(0,0)[r]{\strut{}2.0}}%
      \csname LTb\endcsname%
      \put(408,1566){\makebox(0,0)[r]{\strut{}2.5}}%
      \csname LTb\endcsname%
      \put(408,1889){\makebox(0,0)[r]{\strut{}3.0}}%
      \csname LTb\endcsname%
      \put(408,2213){\makebox(0,0)[r]{\strut{}3.5}}%
      \csname LTb\endcsname%
      \put(408,2536){\makebox(0,0)[r]{\strut{}4.0}}%
      \csname LTb\endcsname%
      \put(408,2860){\makebox(0,0)[r]{\strut{}4.5}}%
      \csname LTb\endcsname%
      \put(510,409){\makebox(0,0){\strut{}-2.00}}%
      \csname LTb\endcsname%
      \put(925,409){\makebox(0,0){\strut{}-1.50}}%
      \csname LTb\endcsname%
      \put(1340,409){\makebox(0,0){\strut{}-1.00}}%
      \csname LTb\endcsname%
      \put(1754,409){\makebox(0,0){\strut{}-0.50}}%
      \csname LTb\endcsname%
      \put(2169,409){\makebox(0,0){\strut{}0.00}}%
      \csname LTb\endcsname%
      \put(2584,409){\makebox(0,0){\strut{}0.50}}%
      \csname LTb\endcsname%
      \put(2999,409){\makebox(0,0){\strut{}1.00}}%
      \csname LTb\endcsname%
      \put(3413,409){\makebox(0,0){\strut{}1.50}}%
      \csname LTb\endcsname%
      \put(3828,409){\makebox(0,0){\strut{}2.00}}%
      \csname LTb\endcsname%
      \put(162,1727){\makebox(0,0){\strut{}$V(x)$}}%
      \csname LTb\endcsname%
      \put(2169,130){\makebox(0,0){\strut{}$x$}}%
    }%
    \gplgaddtomacro\gplfronttext{%
      \csname LTb\endcsname%
      \put(2234,2693){\makebox(0,0)[r]{\strut{}$n=7$ (GN)}}%
      \csname LTb\endcsname%
      \put(2234,2507){\makebox(0,0)[r]{\strut{}$n=5$}}%
      \csname LTb\endcsname%
      \put(2234,2321){\makebox(0,0)[r]{\strut{}$n=3$}}%
      \csname LTb\endcsname%
      \put(2234,2135){\makebox(0,0)[r]{\strut{}$n=1$}}%
    }%
    \gplbacktext
    \put(0,0){\includegraphics{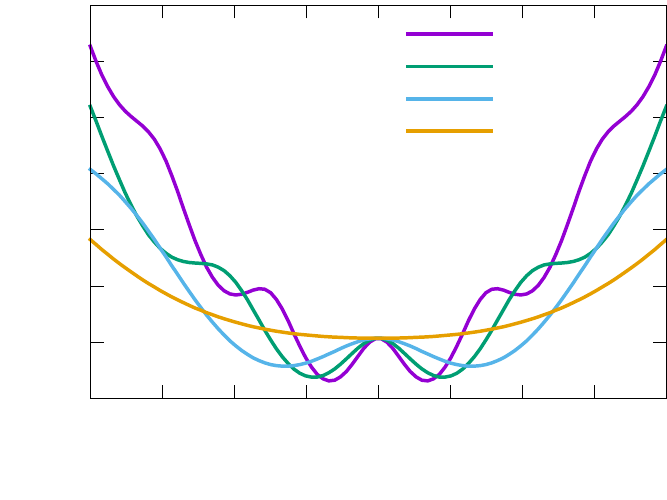}}%
    \gplfronttext
  \end{picture}%
\endgroup

%% file: results.tex
\section{Simulation results}
\label{s:results}
\noindent
 The simulations have been performed with the lattice action
 \begin{equation}
S(\lambda)=\Ni\left(\frac{\lambda}{2} \sum \limits_x v_\mu^2(x)-
 \ln \det (\ii\slashed{D}) \right)
 \end{equation}
for fermions in the irreducible representation of the Clifford algebra with 
 $\gamma_\mu=\sigma_\mu$.
 Note, that we have rescaled $\lambda\to\Ni\lambda$ for our simulations.
 We use the SLAC derivative for fermions with hermitian Dirac operator
 \begin{equation}
\ii\slashed{D}=\sigma_\mu\left(\ii\,\partial_\mu^\text{SLAC}+ v_\mu\right)
 \end{equation} 
 because it preserves the continuum $U(\Ni)$ chiral symmetry and the 
 discrete parity symmetry exactly even
 at finite lattice spacing. Note that Wilson fermions in the irreducible representation also preserve the continuum chiral symmetry but break parity. 
 In order to implement antiperiodic boundary
 conditions for the fermions in time direction, we simulate on lattices with volumes $V=L \times (L-1)^2$ with even $L$. Most of our simulations have been carried out on lattices with $L=8,12,16$ and $20$ with statistics of $1\,000$ to $10\,000$ configurations. For all flavour numbers we use a rational HMC algorithm with 
 \begin{equation}
 \left(\det \big(\slashed{D}\slashed{D}^\dagger\big)^{\Ni/2 N_\text{PF}}
 \right)^{N_\text{PF}}\,,
 \end{equation}
 where the number of pseudofermions is $N_\text{PF}=2 \,\Ni$. In order to calculate
 the expectation values of powers of the condensate, we use $N_{\text{est}}=200 \times \Ni$ stochastic estimators for the fermion propagator on every Monte Carlo configuration.
 
The fermion determinant $\det(\ii \slashed{D})$ is real
but not necessarily positive. Therefore we do not have a sign problem for even flavour numbers. Furthermore, simulations on smaller lattices, where we can compute the fermion determinant numerically, showed that
we only have a sign problem for $\Ni=1$. In this case, the sign problem can be solved with a fermion bag inspired algorithm and appropriate resummations of certain weights.
More details on the sign problem will be published in \cite{Schmidta}.
For $\Ni=2$ to $\Ni=11$ we determined the observables $\sigma_\cQ$ 
in \eqref{e:sigma_condensate}
via Monte Carlo simulations in the vector formulation and calculated the effective potential with the formalism described in the previous sections. 

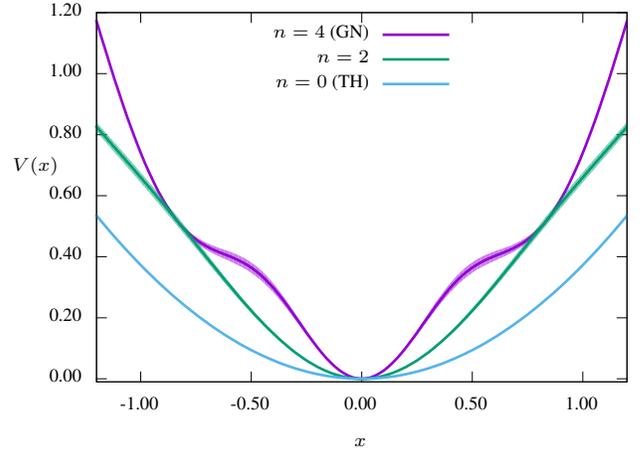
\begin{figure}[htb]
\scalebox{1.00}{\input{Potential_4.tex}}
\caption{Effective potentials for $\Ni=4$ and $\lambda=0.118$ along the different directions labelled by $n=0,2$ and $4$ on a lattice with $L=16$.}
\label{fig:pot4}
\end{figure}

The potentials for $\Ni=4$ and corresponding values of $n$
on a $16 \times 15 \times 15$ lattice are depicted in \autoref{fig:pot4}
for $\lambda=0.118$.
Statistical errors are always obtained with a Jackknife procedure and are indicated by the width of the curves.
It turns out that for every value of $\lambda$ the minimum of the potential is always at the origin $x=0$. 
Therefore we conclude that there is no spontaneous chiral or parity symmetry breaking for $\Ni=4$, at least on the lattice with $L=16$.

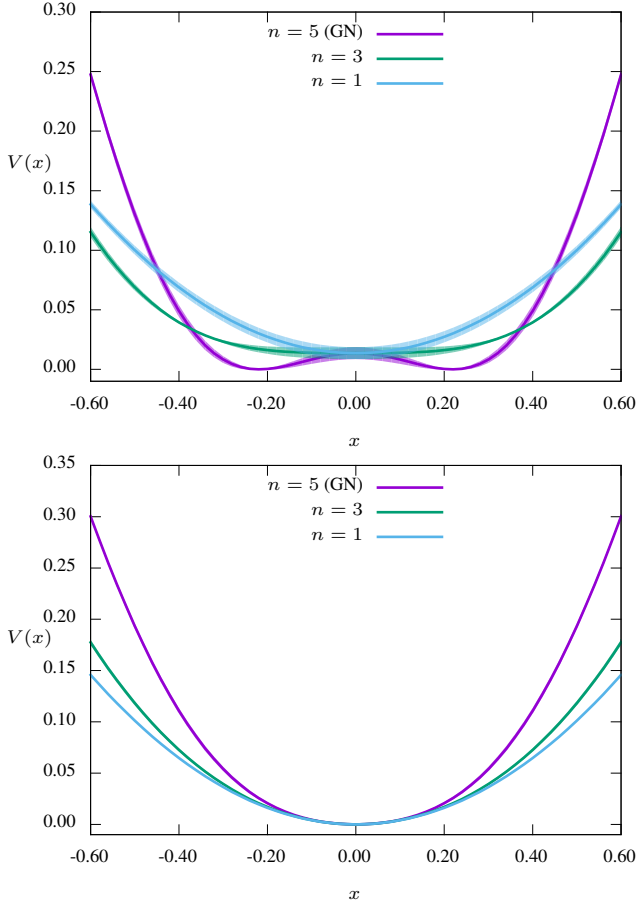
\begin{figure}[tb]
\scalebox{1.00}{\input{Potential_5b.tex}}
\scalebox{1.00}{\input{Potential_5s.tex}}
\caption{Effective potentials for $\Ni=5$ at $\lambda=0.102$ (upper panel) and $\lambda=0.118$ (lower panel) for the different directions $n=1,3$ and $5$ on a lattice with $L=16$.}
\label{fig:pot5}
\end{figure}

For $\Ni=5$ the potentials are depicted in \autoref{fig:pot5} for two
values of $\lambda$. 
For $\lambda=0.102$ the potential has two global minima in the Gross-Neveu direction at $x\approx\pm\, 0.22$ while for the larger value $\lambda=0.118$ the minimum of the potential is at $x=0$. This suggests that for $\Ni=5$ parity is broken at strong couplings
(small $\lambda$).
On larger lattices this result still holds true.

To check, that our conclusions are not blurred by
lattice artefacts we investigated the transition from the strong to weak coupling 
regime more carefully and determined the critical 
coupling where the lattice theory shows a transition from an artificial lattice phase 
at strong coupling to a continuum phase at weak coupling for all simulated $\Ni$.
In the lattice artefact phase and in the infinite volume limit, observables should only
trivially depend on $\lambda$. 
An important quantity to investigate here is the first derivative of the partition function
with respect to $\lambda$. It is connected to the normalized fermion filling factor
in the dual variables approach
 \begin{equation}
 \erw{k_\text{norm}}=\frac{\erw{k}}{2\,V\,\Ni} \in [0,1]\,,
 \end{equation}
 where $k$ has been defined in (\ref{kxi}), and therefore is
 an interesting quantity to investigate lattice artefacts. The relation is
 \begin{equation}
\frac{\lambda}{2 \Ni V}\frac{d \ln Z(\lambda)}{d \lambda}=C+\erw{k_\text{norm}}(\lambda)\,,
 \end{equation}
 where the constant $C$ only depends on the flavour number.
At the transition from the lattice artefact phase to the physical weak coupling phase, we expect a jump or a peak in the first derivative $\partial_\lambda \erw{k_\text{norm}}$. 
The critical value $\lambda^*$ is then obtained as the position of the jump (or peak) in the infinite volume limit. 
\begin{figure}[tb]
\input{k.tex}
\caption{Lattice filling factor $\erw{k_\text{norm}}$ for different $\Ni$ and lattices volumes. Larger volumes are indicated by a darker colour shade.}
\label{fig:k1}
\end{figure}
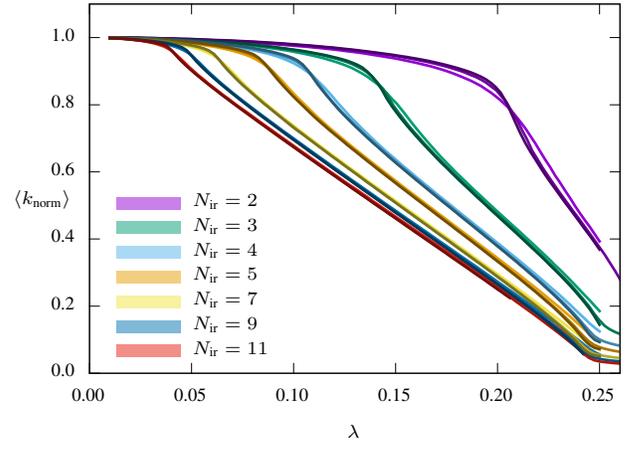
\begin{figure}[tb]
\scalebox{1.00}{\input{dk_9.tex}}
\caption{Derivative of the lattice filling factor $\erw{k_\text{norm}}$ for $\Ni=9$ and different lattice volumes.}
\label{fig:k_der}
\end{figure}
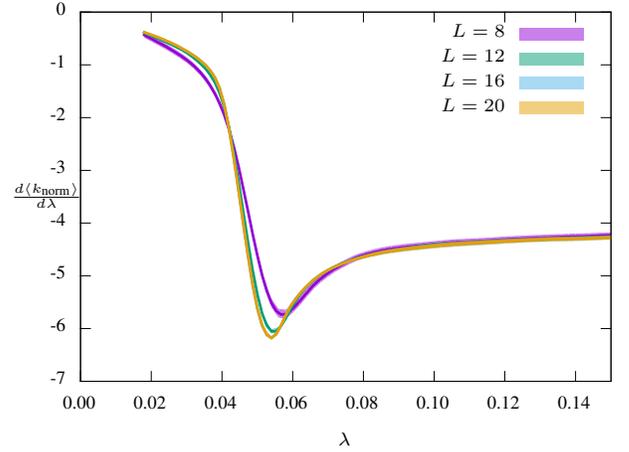
The results for $\erw{k_\text{norm}}$ are shown in \autoref{fig:k1} for different $\Ni$ and volumes.
Below $\lambda^*$, the expectation value $\erw{k_\text{norm}}$ depends only weakly on the volume and increases towards its strong coupling value $\erw{k_\text{norm}}=1$ with increasing lattice volume.
The curves for different volumes intersect close to the critical 
$\lambda^*$. In the physical phase $\erw{k_\text{norm}}$ decreases 
with increasing lattice volume.

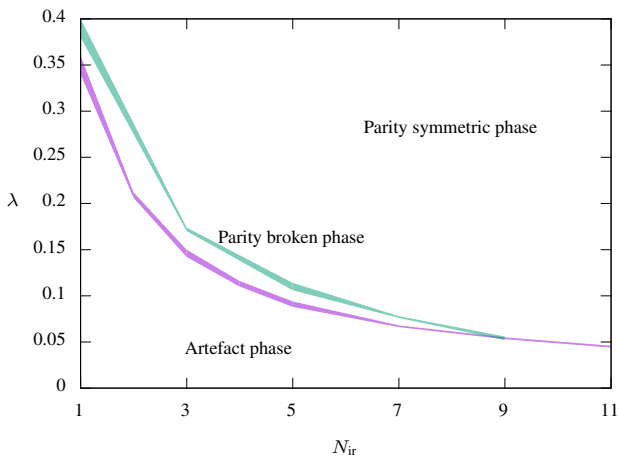
\begin{figure}[htb]
\scalebox{1.00}{\input{phases2.tex}}
\caption{Phase diagram for odd flavour numbers. The lower curve separates the 
strong coupling lattice artefact phase from the physical weak coupling phase
and is known for odd and even flavour numbers. The upper curve for odd 
flavour numbers shows the physical phase transition associated with breaking 
of the discrete parity symmetry.}
\label{fig:phase2}
\end{figure}

The variation of $\erw{k_\text{norm}}$ with $\lambda$ 
for $\Ni=9$ is depicted in \autoref{fig:k_der}.
It stays finite in the infinite volume limit and develops a jump at the critical coupling. 
We observe that for larger flavour numbers the curves for $L=16$ and $L=20$ lie almost 
on top of each other, indicating that finite volume effects are already small on these
relatively small lattices. Even for smaller $\Ni$, finite volume effects are small on the larger lattices. Therefore, we identify the critical coupling on our largest lattice 
as infinite volume coupling $\lambda^*$. The relatively small finite size effects 
are an additional advantage of the SLAC derivative that approximates the continuum derivative for a fixed number of lattice points much better 
than the naive central derivative used for Wilson or staggered 
fermions \cite{Bergner2008}. The results for the critical $\lambda^*$ for 
all $\Ni$ between $1$ and $11$ are displayed in \autoref{lambdastar}.
The results for $\Ni=1$ on lattice size $L=8$ were obtained with a fermion bag algorithm directly calculating the coefficients $a_\cP$ of the effective potential.
The lower curve in the phase diagram \autoref{fig:phase2} shows the phase boundary, 
separating the  strong coupling lattice artefact regime from the physical 
weak coupling regime.  We see that with increasing flavour number the 
critical value $\lambda^*$ decreases monotonically.

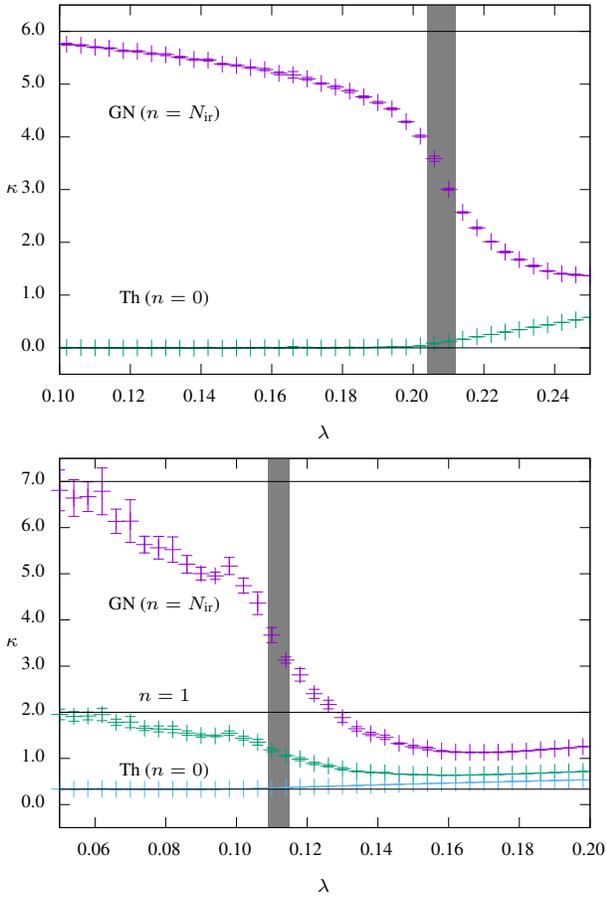
\begin{figure}[htb]
 \scalebox{1.00}{\input{Curvature_2.tex}}
\scalebox{1.00}{\input{Curvature_4.tex}}
\caption{Curvature of the effective potential at the origin for $\Ni=2$ (top) and $\Ni=4$ (bottom) compared to the critical coupling $\lambda^*$ (grey bar). The width of the grey bar indicates the statistical error of $\lambda^*$.}
\label{fig:Curv24}
\end{figure}

 \begin{table*}[htb]
 \begin{tabular}{|c|c|c|c|c|c|c|c|c|}
 \hline
$\Ni$ & $1$ & $2$ & $3$ & $4$ & $5$ & $7$ & $9$ & $11$\\
\hline
$\lambda^*(L=8)$ & $0.35(1)$ & $0.223(6)$ & $0.158(4)$ & $0.122(4)$ & $0.098(2)$ & $0.073(2)$ & $0.058(2)$ & $0.048(2)$\\
$\lambda^*(L=12)$ & -- & $0.214(4)$ & $0.149(4)$ & $0.114(3)$ & $0.094(3)$ & $0.068(2)$ & $0.054(2)$ & $0.046(2)$\\
$\lambda^*(L=16)$ & -- & $0.208(4)$ & $0.146(4)$ & $0.112(3)$ & $0.091(2)$ & $0.067(1)$ & $0.054(1)$ & $0.045(1)$\\
 $\lambda^*(L=20)$ & -- & -- & -- & -- & -- & $0.066(1)$ & $0.053(1)$ & $0.045(1)$\\
\hline
 \end{tabular}
\caption{Critical $\lambda^*$ for different flavour numbers and lattice volumes.
For larger lattices there are small finite size effects.
Simulations for $\Ni=1$ were done with a fermion bag algorithm.
}
 \label{lambdastar}
\end{table*}
After having localized the transition point between the artefact and physical phase, 
we calculate the curvature $\kappa$ of the effective potential 
at the origin as a function of $\lambda$ and compare the 
critical value $\lambda_c$, at which the 
curvature vanishes,
to the critical value $\lambda^*$ of the artefact transition. 
For even $\Ni$ we show $\kappa$ for both the Gross-Neveu-like breaking
as well as the Thirring-like breaking. The results for $\Ni\in\{2,4\}$
are depicted in \autoref{fig:Curv24}, together with the results 
from the strong coupling expansion (solid lines).
The grey bars show the allowed critical values $\lambda^*$.
 For $\Ni=2$ the curvature in the Thirring direction vanishes at strong coupling and 
 increases at the transition from the strong coupling lattice artefact phase to the weak
 coupling physical phase. The potential
 in the Gross-Neveu direction is always positive. Therefore, we conclude, that there is no
 chiral symmetry breaking for $\Ni=2$, which corresponds to \emph{one 
 reducible flavour}. For $\Ni=4$ the curvature for 
 all directions is always positive and there is clearly no chiral symmetry breaking. We also checked this for larger even numbers of flavours 
 with the same result: Chiral symmetry is always unbroken for even flavour numbers.
This implies that there is no spontaneous symmetry breaking for all reducible 
 Thirring models.
This is one important conclusion of our work
 which conflicts with earlier findings but agrees with very recent
 simulations \cite{Karthik:2015sgq,Schmidt2016,Hands2017}.

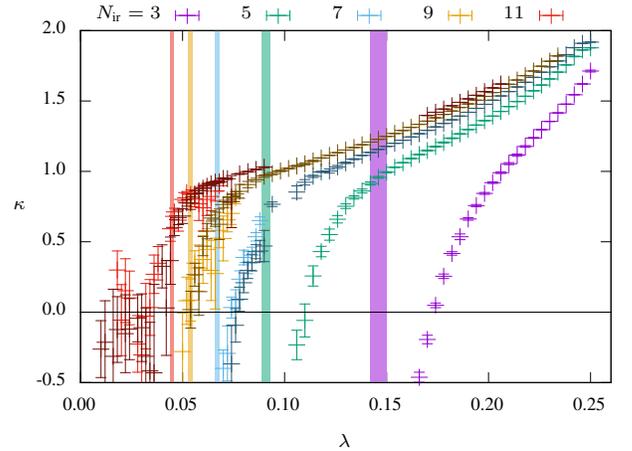
\begin{figure}[htb]
\scalebox{1.00}{\input{CurvatureOdd.tex}}
\caption{Curvature of the effective potential for different odd flavour numbers $\Ni$ on the lattice with $L=16$ (darker shade) and $L=20$ (lighter shade). The coloured bars and their widths denote the corresponding transition from the lattice artefact phase to the physical phase at $\lambda^*$ including the statistical error.}
\label{fig:CurvOdd}
\end{figure}

 For odd flavour numbers there is no Thirring-like potential. Furthermore we checked 
 that the minimum of the full potential is either at the origin or in the Gross-Neveu
 direction. Therefore we show the curvature in the Gross-Neveu direction in  \autoref{fig:CurvOdd} for $\Ni=3,5,7,9$ and $11$ on the $16 \times 15^2$ lattice.
 The critical inverse coupling $\lc$ is defined by vanishing curvature and values are shown in \autoref{t:lc}.
 \begin{table}[htb]
    \begin{tabular}{|c|c|c|c|c|c|c|}
      \hline
	$\Ni$ & $1$ & $3$ & $5$ & $7$ & $9$ & $11$\\
      \hline
	$\lc(L=16)$ &	$0.39(1)$ & $0.172(2)$ & $0.110(4)$ & $0.077(1)$ & $0.054(2)$ & --\\
	$\lc(L=20)$ &	-- & -- & -- & $0.074(2)$ & $0.051(2)$ & --\\
      \hline 
    \end{tabular}
    \caption{Critical inverse coupling $\lc$ on a lattice with $L=16$ and $L=20$ separating the parity broken from the parity symmetric phase.}
    \label{t:lc}
 \end{table}
Again we compare the coupling $\lambda_c$ to the critical value of the strong-coupling transition $\lambda^*$.
 For $\Ni=3,5$ and $7$ we observe that the parity phase transition at $\lambda_c$ lies within the physical phase, i.e. $\lambda_c > \lambda^*$ and
 we conclude that parity symmetry is spontaneously broken for these flavour numbers. For $\Ni=11$ the curvature is always positive and therefore parity symmetry is always unbroken. For $\Ni=9$ both critical couplings coincide within error bars and it is still unclear whether parity symmetry is spontaneously broken or not.

The upper curve in \autoref{fig:phase2} shows the linear interpolation
between the critical values $\lambda_c$ for the Thirring model with 
odd flavour numbers.
We conclude, that the critical flavour number for parity breaking is $\Nic=9$. 
The minimum $x_\text{min}$ of the effective potential is an order parameter for the breaking of parity symmetry and therefore related to a parity condensate $\pi$.
For a second order phase transition, $\pi$ should decrease continuously to zero with increasing coupling
$\lambda$. 
 \begin{figure}[htb]
\scalebox{1.00}{\input{ChiralCondensate.tex}}
\caption{Parity condensate $\pi \sim x_\text{min}$ for odd flavour numbers in the physical phase as a 
function of the renormalized inverse coupling $\lambda_\text{R}=\lambda-\lambda^*$.}
\label{fig:condensate}
\end{figure}
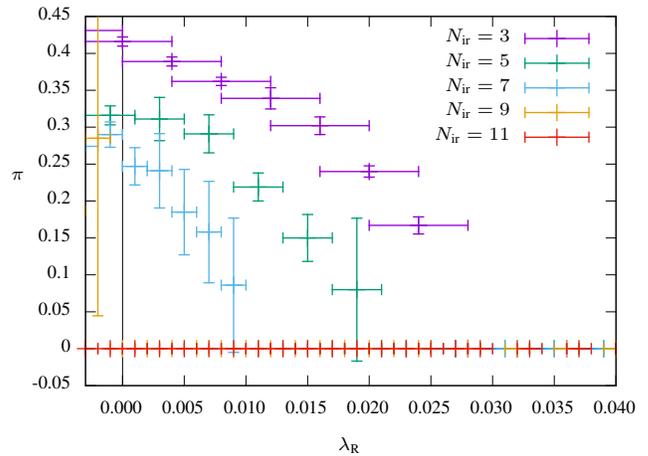
 In \autoref{fig:condensate} we show the condensate for different odd flavour numbers in the physical phase on a lattice with $L=16$. For $\Ni\in\{3,5,7\}$ 
 it decreases monotonically to zero while for $\Ni=9$ the condensate approaches zero at vanishing \emph{renormalized} inverse coupling $\lambda_\text{R}=\lambda-\lambda^*$. This observation is consistent with the scenario proposed in \cite{Kondo1995,Itoh1995,Sugiura1997} where parity at the critical flavour number is only broken at vanishing inverse coupling.
 
 The length dimension of the physical inverse coupling is $[\lambda\srm{phys}]=-1$ and therefore  the dimensionless lattice coupling is related to a physical coupling by $a\lambda\srm{phys} = \lambda$.
 In the cases, where we did not find a second-order phase transition, we can perform the continuum limit by $\lambda\srm{R}\to 0$ for a fixed physical inverse coupling within the physical phase.
 For odd $\Ni<\Nic$, where we spotted a second-order phase transition, we build the continuum limit by $\lambda\to\lc$, corresponding to a non-Gaussian fixed point.
 Both limits coincide for $\Ni=\Nic$.

%% file: Potential_4.tex
\begingroup
\scriptsize
  \makeatletter
  \providecommand\color[2][]{%
    \GenericError{(gnuplot) \space\space\space\@spaces}{%
      Package color not loaded in conjunction with
      terminal option `colourtext'%
    }{See the gnuplot documentation for explanation.%
    }{Either use 'blacktext' in gnuplot or load the package
      color.sty in LaTeX.}%
    \renewcommand\color[2][]{}%
  }%
  \providecommand\includegraphics[2][]{%
    \GenericError{(gnuplot) \space\space\space\@spaces}{%
      Package graphicx or graphics not loaded%
    }{See the gnuplot documentation for explanation.%
    }{The gnuplot epslatex terminal needs graphicx.sty or graphics.sty.}%
    \renewcommand\includegraphics[2][]{}%
  }%
  \providecommand\rotatebox[2]{#2}%
  \@ifundefined{ifGPcolor}{%
    \newif\ifGPcolor
    \GPcolortrue
  }{}%
  \@ifundefined{ifGPblacktext}{%
    \newif\ifGPblacktext
    \GPblacktexttrue
  }{}%
  \let\gplgaddtomacro\g@addto@macro
  \gdef\gplbacktext{}%
  \gdef\gplfronttext{}%
  \makeatother
  \ifGPblacktext
    \def\colorrgb#1{}%
    \def\colorgray#1{}%
  \else
    \ifGPcolor
      \def\colorrgb#1{\color[rgb]{#1}}%
      \def\colorgray#1{\color[gray]{#1}}%
      \expandafter\def\csname LTw\endcsname{\color{white}}%
      \expandafter\def\csname LTb\endcsname{\color{black}}%
      \expandafter\def\csname LTa\endcsname{\color{black}}%
      \expandafter\def\csname LT0\endcsname{\color[rgb]{1,0,0}}%
      \expandafter\def\csname LT1\endcsname{\color[rgb]{0,1,0}}%
      \expandafter\def\csname LT2\endcsname{\color[rgb]{0,0,1}}%
      \expandafter\def\csname LT3\endcsname{\color[rgb]{1,0,1}}%
      \expandafter\def\csname LT4\endcsname{\color[rgb]{0,1,1}}%
      \expandafter\def\csname LT5\endcsname{\color[rgb]{1,1,0}}%
      \expandafter\def\csname LT6\endcsname{\color[rgb]{0,0,0}}%
      \expandafter\def\csname LT7\endcsname{\color[rgb]{1,0.3,0}}%
      \expandafter\def\csname LT8\endcsname{\color[rgb]{0.5,0.5,0.5}}%
    \else
      \def\colorrgb#1{\color{black}}%
      \def\colorgray#1{\color[gray]{#1}}%
      \expandafter\def\csname LTw\endcsname{\color{white}}%
      \expandafter\def\csname LTb\endcsname{\color{black}}%
      \expandafter\def\csname LTa\endcsname{\color{black}}%
      \expandafter\def\csname LT0\endcsname{\color{black}}%
      \expandafter\def\csname LT1\endcsname{\color{black}}%
      \expandafter\def\csname LT2\endcsname{\color{black}}%
      \expandafter\def\csname LT3\endcsname{\color{black}}%
      \expandafter\def\csname LT4\endcsname{\color{black}}%
      \expandafter\def\csname LT5\endcsname{\color{black}}%
      \expandafter\def\csname LT6\endcsname{\color{black}}%
      \expandafter\def\csname LT7\endcsname{\color{black}}%
      \expandafter\def\csname LT8\endcsname{\color{black}}%
    \fi
  \fi
  \setlength{\unitlength}{0.0500bp}%
  \begin{picture}(4520.00,3400.00)%
    \gplgaddtomacro\gplbacktext{%
      \csname LTb\endcsname%
      \put(408,618){\makebox(0,0)[r]{\strut{}0.00}}%
      \csname LTb\endcsname%
      \put(408,1078){\makebox(0,0)[r]{\strut{}0.20}}%
      \csname LTb\endcsname%
      \put(408,1539){\makebox(0,0)[r]{\strut{}0.40}}%
      \csname LTb\endcsname%
      \put(408,1999){\makebox(0,0)[r]{\strut{}0.60}}%
      \csname LTb\endcsname%
      \put(408,2459){\makebox(0,0)[r]{\strut{}0.80}}%
      \csname LTb\endcsname%
      \put(408,2920){\makebox(0,0)[r]{\strut{}1.00}}%
      \csname LTb\endcsname%
      \put(408,3380){\makebox(0,0)[r]{\strut{}1.20}}%
      \csname LTb\endcsname%
      \put(843,409){\makebox(0,0){\strut{}-1.00}}%
      \csname LTb\endcsname%
      \put(1676,409){\makebox(0,0){\strut{}-0.50}}%
      \csname LTb\endcsname%
      \put(2509,409){\makebox(0,0){\strut{}0.00}}%
      \csname LTb\endcsname%
      \put(3342,409){\makebox(0,0){\strut{}0.50}}%
      \csname LTb\endcsname%
      \put(4175,409){\makebox(0,0){\strut{}1.00}}%
      \csname LTb\endcsname%
      \put(60,2210){\makebox(0,0){\strut{}$V(x)$}}%
      \csname LTb\endcsname%
      \put(2509,130){\makebox(0,0){\strut{}$x$}}%
    }%
    \gplgaddtomacro\gplfronttext{%
      \csname LTb\endcsname%
      \put(2574,3213){\makebox(0,0)[r]{\strut{}$n=4$ (GN)}}%
      \csname LTb\endcsname%
      \put(2574,3027){\makebox(0,0)[r]{\strut{}$n=2$}}%
      \csname LTb\endcsname%
      \put(2574,2841){\makebox(0,0)[r]{\strut{}$n=0$ (TH)}}%
    }%
    \gplbacktext
    \put(0,0){\includegraphics{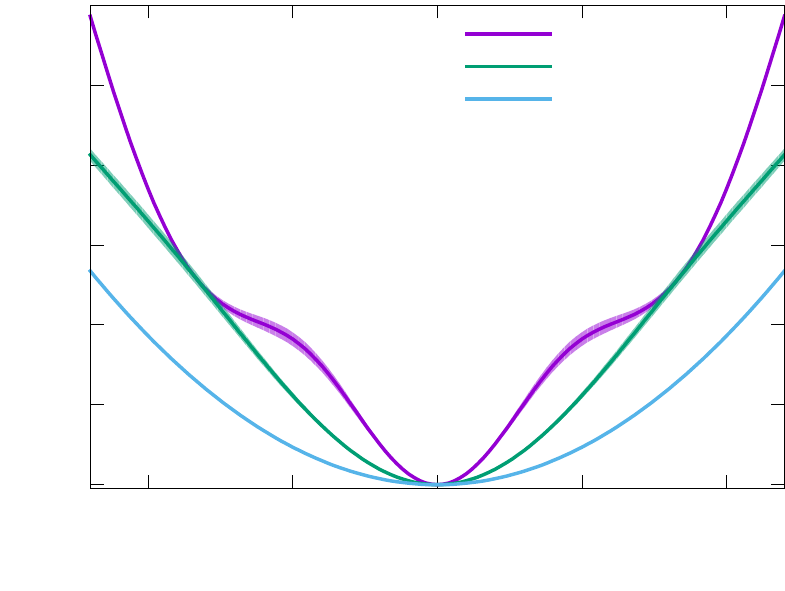}}%
    \gplfronttext
  \end{picture}%
\endgroup

%% file: Potential_5b.tex
\begingroup
\scriptsize
  \makeatletter
  \providecommand\color[2][]{%
    \GenericError{(gnuplot) \space\space\space\@spaces}{%
      Package color not loaded in conjunction with
      terminal option `colourtext'%
    }{See the gnuplot documentation for explanation.%
    }{Either use 'blacktext' in gnuplot or load the package
      color.sty in LaTeX.}%
    \renewcommand\color[2][]{}%
  }%
  \providecommand\includegraphics[2][]{%
    \GenericError{(gnuplot) \space\space\space\@spaces}{%
      Package graphicx or graphics not loaded%
    }{See the gnuplot documentation for explanation.%
    }{The gnuplot epslatex terminal needs graphicx.sty or graphics.sty.}%
    \renewcommand\includegraphics[2][]{}%
  }%
  \providecommand\rotatebox[2]{#2}%
  \@ifundefined{ifGPcolor}{%
    \newif\ifGPcolor
    \GPcolortrue
  }{}%
  \@ifundefined{ifGPblacktext}{%
    \newif\ifGPblacktext
    \GPblacktexttrue
  }{}%
  \let\gplgaddtomacro\g@addto@macro
  \gdef\gplbacktext{}%
  \gdef\gplfronttext{}%
  \makeatother
  \ifGPblacktext
    \def\colorrgb#1{}%
    \def\colorgray#1{}%
  \else
    \ifGPcolor
      \def\colorrgb#1{\color[rgb]{#1}}%
      \def\colorgray#1{\color[gray]{#1}}%
      \expandafter\def\csname LTw\endcsname{\color{white}}%
      \expandafter\def\csname LTb\endcsname{\color{black}}%
      \expandafter\def\csname LTa\endcsname{\color{black}}%
      \expandafter\def\csname LT0\endcsname{\color[rgb]{1,0,0}}%
      \expandafter\def\csname LT1\endcsname{\color[rgb]{0,1,0}}%
      \expandafter\def\csname LT2\endcsname{\color[rgb]{0,0,1}}%
      \expandafter\def\csname LT3\endcsname{\color[rgb]{1,0,1}}%
      \expandafter\def\csname LT4\endcsname{\color[rgb]{0,1,1}}%
      \expandafter\def\csname LT5\endcsname{\color[rgb]{1,1,0}}%
      \expandafter\def\csname LT6\endcsname{\color[rgb]{0,0,0}}%
      \expandafter\def\csname LT7\endcsname{\color[rgb]{1,0.3,0}}%
      \expandafter\def\csname LT8\endcsname{\color[rgb]{0.5,0.5,0.5}}%
    \else
      \def\colorrgb#1{\color{black}}%
      \def\colorgray#1{\color[gray]{#1}}%
      \expandafter\def\csname LTw\endcsname{\color{white}}%
      \expandafter\def\csname LTb\endcsname{\color{black}}%
      \expandafter\def\csname LTa\endcsname{\color{black}}%
      \expandafter\def\csname LT0\endcsname{\color{black}}%
      \expandafter\def\csname LT1\endcsname{\color{black}}%
      \expandafter\def\csname LT2\endcsname{\color{black}}%
      \expandafter\def\csname LT3\endcsname{\color{black}}%
      \expandafter\def\csname LT4\endcsname{\color{black}}%
      \expandafter\def\csname LT5\endcsname{\color{black}}%
      \expandafter\def\csname LT6\endcsname{\color{black}}%
      \expandafter\def\csname LT7\endcsname{\color{black}}%
      \expandafter\def\csname LT8\endcsname{\color{black}}%
    \fi
  \fi
  \setlength{\unitlength}{0.0500bp}%
  \begin{picture}(4520.00,3400.00)%
    \gplgaddtomacro\gplbacktext{%
      \csname LTb\endcsname%
      \put(408,685){\makebox(0,0)[r]{\strut{}0.00}}%
      \csname LTb\endcsname%
      \put(408,1134){\makebox(0,0)[r]{\strut{}0.05}}%
      \csname LTb\endcsname%
      \put(408,1583){\makebox(0,0)[r]{\strut{}0.10}}%
      \csname LTb\endcsname%
      \put(408,2032){\makebox(0,0)[r]{\strut{}0.15}}%
      \csname LTb\endcsname%
      \put(408,2482){\makebox(0,0)[r]{\strut{}0.20}}%
      \csname LTb\endcsname%
      \put(408,2931){\makebox(0,0)[r]{\strut{}0.25}}%
      \csname LTb\endcsname%
      \put(408,3380){\makebox(0,0)[r]{\strut{}0.30}}%
      \csname LTb\endcsname%
      \put(510,409){\makebox(0,0){\strut{}-0.60}}%
      \csname LTb\endcsname%
      \put(1176,409){\makebox(0,0){\strut{}-0.40}}%
      \csname LTb\endcsname%
      \put(1843,409){\makebox(0,0){\strut{}-0.20}}%
      \csname LTb\endcsname%
      \put(2509,409){\makebox(0,0){\strut{}0.00}}%
      \csname LTb\endcsname%
      \put(3175,409){\makebox(0,0){\strut{}0.20}}%
      \csname LTb\endcsname%
      \put(3842,409){\makebox(0,0){\strut{}0.40}}%
      \csname LTb\endcsname%
      \put(4508,409){\makebox(0,0){\strut{}0.60}}%
      \csname LTb\endcsname%
      \put(60,2228){\makebox(0,0){\strut{}$V(x)$}}%
      \csname LTb\endcsname%
      \put(2509,130){\makebox(0,0){\strut{}$x$}}%
    }%
    \gplgaddtomacro\gplfronttext{%
      \csname LTb\endcsname%
      \put(2574,3213){\makebox(0,0)[r]{\strut{}$n=5$ (GN)}}%
      \csname LTb\endcsname%
      \put(2574,3027){\makebox(0,0)[r]{\strut{}$n=3$}}%
      \csname LTb\endcsname%
      \put(2574,2841){\makebox(0,0)[r]{\strut{}$n=1$}}%
    }%
    \gplbacktext
    \put(0,0){\includegraphics{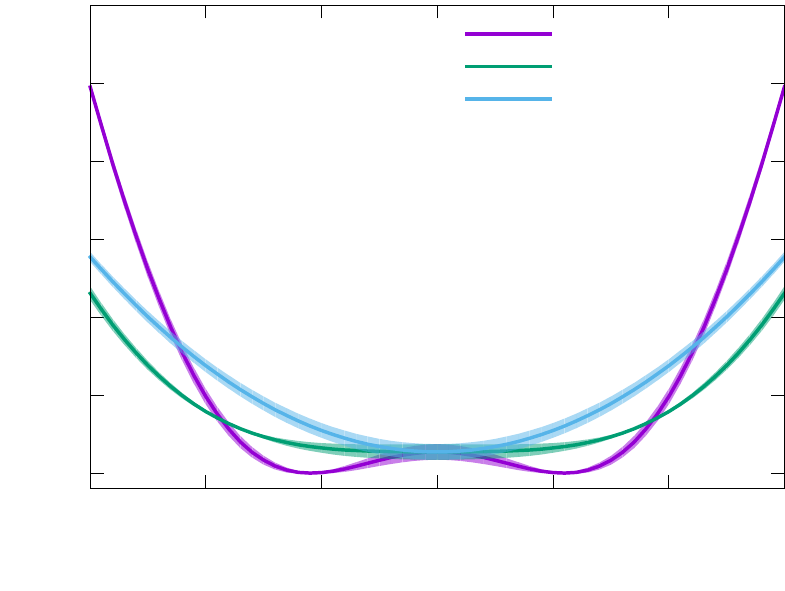}}%
    \gplfronttext
  \end{picture}%
\endgroup

%% file: Potential_5s.tex
\begingroup
\scriptsize
  \makeatletter
  \providecommand\color[2][]{%
    \GenericError{(gnuplot) \space\space\space\@spaces}{%
      Package color not loaded in conjunction with
      terminal option `colourtext'%
    }{See the gnuplot documentation for explanation.%
    }{Either use 'blacktext' in gnuplot or load the package
      color.sty in LaTeX.}%
    \renewcommand\color[2][]{}%
  }%
  \providecommand\includegraphics[2][]{%
    \GenericError{(gnuplot) \space\space\space\@spaces}{%
      Package graphicx or graphics not loaded%
    }{See the gnuplot documentation for explanation.%
    }{The gnuplot epslatex terminal needs graphicx.sty or graphics.sty.}%
    \renewcommand\includegraphics[2][]{}%
  }%
  \providecommand\rotatebox[2]{#2}%
  \@ifundefined{ifGPcolor}{%
    \newif\ifGPcolor
    \GPcolortrue
  }{}%
  \@ifundefined{ifGPblacktext}{%
    \newif\ifGPblacktext
    \GPblacktexttrue
  }{}%
  \let\gplgaddtomacro\g@addto@macro
  \gdef\gplbacktext{}%
  \gdef\gplfronttext{}%
  \makeatother
  \ifGPblacktext
    \def\colorrgb#1{}%
    \def\colorgray#1{}%
  \else
    \ifGPcolor
      \def\colorrgb#1{\color[rgb]{#1}}%
      \def\colorgray#1{\color[gray]{#1}}%
      \expandafter\def\csname LTw\endcsname{\color{white}}%
      \expandafter\def\csname LTb\endcsname{\color{black}}%
      \expandafter\def\csname LTa\endcsname{\color{black}}%
      \expandafter\def\csname LT0\endcsname{\color[rgb]{1,0,0}}%
      \expandafter\def\csname LT1\endcsname{\color[rgb]{0,1,0}}%
      \expandafter\def\csname LT2\endcsname{\color[rgb]{0,0,1}}%
      \expandafter\def\csname LT3\endcsname{\color[rgb]{1,0,1}}%
      \expandafter\def\csname LT4\endcsname{\color[rgb]{0,1,1}}%
      \expandafter\def\csname LT5\endcsname{\color[rgb]{1,1,0}}%
      \expandafter\def\csname LT6\endcsname{\color[rgb]{0,0,0}}%
      \expandafter\def\csname LT7\endcsname{\color[rgb]{1,0.3,0}}%
      \expandafter\def\csname LT8\endcsname{\color[rgb]{0.5,0.5,0.5}}%
    \else
      \def\colorrgb#1{\color{black}}%
      \def\colorgray#1{\color[gray]{#1}}%
      \expandafter\def\csname LTw\endcsname{\color{white}}%
      \expandafter\def\csname LTb\endcsname{\color{black}}%
      \expandafter\def\csname LTa\endcsname{\color{black}}%
      \expandafter\def\csname LT0\endcsname{\color{black}}%
      \expandafter\def\csname LT1\endcsname{\color{black}}%
      \expandafter\def\csname LT2\endcsname{\color{black}}%
      \expandafter\def\csname LT3\endcsname{\color{black}}%
      \expandafter\def\csname LT4\endcsname{\color{black}}%
      \expandafter\def\csname LT5\endcsname{\color{black}}%
      \expandafter\def\csname LT6\endcsname{\color{black}}%
      \expandafter\def\csname LT7\endcsname{\color{black}}%
      \expandafter\def\csname LT8\endcsname{\color{black}}%
    \fi
  \fi
  \setlength{\unitlength}{0.0500bp}%
  \begin{picture}(4520.00,3400.00)%
    \gplgaddtomacro\gplbacktext{%
      \csname LTb\endcsname%
      \put(408,672){\makebox(0,0)[r]{\strut{}0.00}}%
      \csname LTb\endcsname%
      \put(408,1059){\makebox(0,0)[r]{\strut{}0.05}}%
      \csname LTb\endcsname%
      \put(408,1446){\makebox(0,0)[r]{\strut{}0.10}}%
      \csname LTb\endcsname%
      \put(408,1833){\makebox(0,0)[r]{\strut{}0.15}}%
      \csname LTb\endcsname%
      \put(408,2220){\makebox(0,0)[r]{\strut{}0.20}}%
      \csname LTb\endcsname%
      \put(408,2606){\makebox(0,0)[r]{\strut{}0.25}}%
      \csname LTb\endcsname%
      \put(408,2993){\makebox(0,0)[r]{\strut{}0.30}}%
      \csname LTb\endcsname%
      \put(408,3380){\makebox(0,0)[r]{\strut{}0.35}}%
      \csname LTb\endcsname%
      \put(510,409){\makebox(0,0){\strut{}-0.60}}%
      \csname LTb\endcsname%
      \put(1176,409){\makebox(0,0){\strut{}-0.40}}%
      \csname LTb\endcsname%
      \put(1843,409){\makebox(0,0){\strut{}-0.20}}%
      \csname LTb\endcsname%
      \put(2509,409){\makebox(0,0){\strut{}0.00}}%
      \csname LTb\endcsname%
      \put(3175,409){\makebox(0,0){\strut{}0.20}}%
      \csname LTb\endcsname%
      \put(3842,409){\makebox(0,0){\strut{}0.40}}%
      \csname LTb\endcsname%
      \put(4508,409){\makebox(0,0){\strut{}0.60}}%
      \csname LTb\endcsname%
      \put(60,2061){\makebox(0,0){\strut{}$V(x)$}}%
      \csname LTb\endcsname%
      \put(2509,130){\makebox(0,0){\strut{}$x$}}%
    }%
    \gplgaddtomacro\gplfronttext{%
      \csname LTb\endcsname%
      \put(2574,3213){\makebox(0,0)[r]{\strut{}$n=5$ (GN)}}%
      \csname LTb\endcsname%
      \put(2574,3027){\makebox(0,0)[r]{\strut{}$n=3$}}%
      \csname LTb\endcsname%
      \put(2574,2841){\makebox(0,0)[r]{\strut{}$n=1$}}%
    }%
    \gplbacktext
    \put(0,0){\includegraphics{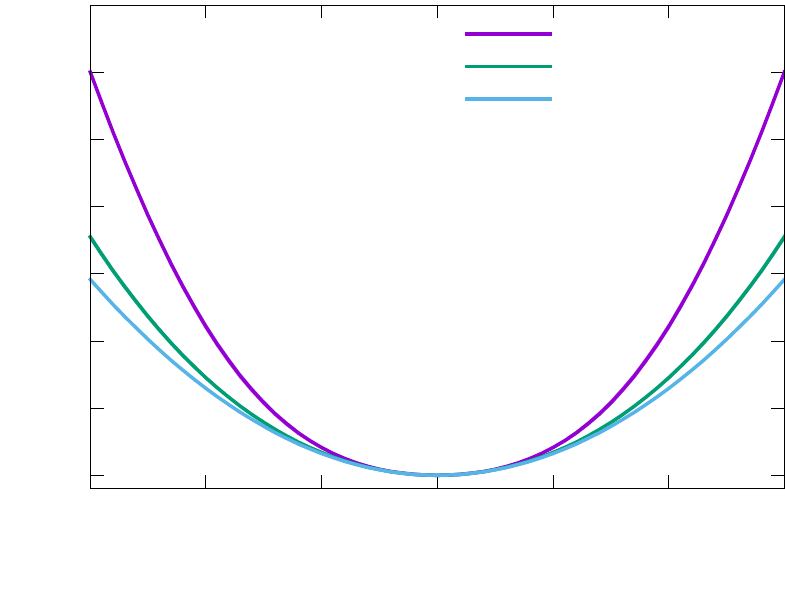}}%
    \gplfronttext
  \end{picture}%
\endgroup

%% file: k.tex
\begingroup
\scriptsize
  \makeatletter
  \providecommand\color[2][]{%
    \GenericError{(gnuplot) \space\space\space\@spaces}{%
      Package color not loaded in conjunction with
      terminal option `colourtext'%
    }{See the gnuplot documentation for explanation.%
    }{Either use 'blacktext' in gnuplot or load the package
      color.sty in LaTeX.}%
    \renewcommand\color[2][]{}%
  }%
  \providecommand\includegraphics[2][]{%
    \GenericError{(gnuplot) \space\space\space\@spaces}{%
      Package graphicx or graphics not loaded%
    }{See the gnuplot documentation for explanation.%
    }{The gnuplot epslatex terminal needs graphicx.sty or graphics.sty.}%
    \renewcommand\includegraphics[2][]{}%
  }%
  \providecommand\rotatebox[2]{#2}%
  \@ifundefined{ifGPcolor}{%
    \newif\ifGPcolor
    \GPcolortrue
  }{}%
  \@ifundefined{ifGPblacktext}{%
    \newif\ifGPblacktext
    \GPblacktexttrue
  }{}%
  \let\gplgaddtomacro\g@addto@macro
  \gdef\gplbacktext{}%
  \gdef\gplfronttext{}%
  \makeatother
  \ifGPblacktext
    \def\colorrgb#1{}%
    \def\colorgray#1{}%
  \else
    \ifGPcolor
      \def\colorrgb#1{\color[rgb]{#1}}%
      \def\colorgray#1{\color[gray]{#1}}%
      \expandafter\def\csname LTw\endcsname{\color{white}}%
      \expandafter\def\csname LTb\endcsname{\color{black}}%
      \expandafter\def\csname LTa\endcsname{\color{black}}%
      \expandafter\def\csname LT0\endcsname{\color[rgb]{1,0,0}}%
      \expandafter\def\csname LT1\endcsname{\color[rgb]{0,1,0}}%
      \expandafter\def\csname LT2\endcsname{\color[rgb]{0,0,1}}%
      \expandafter\def\csname LT3\endcsname{\color[rgb]{1,0,1}}%
      \expandafter\def\csname LT4\endcsname{\color[rgb]{0,1,1}}%
      \expandafter\def\csname LT5\endcsname{\color[rgb]{1,1,0}}%
      \expandafter\def\csname LT6\endcsname{\color[rgb]{0,0,0}}%
      \expandafter\def\csname LT7\endcsname{\color[rgb]{1,0.3,0}}%
      \expandafter\def\csname LT8\endcsname{\color[rgb]{0.5,0.5,0.5}}%
    \else
      \def\colorrgb#1{\color{black}}%
      \def\colorgray#1{\color[gray]{#1}}%
      \expandafter\def\csname LTw\endcsname{\color{white}}%
      \expandafter\def\csname LTb\endcsname{\color{black}}%
      \expandafter\def\csname LTa\endcsname{\color{black}}%
      \expandafter\def\csname LT0\endcsname{\color{black}}%
      \expandafter\def\csname LT1\endcsname{\color{black}}%
      \expandafter\def\csname LT2\endcsname{\color{black}}%
      \expandafter\def\csname LT3\endcsname{\color{black}}%
      \expandafter\def\csname LT4\endcsname{\color{black}}%
      \expandafter\def\csname LT5\endcsname{\color{black}}%
      \expandafter\def\csname LT6\endcsname{\color{black}}%
      \expandafter\def\csname LT7\endcsname{\color{black}}%
      \expandafter\def\csname LT8\endcsname{\color{black}}%
    \fi
  \fi
  \setlength{\unitlength}{0.0500bp}%
  \begin{picture}(4520.00,3400.00)%
    \gplgaddtomacro\gplbacktext{%
      \csname LTb\endcsname%
      \put(408,595){\makebox(0,0)[r]{\strut{}0.0}}%
      \csname LTb\endcsname%
      \put(408,1101){\makebox(0,0)[r]{\strut{}0.2}}%
      \csname LTb\endcsname%
      \put(408,1608){\makebox(0,0)[r]{\strut{}0.4}}%
      \csname LTb\endcsname%
      \put(408,2114){\makebox(0,0)[r]{\strut{}0.6}}%
      \csname LTb\endcsname%
      \put(408,2620){\makebox(0,0)[r]{\strut{}0.8}}%
      \csname LTb\endcsname%
      \put(408,3127){\makebox(0,0)[r]{\strut{}1.0}}%
      \csname LTb\endcsname%
      \put(510,409){\makebox(0,0){\strut{}0.00}}%
      \csname LTb\endcsname%
      \put(1279,409){\makebox(0,0){\strut{}0.05}}%
      \csname LTb\endcsname%
      \put(2048,409){\makebox(0,0){\strut{}0.10}}%
      \csname LTb\endcsname%
      \put(2817,409){\makebox(0,0){\strut{}0.15}}%
      \csname LTb\endcsname%
      \put(3585,409){\makebox(0,0){\strut{}0.20}}%
      \csname LTb\endcsname%
      \put(4354,409){\makebox(0,0){\strut{}0.25}}%
      \csname LTb\endcsname%
      \put(162,1894){\makebox(0,0){\strut{}$\erw{k_\text{norm}}$}}%
      \csname LTb\endcsname%
      \put(2509,130){\makebox(0,0){\strut{}$\lambda$}}%
    }%
    \gplgaddtomacro\gplfronttext{%
      \csname LTb\endcsname%
      \put(1298,1878){\makebox(0,0)[l]{\strut{}$\Ni=2$}}%
      \csname LTb\endcsname%
      \put(1298,1692){\makebox(0,0)[l]{\strut{}$\Ni=3$}}%
      \csname LTb\endcsname%
      \put(1298,1506){\makebox(0,0)[l]{\strut{}$\Ni=4$}}%
      \csname LTb\endcsname%
      \put(1298,1320){\makebox(0,0)[l]{\strut{}$\Ni=5$}}%
      \csname LTb\endcsname%
      \put(1298,1134){\makebox(0,0)[l]{\strut{}$\Ni=7$}}%
      \csname LTb\endcsname%
      \put(1298,948){\makebox(0,0)[l]{\strut{}$\Ni=9$}}%
      \csname LTb\endcsname%
      \put(1298,762){\makebox(0,0)[l]{\strut{}$\Ni=11$}}%
    }%
    \gplbacktext
    \put(0,0){\includegraphics{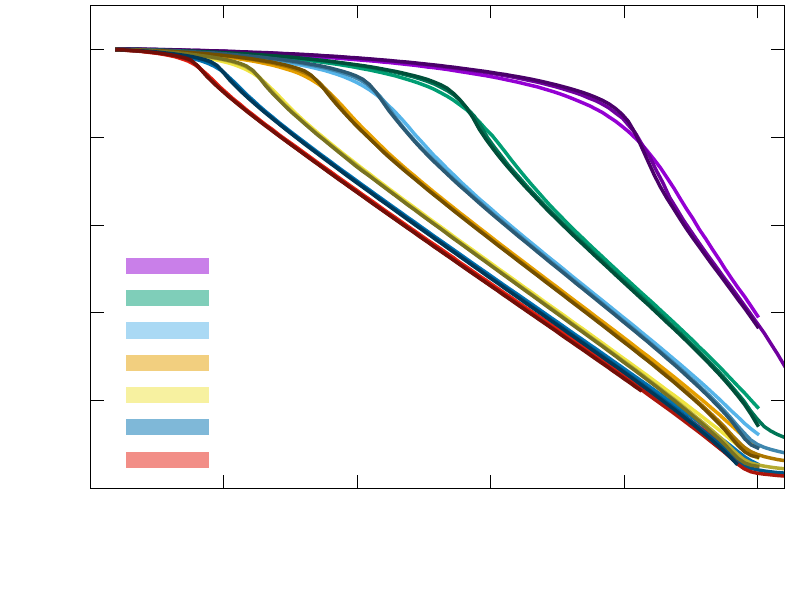}}%
    \gplfronttext
  \end{picture}%
\endgroup

%% file: dk_9.tex
\begingroup
\scriptsize
  \makeatletter
  \providecommand\color[2][]{%
    \GenericError{(gnuplot) \space\space\space\@spaces}{%
      Package color not loaded in conjunction with
      terminal option `colourtext'%
    }{See the gnuplot documentation for explanation.%
    }{Either use 'blacktext' in gnuplot or load the package
      color.sty in LaTeX.}%
    \renewcommand\color[2][]{}%
  }%
  \providecommand\includegraphics[2][]{%
    \GenericError{(gnuplot) \space\space\space\@spaces}{%
      Package graphicx or graphics not loaded%
    }{See the gnuplot documentation for explanation.%
    }{The gnuplot epslatex terminal needs graphicx.sty or graphics.sty.}%
    \renewcommand\includegraphics[2][]{}%
  }%
  \providecommand\rotatebox[2]{#2}%
  \@ifundefined{ifGPcolor}{%
    \newif\ifGPcolor
    \GPcolortrue
  }{}%
  \@ifundefined{ifGPblacktext}{%
    \newif\ifGPblacktext
    \GPblacktexttrue
  }{}%
  \let\gplgaddtomacro\g@addto@macro
  \gdef\gplbacktext{}%
  \gdef\gplfronttext{}%
  \makeatother
  \ifGPblacktext
    \def\colorrgb#1{}%
    \def\colorgray#1{}%
  \else
    \ifGPcolor
      \def\colorrgb#1{\color[rgb]{#1}}%
      \def\colorgray#1{\color[gray]{#1}}%
      \expandafter\def\csname LTw\endcsname{\color{white}}%
      \expandafter\def\csname LTb\endcsname{\color{black}}%
      \expandafter\def\csname LTa\endcsname{\color{black}}%
      \expandafter\def\csname LT0\endcsname{\color[rgb]{1,0,0}}%
      \expandafter\def\csname LT1\endcsname{\color[rgb]{0,1,0}}%
      \expandafter\def\csname LT2\endcsname{\color[rgb]{0,0,1}}%
      \expandafter\def\csname LT3\endcsname{\color[rgb]{1,0,1}}%
      \expandafter\def\csname LT4\endcsname{\color[rgb]{0,1,1}}%
      \expandafter\def\csname LT5\endcsname{\color[rgb]{1,1,0}}%
      \expandafter\def\csname LT6\endcsname{\color[rgb]{0,0,0}}%
      \expandafter\def\csname LT7\endcsname{\color[rgb]{1,0.3,0}}%
      \expandafter\def\csname LT8\endcsname{\color[rgb]{0.5,0.5,0.5}}%
    \else
      \def\colorrgb#1{\color{black}}%
      \def\colorgray#1{\color[gray]{#1}}%
      \expandafter\def\csname LTw\endcsname{\color{white}}%
      \expandafter\def\csname LTb\endcsname{\color{black}}%
      \expandafter\def\csname LTa\endcsname{\color{black}}%
      \expandafter\def\csname LT0\endcsname{\color{black}}%
      \expandafter\def\csname LT1\endcsname{\color{black}}%
      \expandafter\def\csname LT2\endcsname{\color{black}}%
      \expandafter\def\csname LT3\endcsname{\color{black}}%
      \expandafter\def\csname LT4\endcsname{\color{black}}%
      \expandafter\def\csname LT5\endcsname{\color{black}}%
      \expandafter\def\csname LT6\endcsname{\color{black}}%
      \expandafter\def\csname LT7\endcsname{\color{black}}%
      \expandafter\def\csname LT8\endcsname{\color{black}}%
    \fi
  \fi
  \setlength{\unitlength}{0.0500bp}%
  \begin{picture}(4520.00,3400.00)%
    \gplgaddtomacro\gplbacktext{%
      \csname LTb\endcsname%
      \put(408,595){\makebox(0,0)[r]{\strut{}-7}}%
      \csname LTb\endcsname%
      \put(408,993){\makebox(0,0)[r]{\strut{}-6}}%
      \csname LTb\endcsname%
      \put(408,1391){\makebox(0,0)[r]{\strut{}-5}}%
      \csname LTb\endcsname%
      \put(408,1789){\makebox(0,0)[r]{\strut{}-4}}%
      \csname LTb\endcsname%
      \put(408,2186){\makebox(0,0)[r]{\strut{}-3}}%
      \csname LTb\endcsname%
      \put(408,2584){\makebox(0,0)[r]{\strut{}-2}}%
      \csname LTb\endcsname%
      \put(408,2982){\makebox(0,0)[r]{\strut{}-1}}%
      \csname LTb\endcsname%
      \put(408,3380){\makebox(0,0)[r]{\strut{}0}}%
      \csname LTb\endcsname%
      \put(510,409){\makebox(0,0){\strut{}0.00}}%
      \csname LTb\endcsname%
      \put(1043,409){\makebox(0,0){\strut{}0.02}}%
      \csname LTb\endcsname%
      \put(1576,409){\makebox(0,0){\strut{}0.04}}%
      \csname LTb\endcsname%
      \put(2109,409){\makebox(0,0){\strut{}0.06}}%
      \csname LTb\endcsname%
      \put(2642,409){\makebox(0,0){\strut{}0.08}}%
      \csname LTb\endcsname%
      \put(3175,409){\makebox(0,0){\strut{}0.10}}%
      \csname LTb\endcsname%
      \put(3708,409){\makebox(0,0){\strut{}0.12}}%
      \csname LTb\endcsname%
      \put(4241,409){\makebox(0,0){\strut{}0.14}}%
      \csname LTb\endcsname%
      \put(264,1987){\makebox(0,0){\strut{}$\frac{d\erw{k_\text{norm}}}{d\lambda}$}}%
      \csname LTb\endcsname%
      \put(2509,130){\makebox(0,0){\strut{}$\lambda$}}%
    }%
    \gplgaddtomacro\gplfronttext{%
      \csname LTb\endcsname%
      \put(3720,3213){\makebox(0,0)[r]{\strut{}$L=8$}}%
      \csname LTb\endcsname%
      \put(3720,3027){\makebox(0,0)[r]{\strut{}$L=12$}}%
      \csname LTb\endcsname%
      \put(3720,2841){\makebox(0,0)[r]{\strut{}$L=16$}}%
      \csname LTb\endcsname%
      \put(3720,2655){\makebox(0,0)[r]{\strut{}$L=20$}}%
    }%
    \gplbacktext
    \put(0,0){\includegraphics{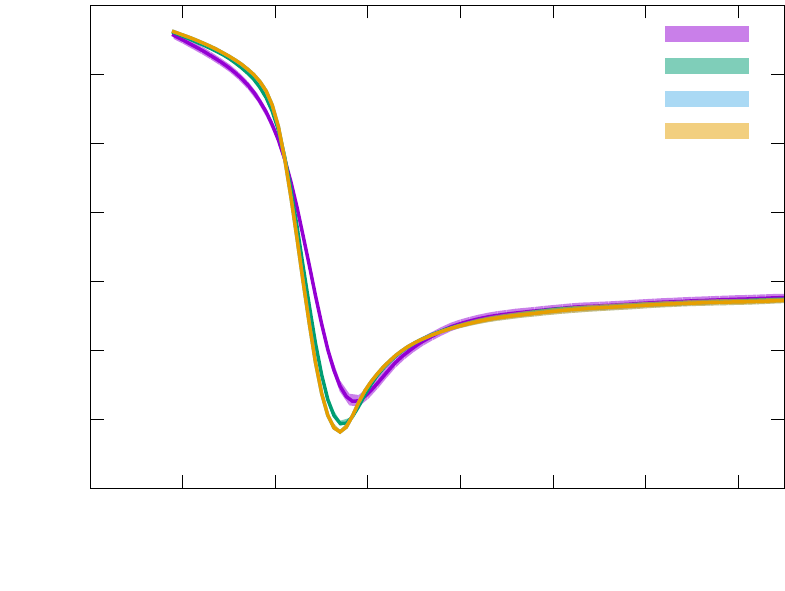}}%
    \gplfronttext
  \end{picture}%
\endgroup

%% file: Curvature_2.tex
\begingroup
\scriptsize
  \makeatletter
  \providecommand\color[2][]{%
    \GenericError{(gnuplot) \space\space\space\@spaces}{%
      Package color not loaded in conjunction with
      terminal option `colourtext'%
    }{See the gnuplot documentation for explanation.%
    }{Either use 'blacktext' in gnuplot or load the package
      color.sty in LaTeX.}%
    \renewcommand\color[2][]{}%
  }%
  \providecommand\includegraphics[2][]{%
    \GenericError{(gnuplot) \space\space\space\@spaces}{%
      Package graphicx or graphics not loaded%
    }{See the gnuplot documentation for explanation.%
    }{The gnuplot epslatex terminal needs graphicx.sty or graphics.sty.}%
    \renewcommand\includegraphics[2][]{}%
  }%
  \providecommand\rotatebox[2]{#2}%
  \@ifundefined{ifGPcolor}{%
    \newif\ifGPcolor
    \GPcolortrue
  }{}%
  \@ifundefined{ifGPblacktext}{%
    \newif\ifGPblacktext
    \GPblacktexttrue
  }{}%
  \let\gplgaddtomacro\g@addto@macro
  \gdef\gplbacktext{}%
  \gdef\gplfronttext{}%
  \makeatother
  \ifGPblacktext
    \def\colorrgb#1{}%
    \def\colorgray#1{}%
  \else
    \ifGPcolor
      \def\colorrgb#1{\color[rgb]{#1}}%
      \def\colorgray#1{\color[gray]{#1}}%
      \expandafter\def\csname LTw\endcsname{\color{white}}%
      \expandafter\def\csname LTb\endcsname{\color{black}}%
      \expandafter\def\csname LTa\endcsname{\color{black}}%
      \expandafter\def\csname LT0\endcsname{\color[rgb]{1,0,0}}%
      \expandafter\def\csname LT1\endcsname{\color[rgb]{0,1,0}}%
      \expandafter\def\csname LT2\endcsname{\color[rgb]{0,0,1}}%
      \expandafter\def\csname LT3\endcsname{\color[rgb]{1,0,1}}%
      \expandafter\def\csname LT4\endcsname{\color[rgb]{0,1,1}}%
      \expandafter\def\csname LT5\endcsname{\color[rgb]{1,1,0}}%
      \expandafter\def\csname LT6\endcsname{\color[rgb]{0,0,0}}%
      \expandafter\def\csname LT7\endcsname{\color[rgb]{1,0.3,0}}%
      \expandafter\def\csname LT8\endcsname{\color[rgb]{0.5,0.5,0.5}}%
    \else
      \def\colorrgb#1{\color{black}}%
      \def\colorgray#1{\color[gray]{#1}}%
      \expandafter\def\csname LTw\endcsname{\color{white}}%
      \expandafter\def\csname LTb\endcsname{\color{black}}%
      \expandafter\def\csname LTa\endcsname{\color{black}}%
      \expandafter\def\csname LT0\endcsname{\color{black}}%
      \expandafter\def\csname LT1\endcsname{\color{black}}%
      \expandafter\def\csname LT2\endcsname{\color{black}}%
      \expandafter\def\csname LT3\endcsname{\color{black}}%
      \expandafter\def\csname LT4\endcsname{\color{black}}%
      \expandafter\def\csname LT5\endcsname{\color{black}}%
      \expandafter\def\csname LT6\endcsname{\color{black}}%
      \expandafter\def\csname LT7\endcsname{\color{black}}%
      \expandafter\def\csname LT8\endcsname{\color{black}}%
    \fi
  \fi
  \setlength{\unitlength}{0.0500bp}%
  \begin{picture}(4520.00,3400.00)%
    \gplgaddtomacro\gplbacktext{%
      \csname LTb\endcsname%
      \put(408,794){\makebox(0,0)[r]{\strut{}0.0}}%
      \csname LTb\endcsname%
      \put(408,1192){\makebox(0,0)[r]{\strut{}1.0}}%
      \csname LTb\endcsname%
      \put(408,1590){\makebox(0,0)[r]{\strut{}2.0}}%
      \csname LTb\endcsname%
      \put(408,1988){\makebox(0,0)[r]{\strut{}3.0}}%
      \csname LTb\endcsname%
      \put(408,2385){\makebox(0,0)[r]{\strut{}4.0}}%
      \csname LTb\endcsname%
      \put(408,2783){\makebox(0,0)[r]{\strut{}5.0}}%
      \csname LTb\endcsname%
      \put(408,3181){\makebox(0,0)[r]{\strut{}6.0}}%
      \csname LTb\endcsname%
      \put(510,409){\makebox(0,0){\strut{}0.10}}%
      \csname LTb\endcsname%
      \put(1043,409){\makebox(0,0){\strut{}0.12}}%
      \csname LTb\endcsname%
      \put(1576,409){\makebox(0,0){\strut{}0.14}}%
      \csname LTb\endcsname%
      \put(2109,409){\makebox(0,0){\strut{}0.16}}%
      \csname LTb\endcsname%
      \put(2642,409){\makebox(0,0){\strut{}0.18}}%
      \csname LTb\endcsname%
      \put(3175,409){\makebox(0,0){\strut{}0.20}}%
      \csname LTb\endcsname%
      \put(3708,409){\makebox(0,0){\strut{}0.22}}%
      \csname LTb\endcsname%
      \put(4241,409){\makebox(0,0){\strut{}0.24}}%
      \csname LTb\endcsname%
      \put(162,1987){\makebox(0,0){\strut{}$\kappa$}}%
      \csname LTb\endcsname%
      \put(2509,130){\makebox(0,0){\strut{}$\lambda$}}%
      \csname LTb\endcsname%
      \put(1310,1152){\makebox(0,0){\strut{}Th ($n=0$)}}%
      \put(1310,2545){\makebox(0,0){\strut{}GN ($n=\Ni$)}}%
    }%
    \gplgaddtomacro\gplfronttext{%
    }%
    \gplbacktext
    \put(0,0){\includegraphics{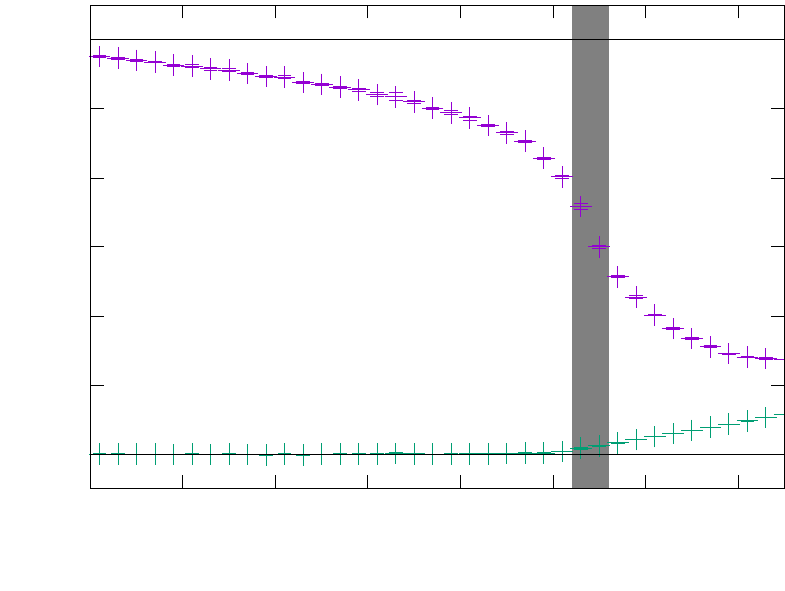}}%
    \gplfronttext
  \end{picture}%
\endgroup

%% file: Curvature_4.tex
\begingroup
\scriptsize
  \makeatletter
  \providecommand\color[2][]{%
    \GenericError{(gnuplot) \space\space\space\@spaces}{%
      Package color not loaded in conjunction with
      terminal option `colourtext'%
    }{See the gnuplot documentation for explanation.%
    }{Either use 'blacktext' in gnuplot or load the package
      color.sty in LaTeX.}%
    \renewcommand\color[2][]{}%
  }%
  \providecommand\includegraphics[2][]{%
    \GenericError{(gnuplot) \space\space\space\@spaces}{%
      Package graphicx or graphics not loaded%
    }{See the gnuplot documentation for explanation.%
    }{The gnuplot epslatex terminal needs graphicx.sty or graphics.sty.}%
    \renewcommand\includegraphics[2][]{}%
  }%
  \providecommand\rotatebox[2]{#2}%
  \@ifundefined{ifGPcolor}{%
    \newif\ifGPcolor
    \GPcolortrue
  }{}%
  \@ifundefined{ifGPblacktext}{%
    \newif\ifGPblacktext
    \GPblacktexttrue
  }{}%
  \let\gplgaddtomacro\g@addto@macro
  \gdef\gplbacktext{}%
  \gdef\gplfronttext{}%
  \makeatother
  \ifGPblacktext
    \def\colorrgb#1{}%
    \def\colorgray#1{}%
  \else
    \ifGPcolor
      \def\colorrgb#1{\color[rgb]{#1}}%
      \def\colorgray#1{\color[gray]{#1}}%
      \expandafter\def\csname LTw\endcsname{\color{white}}%
      \expandafter\def\csname LTb\endcsname{\color{black}}%
      \expandafter\def\csname LTa\endcsname{\color{black}}%
      \expandafter\def\csname LT0\endcsname{\color[rgb]{1,0,0}}%
      \expandafter\def\csname LT1\endcsname{\color[rgb]{0,1,0}}%
      \expandafter\def\csname LT2\endcsname{\color[rgb]{0,0,1}}%
      \expandafter\def\csname LT3\endcsname{\color[rgb]{1,0,1}}%
      \expandafter\def\csname LT4\endcsname{\color[rgb]{0,1,1}}%
      \expandafter\def\csname LT5\endcsname{\color[rgb]{1,1,0}}%
      \expandafter\def\csname LT6\endcsname{\color[rgb]{0,0,0}}%
      \expandafter\def\csname LT7\endcsname{\color[rgb]{1,0.3,0}}%
      \expandafter\def\csname LT8\endcsname{\color[rgb]{0.5,0.5,0.5}}%
    \else
      \def\colorrgb#1{\color{black}}%
      \def\colorgray#1{\color[gray]{#1}}%
      \expandafter\def\csname LTw\endcsname{\color{white}}%
      \expandafter\def\csname LTb\endcsname{\color{black}}%
      \expandafter\def\csname LTa\endcsname{\color{black}}%
      \expandafter\def\csname LT0\endcsname{\color{black}}%
      \expandafter\def\csname LT1\endcsname{\color{black}}%
      \expandafter\def\csname LT2\endcsname{\color{black}}%
      \expandafter\def\csname LT3\endcsname{\color{black}}%
      \expandafter\def\csname LT4\endcsname{\color{black}}%
      \expandafter\def\csname LT5\endcsname{\color{black}}%
      \expandafter\def\csname LT6\endcsname{\color{black}}%
      \expandafter\def\csname LT7\endcsname{\color{black}}%
      \expandafter\def\csname LT8\endcsname{\color{black}}%
    \fi
  \fi
  \setlength{\unitlength}{0.0500bp}%
  \begin{picture}(4520.00,3400.00)%
    \gplgaddtomacro\gplbacktext{%
      \csname LTb\endcsname%
      \put(408,769){\makebox(0,0)[r]{\strut{}0.0}}%
      \csname LTb\endcsname%
      \put(408,1117){\makebox(0,0)[r]{\strut{}1.0}}%
      \csname LTb\endcsname%
      \put(408,1465){\makebox(0,0)[r]{\strut{}2.0}}%
      \csname LTb\endcsname%
      \put(408,1813){\makebox(0,0)[r]{\strut{}3.0}}%
      \csname LTb\endcsname%
      \put(408,2162){\makebox(0,0)[r]{\strut{}4.0}}%
      \csname LTb\endcsname%
      \put(408,2510){\makebox(0,0)[r]{\strut{}5.0}}%
      \csname LTb\endcsname%
      \put(408,2858){\makebox(0,0)[r]{\strut{}6.0}}%
      \csname LTb\endcsname%
      \put(408,3206){\makebox(0,0)[r]{\strut{}7.0}}%
      \csname LTb\endcsname%
      \put(777,409){\makebox(0,0){\strut{}0.06}}%
      \csname LTb\endcsname%
      \put(1310,409){\makebox(0,0){\strut{}0.08}}%
      \csname LTb\endcsname%
      \put(1843,409){\makebox(0,0){\strut{}0.10}}%
      \csname LTb\endcsname%
      \put(2376,409){\makebox(0,0){\strut{}0.12}}%
      \csname LTb\endcsname%
      \put(2909,409){\makebox(0,0){\strut{}0.14}}%
      \csname LTb\endcsname%
      \put(3442,409){\makebox(0,0){\strut{}0.16}}%
      \csname LTb\endcsname%
      \put(3975,409){\makebox(0,0){\strut{}0.18}}%
      \csname LTb\endcsname%
      \put(4508,409){\makebox(0,0){\strut{}0.20}}%
      \csname LTb\endcsname%
      \put(162,1987){\makebox(0,0){\strut{}$\kappa$}}%
      \csname LTb\endcsname%
      \put(2509,130){\makebox(0,0){\strut{}$\lambda$}}%
      \csname LTb\endcsname%
      \put(1310,1013){\makebox(0,0){\strut{}Th ($n=0$)}}%
      \put(1310,1570){\makebox(0,0){\strut{}$n=1$}}%
      \put(1310,2266){\makebox(0,0){\strut{}GN ($n=\Ni$)}}%
    }%
    \gplgaddtomacro\gplfronttext{%
    }%
    \gplbacktext
    \put(0,0){\includegraphics{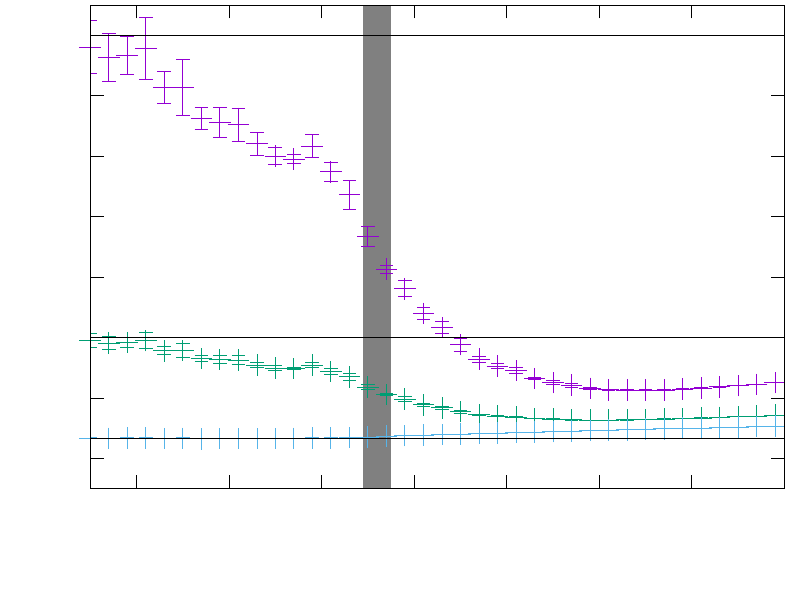}}%
    \gplfronttext
  \end{picture}%
\endgroup

%% file: CurvatureOdd.tex
\begingroup
\scriptsize
  \makeatletter
  \providecommand\color[2][]{%
    \GenericError{(gnuplot) \space\space\space\@spaces}{%
      Package color not loaded in conjunction with
      terminal option `colourtext'%
    }{See the gnuplot documentation for explanation.%
    }{Either use 'blacktext' in gnuplot or load the package
      color.sty in LaTeX.}%
    \renewcommand\color[2][]{}%
  }%
  \providecommand\includegraphics[2][]{%
    \GenericError{(gnuplot) \space\space\space\@spaces}{%
      Package graphicx or graphics not loaded%
    }{See the gnuplot documentation for explanation.%
    }{The gnuplot epslatex terminal needs graphicx.sty or graphics.sty.}%
    \renewcommand\includegraphics[2][]{}%
  }%
  \providecommand\rotatebox[2]{#2}%
  \@ifundefined{ifGPcolor}{%
    \newif\ifGPcolor
    \GPcolortrue
  }{}%
  \@ifundefined{ifGPblacktext}{%
    \newif\ifGPblacktext
    \GPblacktexttrue
  }{}%
  \let\gplgaddtomacro\g@addto@macro
  \gdef\gplbacktext{}%
  \gdef\gplfronttext{}%
  \makeatother
  \ifGPblacktext
    \def\colorrgb#1{}%
    \def\colorgray#1{}%
  \else
    \ifGPcolor
      \def\colorrgb#1{\color[rgb]{#1}}%
      \def\colorgray#1{\color[gray]{#1}}%
      \expandafter\def\csname LTw\endcsname{\color{white}}%
      \expandafter\def\csname LTb\endcsname{\color{black}}%
      \expandafter\def\csname LTa\endcsname{\color{black}}%
      \expandafter\def\csname LT0\endcsname{\color[rgb]{1,0,0}}%
      \expandafter\def\csname LT1\endcsname{\color[rgb]{0,1,0}}%
      \expandafter\def\csname LT2\endcsname{\color[rgb]{0,0,1}}%
      \expandafter\def\csname LT3\endcsname{\color[rgb]{1,0,1}}%
      \expandafter\def\csname LT4\endcsname{\color[rgb]{0,1,1}}%
      \expandafter\def\csname LT5\endcsname{\color[rgb]{1,1,0}}%
      \expandafter\def\csname LT6\endcsname{\color[rgb]{0,0,0}}%
      \expandafter\def\csname LT7\endcsname{\color[rgb]{1,0.3,0}}%
      \expandafter\def\csname LT8\endcsname{\color[rgb]{0.5,0.5,0.5}}%
    \else
      \def\colorrgb#1{\color{black}}%
      \def\colorgray#1{\color[gray]{#1}}%
      \expandafter\def\csname LTw\endcsname{\color{white}}%
      \expandafter\def\csname LTb\endcsname{\color{black}}%
      \expandafter\def\csname LTa\endcsname{\color{black}}%
      \expandafter\def\csname LT0\endcsname{\color{black}}%
      \expandafter\def\csname LT1\endcsname{\color{black}}%
      \expandafter\def\csname LT2\endcsname{\color{black}}%
      \expandafter\def\csname LT3\endcsname{\color{black}}%
      \expandafter\def\csname LT4\endcsname{\color{black}}%
      \expandafter\def\csname LT5\endcsname{\color{black}}%
      \expandafter\def\csname LT6\endcsname{\color{black}}%
      \expandafter\def\csname LT7\endcsname{\color{black}}%
      \expandafter\def\csname LT8\endcsname{\color{black}}%
    \fi
  \fi
  \setlength{\unitlength}{0.0500bp}%
  \begin{picture}(4520.00,3400.00)%
    \gplgaddtomacro\gplbacktext{%
      \csname LTb\endcsname%
      \put(408,595){\makebox(0,0)[r]{\strut{}-0.5}}%
      \csname LTb\endcsname%
      \put(408,1126){\makebox(0,0)[r]{\strut{}0.0}}%
      \csname LTb\endcsname%
      \put(408,1657){\makebox(0,0)[r]{\strut{}0.5}}%
      \csname LTb\endcsname%
      \put(408,2188){\makebox(0,0)[r]{\strut{}1.0}}%
      \csname LTb\endcsname%
      \put(408,2719){\makebox(0,0)[r]{\strut{}1.5}}%
      \csname LTb\endcsname%
      \put(408,3250){\makebox(0,0)[r]{\strut{}2.0}}%
      \csname LTb\endcsname%
      \put(510,409){\makebox(0,0){\strut{}0.00}}%
      \csname LTb\endcsname%
      \put(1279,409){\makebox(0,0){\strut{}0.05}}%
      \csname LTb\endcsname%
      \put(2048,409){\makebox(0,0){\strut{}0.10}}%
      \csname LTb\endcsname%
      \put(2817,409){\makebox(0,0){\strut{}0.15}}%
      \csname LTb\endcsname%
      \put(3585,409){\makebox(0,0){\strut{}0.20}}%
      \csname LTb\endcsname%
      \put(4354,409){\makebox(0,0){\strut{}0.25}}%
      \csname LTb\endcsname%
      \put(60,1922){\makebox(0,0){\strut{}$\kappa$}}%
      \csname LTb\endcsname%
      \put(2509,130){\makebox(0,0){\strut{}$\lambda$}}%
    }%
    \gplgaddtomacro\gplfronttext{%
      \csname LTb\endcsname%
      \put(1124,3364){\makebox(0,0)[r]{\strut{}$\Ni=3$}}%
      \csname LTb\endcsname%
      \put(1810,3364){\makebox(0,0)[r]{\strut{}$5$}}%
      \csname LTb\endcsname%
      \put(2496,3364){\makebox(0,0)[r]{\strut{}$7$}}%
      \csname LTb\endcsname%
      \put(3182,3364){\makebox(0,0)[r]{\strut{}$9$}}%
      \csname LTb\endcsname%
      \put(3868,3364){\makebox(0,0)[r]{\strut{}$11$}}%
    }%
    \gplbacktext
    \put(0,0){\includegraphics{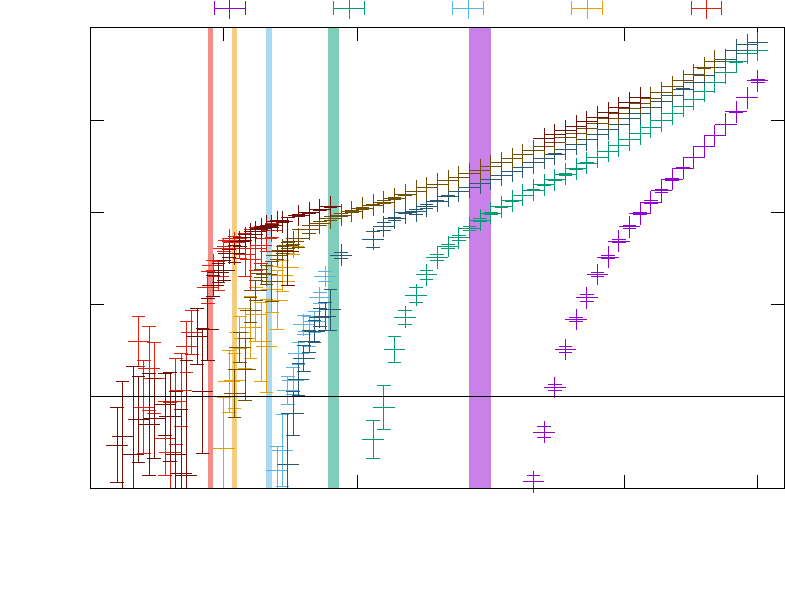}}%
    \gplfronttext
  \end{picture}%
\endgroup

%% file: phases2.tex
\begingroup
\scriptsize
  \makeatletter
  \providecommand\color[2][]{%
    \GenericError{(gnuplot) \space\space\space\@spaces}{%
      Package color not loaded in conjunction with
      terminal option `colourtext'%
    }{See the gnuplot documentation for explanation.%
    }{Either use 'blacktext' in gnuplot or load the package
      color.sty in LaTeX.}%
    \renewcommand\color[2][]{}%
  }%
  \providecommand\includegraphics[2][]{%
    \GenericError{(gnuplot) \space\space\space\@spaces}{%
      Package graphicx or graphics not loaded%
    }{See the gnuplot documentation for explanation.%
    }{The gnuplot epslatex terminal needs graphicx.sty or graphics.sty.}%
    \renewcommand\includegraphics[2][]{}%
  }%
  \providecommand\rotatebox[2]{#2}%
  \@ifundefined{ifGPcolor}{%
    \newif\ifGPcolor
    \GPcolortrue
  }{}%
  \@ifundefined{ifGPblacktext}{%
    \newif\ifGPblacktext
    \GPblacktexttrue
  }{}%
  \let\gplgaddtomacro\g@addto@macro
  \gdef\gplbacktext{}%
  \gdef\gplfronttext{}%
  \makeatother
  \ifGPblacktext
    \def\colorrgb#1{}%
    \def\colorgray#1{}%
  \else
    \ifGPcolor
      \def\colorrgb#1{\color[rgb]{#1}}%
      \def\colorgray#1{\color[gray]{#1}}%
      \expandafter\def\csname LTw\endcsname{\color{white}}%
      \expandafter\def\csname LTb\endcsname{\color{black}}%
      \expandafter\def\csname LTa\endcsname{\color{black}}%
      \expandafter\def\csname LT0\endcsname{\color[rgb]{1,0,0}}%
      \expandafter\def\csname LT1\endcsname{\color[rgb]{0,1,0}}%
      \expandafter\def\csname LT2\endcsname{\color[rgb]{0,0,1}}%
      \expandafter\def\csname LT3\endcsname{\color[rgb]{1,0,1}}%
      \expandafter\def\csname LT4\endcsname{\color[rgb]{0,1,1}}%
      \expandafter\def\csname LT5\endcsname{\color[rgb]{1,1,0}}%
      \expandafter\def\csname LT6\endcsname{\color[rgb]{0,0,0}}%
      \expandafter\def\csname LT7\endcsname{\color[rgb]{1,0.3,0}}%
      \expandafter\def\csname LT8\endcsname{\color[rgb]{0.5,0.5,0.5}}%
    \else
      \def\colorrgb#1{\color{black}}%
      \def\colorgray#1{\color[gray]{#1}}%
      \expandafter\def\csname LTw\endcsname{\color{white}}%
      \expandafter\def\csname LTb\endcsname{\color{black}}%
      \expandafter\def\csname LTa\endcsname{\color{black}}%
      \expandafter\def\csname LT0\endcsname{\color{black}}%
      \expandafter\def\csname LT1\endcsname{\color{black}}%
      \expandafter\def\csname LT2\endcsname{\color{black}}%
      \expandafter\def\csname LT3\endcsname{\color{black}}%
      \expandafter\def\csname LT4\endcsname{\color{black}}%
      \expandafter\def\csname LT5\endcsname{\color{black}}%
      \expandafter\def\csname LT6\endcsname{\color{black}}%
      \expandafter\def\csname LT7\endcsname{\color{black}}%
      \expandafter\def\csname LT8\endcsname{\color{black}}%
    \fi
  \fi
  \setlength{\unitlength}{0.0500bp}%
  \begin{picture}(4520.00,3400.00)%
    \gplgaddtomacro\gplbacktext{%
      \csname LTb\endcsname%
      \put(408,595){\makebox(0,0)[r]{\strut{} 0}}%
      \csname LTb\endcsname%
      \put(408,943){\makebox(0,0)[r]{\strut{} 0.05}}%
      \csname LTb\endcsname%
      \put(408,1291){\makebox(0,0)[r]{\strut{} 0.1}}%
      \csname LTb\endcsname%
      \put(408,1639){\makebox(0,0)[r]{\strut{} 0.15}}%
      \csname LTb\endcsname%
      \put(408,1988){\makebox(0,0)[r]{\strut{} 0.2}}%
      \csname LTb\endcsname%
      \put(408,2336){\makebox(0,0)[r]{\strut{} 0.25}}%
      \csname LTb\endcsname%
      \put(408,2684){\makebox(0,0)[r]{\strut{} 0.3}}%
      \csname LTb\endcsname%
      \put(408,3032){\makebox(0,0)[r]{\strut{} 0.35}}%
      \csname LTb\endcsname%
      \put(408,3380){\makebox(0,0)[r]{\strut{} 0.4}}%
      \csname LTb\endcsname%
      \put(510,409){\makebox(0,0){\strut{}1}}%
      \csname LTb\endcsname%
      \put(1310,409){\makebox(0,0){\strut{}3}}%
      \csname LTb\endcsname%
      \put(2109,409){\makebox(0,0){\strut{}5}}%
      \csname LTb\endcsname%
      \put(2909,409){\makebox(0,0){\strut{}7}}%
      \csname LTb\endcsname%
      \put(3708,409){\makebox(0,0){\strut{}9}}%
      \csname LTb\endcsname%
      \put(4508,409){\makebox(0,0){\strut{}11}}%
      \csname LTb\endcsname%
      \put(9,1987){\makebox(0,0){\strut{}$\lambda$}}%
      \csname LTb\endcsname%
      \put(2509,130){\makebox(0,0){\strut{}$\Ni$}}%
      \put(1709,874){\makebox(0,0){\strut{}Artefact phase}}%
      \put(3309,2545){\makebox(0,0){\strut{}Parity symmetric phase}}%
      \put(2109,1709){\makebox(0,0){\strut{}Parity broken phase}}%
    }%
    \gplgaddtomacro\gplfronttext{%
    }%
    \gplbacktext
    \put(0,0){\includegraphics{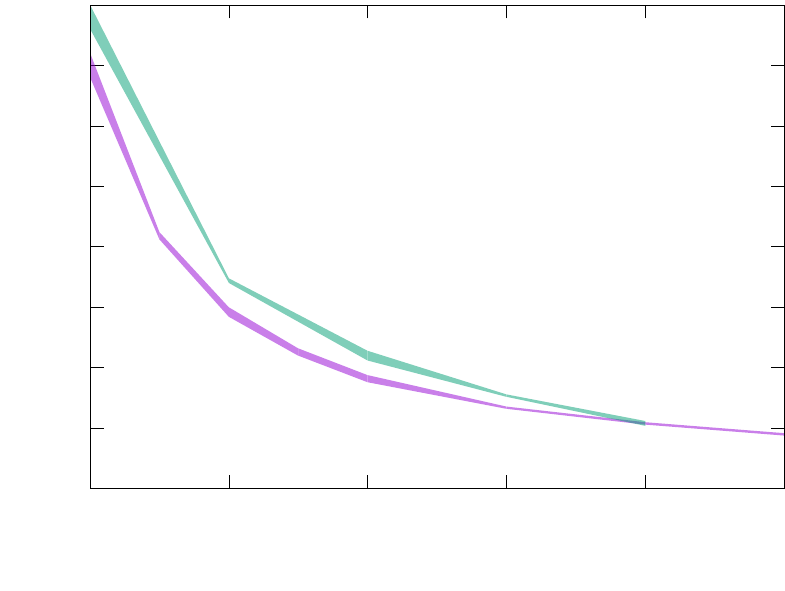}}%
    \gplfronttext
  \end{picture}%
\endgroup

%% file: ChiralCondensate.tex
\begingroup
\scriptsize
  \makeatletter
  \providecommand\color[2][]{%
    \GenericError{(gnuplot) \space\space\space\@spaces}{%
      Package color not loaded in conjunction with
      terminal option `colourtext'%
    }{See the gnuplot documentation for explanation.%
    }{Either use 'blacktext' in gnuplot or load the package
      color.sty in LaTeX.}%
    \renewcommand\color[2][]{}%
  }%
  \providecommand\includegraphics[2][]{%
    \GenericError{(gnuplot) \space\space\space\@spaces}{%
      Package graphicx or graphics not loaded%
    }{See the gnuplot documentation for explanation.%
    }{The gnuplot epslatex terminal needs graphicx.sty or graphics.sty.}%
    \renewcommand\includegraphics[2][]{}%
  }%
  \providecommand\rotatebox[2]{#2}%
  \@ifundefined{ifGPcolor}{%
    \newif\ifGPcolor
    \GPcolortrue
  }{}%
  \@ifundefined{ifGPblacktext}{%
    \newif\ifGPblacktext
    \GPblacktexttrue
  }{}%
  \let\gplgaddtomacro\g@addto@macro
  \gdef\gplbacktext{}%
  \gdef\gplfronttext{}%
  \makeatother
  \ifGPblacktext
    \def\colorrgb#1{}%
    \def\colorgray#1{}%
  \else
    \ifGPcolor
      \def\colorrgb#1{\color[rgb]{#1}}%
      \def\colorgray#1{\color[gray]{#1}}%
      \expandafter\def\csname LTw\endcsname{\color{white}}%
      \expandafter\def\csname LTb\endcsname{\color{black}}%
      \expandafter\def\csname LTa\endcsname{\color{black}}%
      \expandafter\def\csname LT0\endcsname{\color[rgb]{1,0,0}}%
      \expandafter\def\csname LT1\endcsname{\color[rgb]{0,1,0}}%
      \expandafter\def\csname LT2\endcsname{\color[rgb]{0,0,1}}%
      \expandafter\def\csname LT3\endcsname{\color[rgb]{1,0,1}}%
      \expandafter\def\csname LT4\endcsname{\color[rgb]{0,1,1}}%
      \expandafter\def\csname LT5\endcsname{\color[rgb]{1,1,0}}%
      \expandafter\def\csname LT6\endcsname{\color[rgb]{0,0,0}}%
      \expandafter\def\csname LT7\endcsname{\color[rgb]{1,0.3,0}}%
      \expandafter\def\csname LT8\endcsname{\color[rgb]{0.5,0.5,0.5}}%
    \else
      \def\colorrgb#1{\color{black}}%
      \def\colorgray#1{\color[gray]{#1}}%
      \expandafter\def\csname LTw\endcsname{\color{white}}%
      \expandafter\def\csname LTb\endcsname{\color{black}}%
      \expandafter\def\csname LTa\endcsname{\color{black}}%
      \expandafter\def\csname LT0\endcsname{\color{black}}%
      \expandafter\def\csname LT1\endcsname{\color{black}}%
      \expandafter\def\csname LT2\endcsname{\color{black}}%
      \expandafter\def\csname LT3\endcsname{\color{black}}%
      \expandafter\def\csname LT4\endcsname{\color{black}}%
      \expandafter\def\csname LT5\endcsname{\color{black}}%
      \expandafter\def\csname LT6\endcsname{\color{black}}%
      \expandafter\def\csname LT7\endcsname{\color{black}}%
      \expandafter\def\csname LT8\endcsname{\color{black}}%
    \fi
  \fi
  \setlength{\unitlength}{0.0500bp}%
  \begin{picture}(4520.00,3400.00)%
    \gplgaddtomacro\gplbacktext{%
      \csname LTb\endcsname%
      \put(408,595){\makebox(0,0)[r]{\strut{}-0.05}}%
      \csname LTb\endcsname%
      \put(408,874){\makebox(0,0)[r]{\strut{} 0}}%
      \csname LTb\endcsname%
      \put(408,1152){\makebox(0,0)[r]{\strut{} 0.05}}%
      \csname LTb\endcsname%
      \put(408,1431){\makebox(0,0)[r]{\strut{} 0.1}}%
      \csname LTb\endcsname%
      \put(408,1709){\makebox(0,0)[r]{\strut{} 0.15}}%
      \csname LTb\endcsname%
      \put(408,1988){\makebox(0,0)[r]{\strut{} 0.2}}%
      \csname LTb\endcsname%
      \put(408,2266){\makebox(0,0)[r]{\strut{} 0.25}}%
      \csname LTb\endcsname%
      \put(408,2545){\makebox(0,0)[r]{\strut{} 0.3}}%
      \csname LTb\endcsname%
      \put(408,2823){\makebox(0,0)[r]{\strut{} 0.35}}%
      \csname LTb\endcsname%
      \put(408,3101){\makebox(0,0)[r]{\strut{} 0.4}}%
      \csname LTb\endcsname%
      \put(408,3380){\makebox(0,0)[r]{\strut{} 0.45}}%
      \csname LTb\endcsname%
      \put(789,409){\makebox(0,0){\strut{}0.000}}%
      \csname LTb\endcsname%
      \put(1254,409){\makebox(0,0){\strut{}0.005}}%
      \csname LTb\endcsname%
      \put(1719,409){\makebox(0,0){\strut{}0.010}}%
      \csname LTb\endcsname%
      \put(2184,409){\makebox(0,0){\strut{}0.015}}%
      \csname LTb\endcsname%
      \put(2648,409){\makebox(0,0){\strut{}0.020}}%
      \csname LTb\endcsname%
      \put(3113,409){\makebox(0,0){\strut{}0.025}}%
      \csname LTb\endcsname%
      \put(3578,409){\makebox(0,0){\strut{}0.030}}%
      \csname LTb\endcsname%
      \put(4043,409){\makebox(0,0){\strut{}0.035}}%
      \csname LTb\endcsname%
      \put(4508,409){\makebox(0,0){\strut{}0.040}}%
      \csname LTb\endcsname%
      \put(9,2173){\makebox(0,0){\strut{}$\pi$}}%
      \csname LTb\endcsname%
      \put(2509,130){\makebox(0,0){\strut{}$\lambda_\text{R}$}}%
    }%
    \gplgaddtomacro\gplfronttext{%
      \csname LTb\endcsname%
      \put(3720,3213){\makebox(0,0)[r]{\strut{}$\Ni=3$}}%
      \csname LTb\endcsname%
      \put(3720,3027){\makebox(0,0)[r]{\strut{}$\Ni=5$}}%
      \csname LTb\endcsname%
      \put(3720,2841){\makebox(0,0)[r]{\strut{}$\Ni=7$}}%
      \csname LTb\endcsname%
      \put(3720,2655){\makebox(0,0)[r]{\strut{}$\Ni=9$}}%
      \csname LTb\endcsname%
      \put(3720,2469){\makebox(0,0)[r]{\strut{}$\Ni=11$}}%
    }%
    \gplbacktext
    \put(0,0){\includegraphics{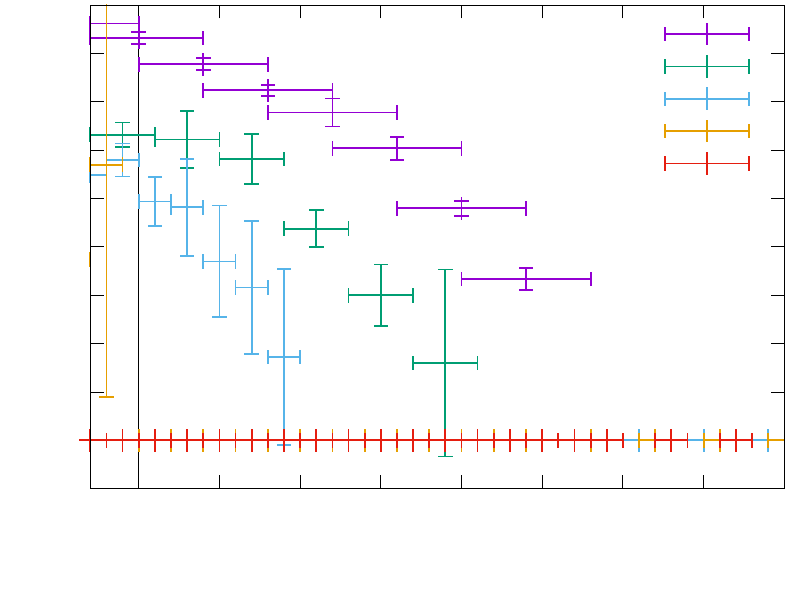}}%
    \gplfronttext
  \end{picture}%
\endgroup

%% file: conclusions.tex
\section{Conclusions and Discussion}
\label{s:conclusions}
\noindent
Our main observation is, that the irreducible Thirring model with
an odd number of irreducible flavours behaves different 
compared to the model with an even number of flavours. For massless
fermions the latter class is equivalent to the class of 
well-studied reducible models with $\Nr=2\Ni$.
While for odd flavour numbers we find a critical flavour number $\Nic=9$ below which the Thirring model shows spontaneous parity breaking, for even flavour numbers, neither parity nor chiral symmetry is broken.
This implies that \emph{no spontaneous breaking of chiral symmetry exists for 
all reducible models}, which are usually discussed in the literature.
Our earlier, more straightforward simulations with SLAC fermions 
already pointed to this result \cite{Schmidt2015,Schmidt2016}.
Furthermore our conclusions are also consistent with recent simulations 
with domain wall fermions \cite{Hands2016,Hands2017}.

Regarding the older lattice simulations with staggered fermions \cite{DelDebbio1997,Kim1996,DelDebbio1996,Christofi2007,DelDebbio1999,Hands1999}, including the fermion bag formulation \cite{Chandrasekharan2012}, their results seem not to be valid for the reducible Thirring model, likely because their lattice formulation 
does not have the correct symmetry. To see this more clearly
we should recall that for massive fermions the reducible
models are no longer equivalent to the irreducible models with $\Ni=2\Nr$
flavours. For reducible massive models $\log\det(\ii\slashed{D}+\ii m)$
is real and does not contain any imaginary Chern-Simons-type term as it does
for all irreducible massive models. After the infinite volume limit
has been taken the zero-mass limits of the reducible models are not equal to 
the zero-mass limits of the irreducible models with $\Ni=2\Nr$. Thus any 
lattice simulation (or any other regularization) which needs a 
fine-tuning to reach the chiral limit may yield erroneous results. It 
may very well happen that instead of the massless reducible model one simulates 
an irreducible model at small masses. This could be a partial explanation 
why the earlier prediction $8 < \Nic < 12$ is consistent with ours, 
but only for odd flavour numbers in the irreducible representation.

Also note, that most of the previous analytical studies focus 
on csb in the reducible representation assuming conservation of 
the reducible parity \eqref{e:red_parity} \cite{Hong1994, Itoh1995, Sugiura1997, Kondo1995} or do not distinguish between irreducible and reducible models \cite{Gomes1991, Hyun1994}.
For example, the authors of \cite{Gomes1991} use DSEs to investigate mass generation either from parity or chiral symmetry breaking and find $\Nrc\approx 12.97/D$ where $D$ is the dimension of the representation of the Clifford algebra.
These earlier results are not consistent with ours and other 
ongoing simulations with chiral fermions \cite{Hands2016,Hands2017}, 
where csb is not present at all.

Not much emphasis was put on parity breaking for odd flavour numbers in 
the irreducible representation, but most studies found a cancellation of 
the Chern-Simons terms for even $\Ni$ \cite{Hong1994, Itoh1995, Kondo1995}.
As discussed above, this is a delicate issue and the answer 
depends on the order of limits $\lim_{V\to\infty}$ and $\lim_{m\to 0}$.
We obtained our results for zero masses in a finite volume 
in which case the fermion determinant is real and no (imaginary) Chern-Simons 
term can be generated. In case one considers the Thirring model on $\mathbbm{R}^3$
then such a term can show up for even $\Ni$ \cite{Rossini1994, Ahn1994, Ahn1998}.
It also can show up if one uses a regularization which breaks chiral
symmetry explicitly.

To obtain our novel results it was essential to employ chiral fermions.
But with massless chiral fermions it seems impossible to calculate the chiral
condensates directly \cite{Schmidt2015}. The main ingredient to circumvent
this difficulty was to use both the vector and the matrix formulations
of the Thirring models. By introducing auxiliary (local) masses we could
relate expansion coefficients for effective potentials of a massless model
in the matrix formulation to expectation values of condensates 
in the vector formulation of the same model. The actual proof and 
explicit mapping from coefficients to condensates are based on a 
reformulation of the matrix models in terms of dual spin variables
$k_{xi}^{\alpha\beta}$. They are introduced to represent the result
of the integration over the fermionic variables.

Our analytic results hold for other types of chiral fermions.  
Actually, at present we replace SLAC fermions by overlap fermions in 
our simulation code to calculate the condensates related to
the coefficients of effective potentials\footnote{In collaboration
with Rajamani Narayanan.}. We expect to find comparable
results as for SLAC fermions and in particular a similar value for $\Nic$.

We already mentioned that the irreducible one-flavour model (which is
equivalent to the irreducible one-flavour Gross-Neveu model) has a
severe sign problem. We could show that in the dual formulation there
are subtle cancellations of terms such that the sign problem 
actually goes away \cite{Schmidta}.
It would be interesting to show that a similar fermion bag type
algorithm without sign problems exists for multi-flavour Thirring
models.

%

%% file: appendix.tex
\section{Fermion bag}
\label{s:fermion_bag}
\noindent

Here we present further details concerning the functional integral
in the dual formulation introduced in \autoref{ss:dual_variables}.
We summarize relevant results up to the point, where simulations with a 
fermion  bag algorithm are possible. For that purpose it 
is convenient to use the Lagrangian 
\begin{equation}
 \mathcal{L}=\bar{\psi} \left( \ii \,\slashed{\partial} +\ii \, T + \ii \, \phi \right)\psi+\frac{\lambda}{2} \tr T^2+\lambda \phi^2,
 \label{e:L_redundant}
\end{equation}
with an additional scalar field $\phi$ in place of the equivalent 
Lagrangian \eqref{e:L-Fierz}. In the formulation with scalar field the 
occurring integrals over the hermitian matrix $T$ are more readily calculated.
The equivalence of the two formulations can be seen after splitting $T$ 
into its trace free contribution and a multiple of the identity. Then 
one observes that the integration over $\tr T$ in the formulation \eqref{e:L-Fierz} 
and over $\tr T$ and $\phi$ in the formulation \eqref{e:L_redundant} yield the 
same results, up to an overall factor $\propto \lambda^{V/2}$.
The transition to dual variables is the same as for the Lagrangian
without scalar field, the only difference being that the matrix $(H^{\alpha\beta}_x)$ in the 
interaction term is now
\begin{equation}
  H^{\alpha\beta}_x = T^{\alpha\beta}_x + \phi_x\delta^{\alpha\beta} +M_x^{\alpha\beta}\,,
\end{equation}
and this gives rise to a slightly different local weight and an additional integration 
over $\phi$. Instead of \eqref{wloc1} one obtains
  \begin{equation}
  \begin{aligned}
   W_\text{loc}(k,M)=&\int \prod \limits_{i=1}^{{\Ni}^2}\left(\frac{\dx t_{i}}{\sqrt{\pi }}\right)
   \int \dx\phi
   \,\ex{-\frac{1}{2} \tr T^2 - \phi^2} \\
   &\cdot\prod \limits_{\alpha\beta}\left(T^{\alpha\beta}+ \phi\,\delta^{\alpha\beta}+M^{\alpha\beta}\right)^{k^{\alpha\beta}}\,,
  \end{aligned}
  \end{equation}
up to an irrelevant overall factor $\propto \sqrt{\lambda}$. Since $T$ is hermitian,
the exponential function factorizes as follows:
\begin{equation}
  \ex{-\frac{1}{2}\tr T^2} = 
  \prod_{\alpha<\beta} \ex{-\left| T^{\alpha\beta}\right|^2} \prod_\alpha \ex{-(T^{\alpha\alpha})^2}\,.
\end{equation}
It implies the following factorization of the local weight,
\begin{equation}
  W_\text{loc}(k,  M) = \int \dx\phi\, \ex{-\phi^2} W_{\vecf{p}(k)}(\phi, M)
  w_\text{o}(k_\perp)\,,\label{fact_lw}
\end{equation}
where the integral over the off-diagonal matrix elements produces the $M$- 
and $\phi$-independent factor
\begin{equation}
w_\text{o}(k_\perp)=\prod_{\alpha < \beta} W_\perp(k^{\alpha\beta},k^{\beta\alpha})\,.
\end{equation}
The function $W_\perp$ is determined by a complex Gaussian integral
\begin{equation}
  W_\perp(k, k') = \intop \frac{\dx z \dx \bar z}{\pi}\, \ex{-|z|^2} z^{k} (z^*)^{k'}
  =k!\,\delta_{kk'}
\end{equation}
leading to the local constraints $k^{\alpha\beta} = k^{\beta\alpha}$.
Thus we obtain
\begin{equation}
w_\text{o}(k_\perp)=\prod_{\alpha<\beta} (k^{\alpha\beta})\,!
\end{equation}
Recall that the entries of the symmetric matrix $(k^{\alpha\beta})$ must obey the
local constraints in \eqref{constr2}.

The integral over the Cartan variables leads to a term similar to \eqref{e:wnkm} and 
is given by
 \begin{equation}
 \begin{aligned}
  W_{\vecf{p}(k)}(\phi,M)=& \int \prod \limits_{i=1}^{{\Ni}}
  \left(\frac{\dx t_{i}}{\sqrt{\pi }}\right) 
  \,e^{-t A t} \\
  &\cdot \prod \limits_{\alpha}\left(T^{\alpha\alpha}+\phi + m_\alpha\right)^{p_\alpha(k)}
 \end{aligned}
 \end{equation}
 with $p_\alpha(k)=k^{\alpha\alpha}\in\{0,1,2\}$. Note that the diagonal
 matrix elements $T^{\alpha\alpha}$
are linear functions in the integration variables $t_i$.
The  symmetric positive matrix $A$ has matrix elements
\begin{equation}
  A_{ij}=\frac{1}{2} \tr \left(H_i H_j\right)
 \end{equation}
and differs slightly from the matrix \eqref{ee:wnkm} in the formulation without
 scalar field $\phi$. 
 The normalization of the generators $H_i$ is such that
 \begin{equation}
 \tr (H_i H_j)  = 
  \begin{pmatrix}
  \frac{4}{\Ni+2} &  0 & 0 & 0 & \dots & 0\\
  0 &  2 & -1 & 0 & \dots &  0\\
  0 & -1 & 2 & -1 & \ddots & \vdots\\
  \vdots &  \ddots & \ddots & \ddots & \ddots & 0 \\
  0 &  \dots &  0& -1 & 2 & -1 \\
  0 &  \dots &  0&  0& -1 & 2 \\
  \end{pmatrix}.
\end{equation}
 The final integration over the variables $t_i$ yields
 the factor
 \begin{equation}
 W_{\vecf{p}}(\phi, M)=\hskip-1.5mm \prod_{\alpha:p_\alpha=1}\hskip-1mm(\phi+m_\alpha)
 \!\!\!\prod_{\alpha:p_\alpha=2}\hskip-1mm\left(1+(\phi+m_\alpha)^2\right).
 \end{equation}
In the limit of $m\to0$, the  $\phi$-integral \eqref{fact_lw} vanishes for odd $P_1$
and for even $P_1$  is given by the confluent hypergeometric function,
\begin{equation}
W_\text{loc}(k)=w_\text{o}(k_\perp) \,
 \Gamma\Big(\frac{1\!+\!P_1}{2}\Big)\; U\Big(\frac{1\!+\!P_1}{2},\frac{3\!+\!P_1}{2}+P_2,1\Big),
\end{equation}
where, as in the main body of the text, the number $P_k$ with $k\in\{0,1,2\}$ 
counts the number of indices $p_\alpha$ with $p_\alpha=k$.
In this form, the local weights are suitable for simulations 
with a fermion bag algorithm. Unfortunately, this formulation does 
not solve the sign problem that was introduced by the Fierz transformation.

\section{Strong coupling expansion}
\label{s:strong_details}
\noindent
The partition function in the vector formulation of the Thirring model with fermionic sources is given by \eqref{e:strongCouplingPartition}.
Here we perform the integration over the vector field $v_\mu$ and the fermions. 
After rescaling the vector field according to $\tilde{v}_\mu=\sqrt{\lambda}v_\mu$ 
(and afterwards dropping the tilde)
the integration over the fermions yields
\begin{equation}
\begin{aligned}
Z[\eta,\bar{\eta}]&=\left(\lambda\right)^{-(3/2+\Ni) V}K\left[\frac{\delta}{\delta \eta},\frac{\delta}{\delta \bar{\eta}}\right]\\
&\;\;\int \mathcal{D} v \,\left(\det \slashed{v}_x\right)^{\Ni} e^{-\sum\limits_x v^2_x} e^{\sum \limits_x \,\sqrt{\lambda}\, \bar{\eta}_x\frac{1}{\slashed{v}_x}\eta_x}\,.
\end{aligned}
\end{equation}
The integral over the vector field factorizes and we can expand 
in powers of the fermionic bilinear in the exponent,
\begin{equation}
\begin{aligned}
 Z[\eta,\bar{\eta}]=&\left(\lambda\right)^{-(3/2+\Ni) V} K\left[\frac{\delta}{\delta \eta},\frac{\delta}{\delta \bar{\eta}}\right] \\ 
 &\int \mathcal{D}v\, e^{-\sum\limits_x\left(v_x^2+\Ni \,\ln v_x^2\right)} 
 e^{\sqrt{\lambda}\sum \limits_x \bar{\eta}_x\slashed{v}_x\eta_x/v_x^2}\\
 =&\left(\lambda\right)^{-(\frac{3}{2}+\Ni) V} K\left[\frac{\delta}{\delta \eta},\frac{\delta}{\delta \bar{\eta}}\right] \prod \limits_x \sum \limits_{k=0}^{\Ni} I_k(\eta_x,\bar{\eta}_x),
\end{aligned}
\end{equation}
where $I_k$ is the $3$-dimensional one-site integral
\begin{equation}
 I_k(\eta,\bar{\eta})=\frac{\lambda^k}{(2k)!} \int \dx^3 v\, e^{-v^2} (v^2)^{\Ni-2 k} \left(\bar{\eta}\,\slashed{v}\,\eta\right)^{2k}.
\end{equation}
Here, we already used that only even powers of $\slashed{v}$ contribute to 
the integrals over $v_\mu$.
In spherical coordinates $v=r\hat v$, the integration over the radial direction is simple and we obtain
\begin{equation}
\begin{aligned}
 I_k(\eta,\bar{\eta})=&\frac{\lambda^k}{(2k)!} \int \dx r \, \dx^2 \hat{v}\, e^{-r^2} {r}^{2(\Ni+1-k)} \left(\bar{\eta}\,\slashed{\hat v}\,\eta\right)^{2k}\\
 =&\frac{\lambda^k \Gamma\mleft(\frac{3}{2}+\Ni-k\mright)}{2(2k)!} \int_{S^2} \dx^2 \hat{v}\left(\bar{\eta}\,\slashed{\hat v}\,\eta\right)^{2k}
\end{aligned}
 \end{equation}
 The remaining integral can be calculated from the generating function
 \begin{equation}
  z(j)=\int \dx^2 \hat v \, e^{\,\sum j_\mu\hat v_\mu}=4\pi \frac{\sinh \abs{j}}{\abs{j}},
  \quad j\in \mathbbm{R}^3\,,
 \end{equation}
 by taking derivatives. This way we find
 \begin{equation}
 \begin{aligned}
  \int \dx^2 \hat v\left(\bar{\eta}\gamma_\mu\eta\,\hat v_\mu\right)^{2k}
  &=\Bigg(\bar\eta\gamma_\mu\eta\,\frac{\partial}{\partial j_{\mu}}\Bigg)^{2k} \,z(j)\Big\vert_{j=0}\\
  &=\frac{4 \pi}{2k+1}\left((\bar{\eta}\,\gamma_\mu\,\eta)^2\right)^k.
 \end{aligned}
 \end{equation}
 Hence, the integral over the vector field yields
 \begin{equation}
  I_k(\eta,\bar{\eta})=2\pi \frac{\lambda^k \, \Gamma(\frac{3}{2}+\Ni-k)}{\Gamma(2k+2)}\left((\bar{\eta}\,\gamma_\mu\,\eta)^2\right)^k
 \end{equation}
 such that the final expression for the partition function after integration over 
 vector and fermion fields is given by \eqref{e:strongCouplingPartition1}.
 
 In the main body of the paper, just after equation \eqref{e:strongCouplingPartition2}, 
 we remark that in the infinite volume limit the partition function only has a trivial dependence on the inverse coupling $\lambda$ and that local expectation
 values do not depend on $\lambda$ at all. In order to show this, we write the $n$-th 
 order in the corresponding expansion as a sum over the configurations 
 $\check{k}=\{k_x\vert x\neq x_0\}$, where $k_x$ is the order of the function $F^{(k_x)}(x)$. We obtain for the partition function
 \begin{align}
  Z^{(n)}[\chi_{x_0},\bar{\chi}_{x_0}]&=K^{(n)} \prod \limits_x 
  \sum \limits_{k_x} F^{(k_x)}(x) \Big\vert_{\chi_{x \neq x_0}=0}\\
  &\hskip-12mm =\,\sum\limits_{\check{k}}K^{(n)} \prod \limits_{x\neq x_0} F^{(k_x)}(x) 
  \sum \limits_{k_{x_0}} F^{(k_{x_0})}(x_0)\Big\vert_{\chi_{x \neq x_0}=0}\,,\nonumber
 \end{align}
 where a particular the point $x=x_0$ was singled out, because we later differentiate
the partition function with respect to the source at this point.
 The operator $K^{(n)}$ contains $(n,n)$ derivatives with respect to $(\chi,\bar{\chi})$ at all lattice points while the function $F^{(k)}(x)$ contains $(2k,2k)$ fermion sources $(\chi,\bar{\chi})$ at the lattice point $x$. Symbolically, we introduce the operator $D^{(i,i)}$, that collects $(i,i)$ derivatives together with the $2i$ sums over the lattice points. This allows us to write the partition function as
 \begin{equation}
  Z^{(n)}[\chi_{x_0},\bar{\chi}_{x_0}]=\sum \limits_{i=1}^n A^{(i)} B^{(i)} 
 \end{equation}
 with the functions $A$ and $B$ defined as
 \begin{equation}
 \begin{aligned}
  A^{(i)}=&\Bigg(D^{(i,i)} \sum \limits_{\check{k}}\prod \limits_{x\neq x_0} F^{(k_x)}(x)\Bigg)\Bigg\vert_{\chi=0}\,,\\
  B^{(i)}=&D^{(n-i,n-i)}\sum \limits_{k_{x_0}} F^{(k_{x_0})}(x_0).
 \end{aligned}
 \end{equation}
 To investigate the volume dependence of the $A$ functions we act with $i$ derivatives on the $F$ functions and afterwards set the sources to zero. Only terms 
 with $2\sum k_x=i$ yield a non-vanishing contribution to the partition function. Furthermore, we need more than one lattice point, because the massless inverse fermion propagator vanishes for $x=y$ (this is true for SLAC-fermions on lattices with 
 even $L$).
  Therefore, the number $n_x$ of lattice points with sources in the product of the $F$ function is $n_x=2\,, \dots\,, i/2$. For the first lattice point, we have $V-1$ possibilities, for the second lattice point $V-2$ etc.
 Thus the volume dependence of $A^{(i)}$ is 
\begin{equation}
   A^{(i)} \sim \sum_{n_x=2}^{i/2} a_{n_x} \binom{V-1}{n_x}
   \underset{V\gg i}{\longrightarrow} \sum_{n_x=2}^{i/2}a_{n_x} \frac{V^{n_x}}{n_x!}.
\end{equation}
We conclude that the dominant contribution to the partition function in the 
infinite volume limit and for a fixed order of the expansion is
 \begin{equation}
  Z^{(n)}[\chi_{x_0},\bar{\chi}_{x_0}]=A^{(n)} B^{(n)} \sim C(\lambda) \sum \limits_k F^{(k)}(x_0).
 \end{equation}
 This leads to the form of the partition function given in \eqref{e:strongCouplingPartition3}.